%% file: main.tex
\documentclass[preprint,times,12pt]{elsarticle}
\usepackage{graphicx} 
\usepackage{caption}
\usepackage{algpseudocode}
\usepackage{algorithm}
\usepackage{xcolor}
\usepackage{soul}
\usepackage{amsmath}
\usepackage{gensymb}
\usepackage{lineno}
\usepackage{bm}
\usepackage{physics}
\usepackage{booktabs}
\usepackage{subcaption}
\usepackage{nicefrac}
\usepackage{tabularx}
\usepackage{enumitem}
\usepackage{siunitx}
\DeclareSIUnit{\bpm}{bpm}
\usepackage{pgfplots}
\pgfplotsset{compat=1.18}
\usepgfplotslibrary{groupplots}
\usetikzlibrary{decorations.pathmorphing}
\usetikzlibrary{calc,patterns,matrix,shadings,fadings,shadows,positioning}
\usetikzlibrary{spy}
\journal{Journal of Computational Physics}

\input{aux_plots}

\begin{document}

\begin{frontmatter}
\title{A fast and consistent sharp-interface immersed boundary method for moving bodies of arbitrary thickness}
\author[1]{Giovanni Vagnoli}
\author[1]{Martino A. Scarpolini}
\author[1,2]{Roberto Verzicco}        
\author[1]{Francesco Viola\corref{cor1}}
\ead{francesco.viola@gssi.it}
\cortext[cor1]{Corresponding author}
\affiliation[1]{organization={Gran Sasso Science Institute (GSSI)},
	addressline={Viale Luigi Rendina, 26}, 
	city={L'Aquila},
	postcode={67100}, 
	country={Italy}}
\affiliation[2]{organization={Department of Physics, Mesa+ Institute, and J. M. Burgers Centre for Fluid Dynamics, University of Twente},
	addressline={AE Enschede}, 
	city={Twente},
	postcode={7500}, 
	country={The Netherlands}}
    
\begin{abstract}
\input{abstract}
\end{abstract}


\end{frontmatter}


\input{body}

\section{Acknowledgements}
This project has received funding from the European Research Council (ERC) under the European Union’s Horizon Europe research and innovation program
(Grant No. 101039657, CARDIOTRIALS to F.V.).

\section{Declaration of Competing Interest}
The authors declare that they have no known competing financial interests or personal relationships that could have appeared to influence the work reported in this document

\appendix
\input{appendix}

\bibliographystyle{elsarticle-num} 
\bibliography{refs}
\end{document}

%% file: aux_plots.tex
\definecolor{mycolor1}{rgb}{0.34667,0.53600,0.69067}%
\definecolor{mycolor2}{rgb}{0.91529,0.28157,0.28784}%
\definecolor{mycolor3}{rgb}{0.44157,0.74902,0.43216}%
\definecolor{mycolor4}{rgb}{1.00000,0.59843,0.20000}%

\definecolor{region1}{rgb}{0.231373, 0.298039 , 0.752941}%
\definecolor{region2}{rgb}{0.865003, 0.865003 , 0.865003}%
\definecolor{region3}{rgb}{0.705882, 0.0156863, 0.14902 }%

%% file: abstract.tex
Immersed boundary methods (IBMs) are widely used to simulate flows around complex geometries and moving bodies, but they often involve a trade-off between precision and computational efficiency.
Eulerian formulations require special treatments for moving walls and may generate spurious force oscillations, whereas Lagrangian formulations can suffer from slip errors at the immersed surfaces.
We propose a novel sharp-interface IBM for incompressible flows involving moving, deformable, and arbitrary-thickness bodies.
The method combines a fast tagging algorithm, a two-sided Eulerian forcing strategy, and a consistent mass correction that reduces the splitting error of fractional-step schemes, while preserving the structure of the discrete Laplacian operator.
This formulation retains the efficiency of direct Poisson solvers, thus avoiding the overhead of cut-cell, multigrid, and projection-based approaches.
The method naturally handles moving boundaries, and yields small transpiration errors with second-order accuracy in the enforcement of the no-slip condition.
Numerical tests using rigid, deformable, turbulent, and biologically inspired flows demonstrate the accuracy, robustness, and efficiency of the method, without compromising computational cost.

%% file: body.tex
\section{Introduction}
The immersed boundary method (IBM) has become a widely adopted approach for the analysis of complex flow configurations \cite{mittalImmersedBoundaryMethods2005,verziccoImmersedBoundaryMethods2023,verziccoIntroductionImmersedBoundary2025}, and it is extensively applied to biological \cite{violaHighfidelityModelHuman2023} and fluid-structure interaction (FSI) problems \cite{huangThreedimensionalSimulationFlapping2010}, due to its capability to handle moving boundaries.
The core idea of IBMs is the discretization of the Navier-Stokes equation on a grid that is not conformal to the solid body, and the no-slip condition is enforced on the immersed solid surface using ad-hoc volume forces that mimic the effects of the wall on the flow.
Typically, IMBs are coupled with the fractional step method and low-order finite difference operators, allowing the use of highly-efficient numerical solvers, such as the approximate factorisation of the implicit viscous term and fast Poisson solver, based on the fast Fourier transform (FFT), to enforce the incompressibility constraint \cite{kimApplicationFractionalstepMethod1985, zhuAFiDGPUVersatileNavier2018}.

Multiple forcing strategies have been proposed to impose the no-slip condition on solid walls, but among many, two main families have emerged.
The first one encompasses \emph{Eulerian} methods, in which the momentum forcing is evaluated on the fluid grid using the velocity boundary condition and the surrounding bulk values \cite{fadlunCombinedImmersedBoundaryFiniteDifference2000,tsengGhostcellImmersedBoundary2003, mittalVersatileSharpInterface2008}.
In this approach, the shape of the wall is represented sharply and the boundary condition on the wet side(s) is enforced with the same order of accuracy as the interpolator used to reconstruct the velocity field near the wall.
It should be noted that in presence of moving boundaries, these methods require additional treatment to handle cells that change side of the interface (so-called \emph{fresh cells} \cite{mittalVersatileSharpInterface2008}), in order to evaluate properly the non-linear terms.
Furthermore, the occurrence of spurious pressure and viscous oscillations on solid walls produces non-smooth hydrodynamic loads acting on solid bodies \cite{seoSharpinterfaceImmersedBoundary2011}, hence limiting their application to FSI problems.
Moreover, Eulerian methods are typically tailored for bodies enclosing a finite solid volume, as the treatment of the fluid-solid interface often requires the modification of the flow field inside the solid region; this makes difficult to tackle zero-thickness bodies like membranes, shells or thin plates.

The other family of forcings is represented by \emph{Lagrangian} methods \cite{peskinImmersedBoundaryMethod2002}, where the former are evaluated on a set of points located on the solid walls (called Lagrangian markers).
The Lagrangian forcing is then distributed to the surrounding fluid points using interpolator and spreading operators, built upon proper transfer functions \cite{uhlmannImmersedBoundaryMethod2005, vanellaMovingleastsquaresReconstructionEmbeddedboundary2009}, resulting in a diffusion of the solid interface among a thin layer of Eulerian points.
Such smearing of the fluid-solid interface allows to deal with moving walls without any special treatments of the boundary points, but it may produce large slip errors \cite{gsellExplicitViscosityindependentImmersedboundary2019} because: (i) the spreading and interpolation operators are not reciprocal, and (ii) the support regions associated with different Lagrangian markers overlap, thus perturbing each other \cite{vagnoliLocalExplicitForcing2025,capuanoAnalysisSplittingErrors2025}.
To address these issues, implicit formulations have been proposed \cite{suImmersedBoundaryTechnique2007}, enabling in principle the imposition of the no-slip condition on immersed surfaces up to machine precision.
However, such approaches significantly increase the computational cost and lead to poorly conditioned linear systems, making three-dimensional simulations computationally prohibitive \cite{vagnoliLocalExplicitForcing2025}.

In addition, when applied in the framework of the fractional step method \cite{kimApplicationFractionalstepMethod1985}, IBMs suffer from the \emph{splitting error} \cite{capuanoAnalysisSplittingErrors2025}.
This stems from the fact that the elliptic equation arising from the enforcement of mass conservation is typically solved as if the solid body were absent.
Hence, mass conservation is enforced also in the cells intersected by the immersed surface (so-called \emph{boundary cells}), thus, the velocity at the nodes forced by the IBM is altered by the solenoidal projection.
Consequently, the enforcement of the no-slip condition on solid walls may be inaccurate, particularly in presence of strong pressure gradients \cite{yildiranPressureBoundaryConditions2024}.
To mitigate this issue, ad hoc mass sources can be introduced near the wall within the flow field to improve mass conservation on the boundary cells \cite{kimImmersedBoundaryFiniteVolumeMethod2001}.
Similarly, \emph{cut-cells} methods \cite{mittalVersatileSharpInterface2008} modify the shape of the boundary cells accordingly to the wall shape to evaluate the mass flux using a body conformal cell.
However, such methods entails a noteworthy overhead to determine the shape of the cell near the wall.
Moreover, the structure of the discrete Laplacian operator in the elliptic equation is modified, hence preventing the use of efficient FFT-based direct solvers in favor of less efficient iterative multigrid methods, which may lead to slow convergence in complex flow configurations.

As an alternative, a homogeneous Neumann boundary condition for the pseudo-pressure on the immersed surface can be embedded in the elliptic problem to reduce the modification of the provisional velocity near the wall \cite{heNumericalSimulationInteraction2022}.
Nevertheless, also this condition alters the system matrix of the elliptic equation, requiring multigrid or other iterative methods.
Furthermore, imposing a vanishing normal derivative of the pseudo-pressure, only preserves the normal component of the velocity, while perturbing the tangential ones.
Finally, in projection-based approaches \cite{tairaImmersedBoundaryMethod2007}, the immersed boundary forcing is interpreted as a Lagrange multiplier, enforcing the no-slip constraint on the immersed surfaces, hence also avoiding the splitting error.
However, the resulting system of equations to be solved at each time step is significantly larger than the elliptic equation in the standard fractional-step method, thus making such approaches computationally prohibitive for three-dimensional cases.

In this work, we propose a fast and consistent IB formulation that circumvents all the above limitations of traditional IBMs.
The method is built using a two-sided sharp Eulerian treatment of the interface along with a parallel tagging algorithm to identify boundary cells.
Furthermore, in order to satisfy numerical consistency in the boundary cells, a modified elliptic equation is obtained, which reduces splitting error and preserves the structure of the Laplacian operator.
The proposed IB forcing naturally accommodates moving boundaries and zero-thickness structures, without additional complexity.
Moving-least squares (MLS) kernels \cite{liuIntroductionMeshfreeMethods2010} are then employed for the smooth evaluation of the hydrodynamic loads acting on the solid wall, reducing spurious oscillations and enabling the investigation of FSI problems.

The rest of the paper is organized as follows: in Section~\ref{sec:method} the details of the underlying numerical solver are introduced, and the novel IB formulation is detailed in Section~\ref{sec:newIB}.
In Section~\ref{sec:numerical_exp}, the method is tested against several benchmarks including fixed domain geometries, rigid body dynamics and FSI.
The numerical tests are designed to demonstrate and quantify the precision of the method, especially in the enforcement of the no-sip condition on pressurized moving walls with zero-thickness.
Conclusions are drawn in Section~\ref{sec:conclusion}.

\section{Numerical method: fractional step with a provisional IB step}
\label{sec:method}
\begin{figure}
    \centering
    \input{_figures/general/problem_and_grid}
    \caption{a) Sketch of the three-dimensional computational domain $\Omega$ and of its spatial discretization.
    b) Magnification of a computational cell with the staggered location of the flow quantities.}
    \label{fig:problem_and_grid}
\end{figure}
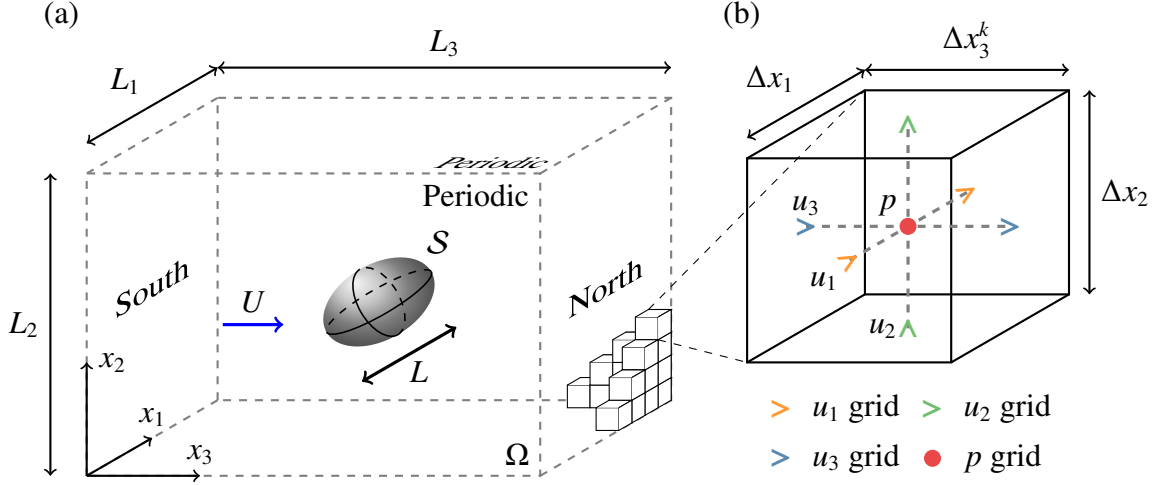
The proposed IB method is incorporated within a standard fractional-step framework, with the key distinction, relatively to existing methods in the literature, that the IB forcing step is performed before the evaluation of the provisional velocity.
As detailed later on, this modification turns out to be useful to treat fresh cells occurring in moving boundary problems.

Let us consider the incompressible Navier-Stokes equations in non-dimensional form:
\begin{subequations}\label{eq:ns}
    \begin{align}
        \pdv{\vb{u}}{t} + \div{\qty(\vb{u}\vb{u}}) &= -\grad{p} + \frac{1}{Re}\laplacian{\vb{u}} + \vb{f},\label{eq:ns1}\\
        \div{\vb{u}} &= 0,\label{eq:ns2}
	\end{align}
\end{subequations}
where $\vb{u}=\qty[u_1, u_2, u_3]^T$ and $p$ are velocity and pressure fields, $\vb{f}$ contains the external volume forces (including the IB forcing), $Re=\rho UL/\mu$ is the Reynolds number based on density $\rho$, dynamic viscosity $\mu$ of the fluid, a reference length $L$ and a reference velocity $U$.
The system of equation is completed by an initial condition $\vb{u} = \vb{u}_0$ at initial time $t = 0$, suitable far-field conditions and the no-slip condition $\vb{u} = \vb{U}\qty(\vb{X}\qty(t))$ on each solid wall $\mathcal{S}$, being $\vb{U}$ the wall velocity depending on the instantaneous boundary position $\vb{X}\qty(t)$.

The computational domain is a box $\Omega$ of size $L_1\times L_2 \times L_3$, partitioned into $N_1 \times N_2 \times N_3$ cells (Figure~\ref{fig:problem_and_grid}a).
In the periodic $x_1$ and $x_2$ directions the cells have uniform spacing $\Delta x_1$ and $\Delta x_2$, while a variable cell size $\Delta x_3^k$, $k\in \qty[1,N_3]$, is used in the $x_3$ direction.
The discrete flow variables are arranged on a staggered grid: the pressure $p$ (as well as all scalars) is located on the centers of the cells, while the velocity components $u_1$, $u_2$ and $u_3$ are located on the cell faces (see Figure~\ref{fig:problem_and_grid}b).
Second-order accurate centered finite-difference schemes are employed to discretize the spatial differential operators.
At the south wall, a Dirichlet boundary condition is applied on $\vb{u}$, while on the north wall a radiative or Dirichlet boundary condition is prescribed.

The time marching is performed using a fractional-step method, possibly with multiple sub-stages within each time step \cite{kimApplicationFractionalstepMethod1985}.
The temporal index is denoted by $n$, while $s \in \qty[1,n_s]$ represents the sub-stage index, with $n_s$ denoting the total number of sub-stages per time step.
In the following, the Adams-Bashfort (AB) multi-step method ($n_s=1$) and a two-stage Runge-Kutta (RK2) scheme ($n_s=2$) are used to advance the non-linear terms, while the viscous terms are advanced using the Crank-Nicholson (CN) scheme.

Each sub-stage starts with the identification of the \emph{forcing nodes}, i.e., the set of Eulerian points on which the IB forcing is applied.
Only on the forcing nodes, the provisional velocity $\hat{\vb{u}}$ is determined by the IB forcing as follows:
\begin{equation}
    \label{eq:IB_step}
    \hat{\vb{u}} = \vb{u}^{s} + \vb{f}\qty[\vb{u}^s, \vb{U}^n]\Delta t^n \qq{on the forcing nodes,}
\end{equation}
where $\Delta t^n$ is the current time-step size, while $\vb{f}\qty[\vb{u}^s, \vb{U}^n] = \qty(\vb{u}^f\qty[\vb{u}^s, \vb{U}^n]- \vb{u}^{s})/\Delta t^n$ denotes the IB forcing, being $\vb{u}^f\qty[\vb{u}^s, \vb{U}^n]$ the velocity imposed on the forcing nodes evaluated from the fluid velocity at the previous sub-stage, $\vb{u}^s$, and the wall velocity at the previous time step, $\vb{U}^n$, as detailed is Section~\ref{sec:newIB_meth}.
Anticipating the IB step before the evaluation of the momentum equation, enables a convenient treatment of fresh cells and improves the precision of the velocity gradients near the wall, as it will be detailed in Section~\ref{sec:newIB_meth} (see also Figures~\ref{fig:stencil} and \ref{fig:fresh_cell}).

On the remaining nodes, denoted \emph{fluid nodes}, the provisional velocity is obtained from the momentum equation:
\begin{equation}
    \label{eq:NS_prov}
    \frac{\hat{\vb{u}} - \vb{u}^s}{\Delta t^n} = \gamma^s \vb{H}^s + \rho^s \vb{H}^{s-1} - \alpha^s Gp^s+ \frac{\alpha^s}{2 Re}L\qty(\hat{\vb{u}} + \vb{u}^s) \qq{on the fluid nodes,}
\end{equation}
where $\rho^s$, $\gamma^s$ and $\alpha^s = \rho^s + \gamma^s$ are the coefficients of the time advancement scheme.
$\vb{H}$ indicates the non-linear terms written in conservative form, while $G$ and $L$ are the finite differences gradient and Laplacian operators.
For the AB+CN scheme, $s=1$ and $s=0$ correspond to $n$ and $n-1$, respectively, with $\gamma^1 = \nicefrac{3}{2}$ and $\rho^1 = -\nicefrac{1}{2}$.
On the other hand, $\gamma^1 = \nicefrac{1}{2}$, $\gamma^2 = 1$, $\rho^1 = 0$ and $\rho^2 = -\nicefrac{1}{2}$ are used for the RK2+CN scheme.
The provisional equation is solved using an approximate factorization of $L$, yielding three tridiagonal systems per-direction that are solved using the Thomas algorithm \cite{kimApplicationFractionalstepMethod1985, zhuAFiDGPUVersatileNavier2018}.

To satisfy mass conservation, $\hat{\vb{u}}$ is projected onto a solenoidal space by means of the pseudo-pressure $\varphi$ as follows:
\begin{equation}
    \label{eq:projection}
    \vb{u}^{s+1} = \hat{\vb{u}} - \alpha^s\Delta t^n G\varphi, 
\end{equation}
being $\vb{u}^{s+1}$ the divergence-free velocity field.
Applying the finite difference divergence operator $D$ to Equation~\eqref{eq:projection} and imposing the discrete divergence-free condition $D\vb{u}^{s+1} = 0$, the following elliptic equation for the pseudo-pressure is obtained:
\begin{equation}
    \label{eq:elliptic}
    L_c\varphi = \frac{1}{\alpha^s \Delta t^n}D\hat{\vb{u}},
\end{equation}
where $L_c = DG$ is the discrete Laplacian operator on the pressure grid with the homogeneous Neumann condition on all the sides of the computational box.
The linear system of Equations~\eqref{eq:elliptic} is solved using FFTs along the periodic directions, and the Thomas algorithm to invert the resulting tridiagonal matrix in the non-periodic direction \cite{kimApplicationFractionalstepMethod1985,zhuAFiDGPUVersatileNavier2018}.
Finally, the pressure field is updated as:
\begin{equation}
    \label{eq:update_pressure}
    p^{s+1} = p^s + \varphi - \frac{\alpha^s \Delta t^n}{2Re}L_c\varphi.
\end{equation}

\section{A fast and consistent sharp-interface immersed boundary method}
\label{sec:newIB}
\begin{algorithm}[t]
\caption{The tagging algorithm}\label{alg:tagging}
\begin{algorithmic}[1]
    \Require Body surface discretized into $N_t$ triangular elements;
    \Require One of the three staggered grids of the velocity;
    \ForAll{triangles with vertices $\qty{\vb{v}^1,\vb{v}^2,\vb{v}^3}$}
        \State \parbox[t]{0.85\linewidth}{%
        Build the Cartesian bounding box:\\
        $\mathcal{B} = \qty{\vb{x} \bigg| \min{\qty{v_i^1, v_i^2, v_i^3}} - \Delta x_i< x_i < \max{\qty{v_i^1, v_i^2, v_i^3}} + \Delta x_i, i = 1,2,3}$;
        }
        \State Identify the $N_e$ Eulerian points within the bounding box;
        \ForAll{Eulerian point $\vb{x}^k$, $k\in \qty[1,N_e]$}
            \State \parbox[t]{0.85\linewidth}{%
            Identify the closest point on the triangle to $\vb{x}^k$, denoted $\vb{X}^k$,
            using a fast point-triangle distance algorithm
            \cite{ericsonRealTimeCollisionDetection2004};%
            }
            \State Define the normal-to-the-surface vector $\vb{N}^k = \vb{x}^k - \vb{X}^k$;
            \If {$\abs{\vb{N}^k}\leq$ diagonal length of the local Eulerian cell}
                \State Identify the six neighboring nodes $\vb{x}^{k,n}$ of $\vb{x}^k$ on the lattice;
                \State Identify the vectors $\vb{v}^{k,n} = \vb{x}^{k,n} - \vb{X}^k$;
                \State \parbox[t]{0.85\linewidth}{%
                Classify the Eulerian node $\vb{x}^k$ as a forcing node if at least one of the scalar products $\vb{N}^k \vdot \vb{v}^{k,n}$ is negative.
                }
            \EndIf
        \EndFor
    \EndFor
\end{algorithmic}
\end{algorithm}
The IBM proposed in this work is built upon \cite{fadlunCombinedImmersedBoundaryFiniteDifference2000}, where, for a fixed body enclosing a solid volume, the velocity field was linearly reconstructed at the nearest point to the surface on the fluid side to enforce the no-slip condition.

However, several key differences distinguish the proposed formulation. 
First, the no-slip condition is here enforced through IB forcing on both sides of the wet surface, with the interpolation performed along the local normal direction rather than along the Cartesian directions \cite{yangEmbeddedboundaryFormulationLargeeddy2006}, thus enabling the treatment of zero-thickness structures.
Furthermore, to efficiently cope with moving walls, a novel fast and parallel tagging algorithm to identify the forcing nodes at each time step is introduced.
As anticipated in Section~\ref{sec:method}, the spurious effect of fresh cells in the presence of moving walls is prevented by performing the IB step prior the solution of the momentum equation in the fluid cells.
Importantly, the hydrodynamic loads are evaluated using MLS kernels, yielding a smooth stress distribution over the wet surface and enabling FSI problems.
Finally, the elliptic equation is modified to ensure a consistent treatment of boundary cells, hence reducing the splitting error.

\subsection{Fast tagging and IB on provisional velocity}\label{sec:newIB_meth}
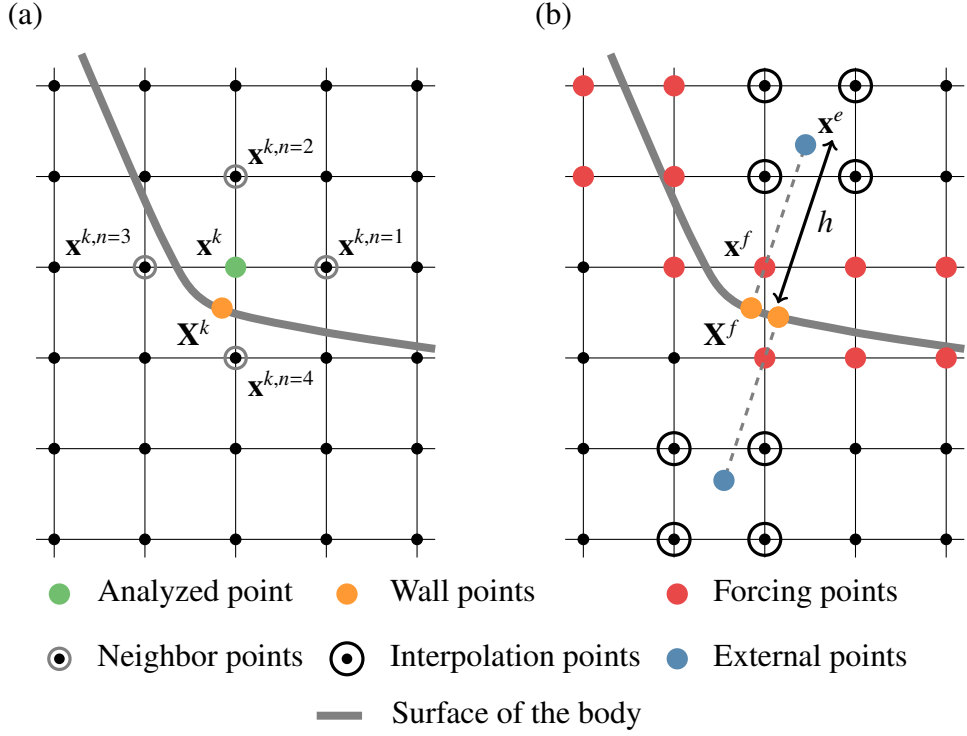
\begin{figure}[t]
    \centering
    \input{_figures/general/tagging_double}
    \caption{Two-dimensional sketch of (a) the tagging procedure and (b) the interpolation step.
    The point $\vb{x}^k$ in (a) is identified as a forcing point, denoted as $\vb{x}^f$ in (b).} 
    \label{fig:tagging}
\end{figure}
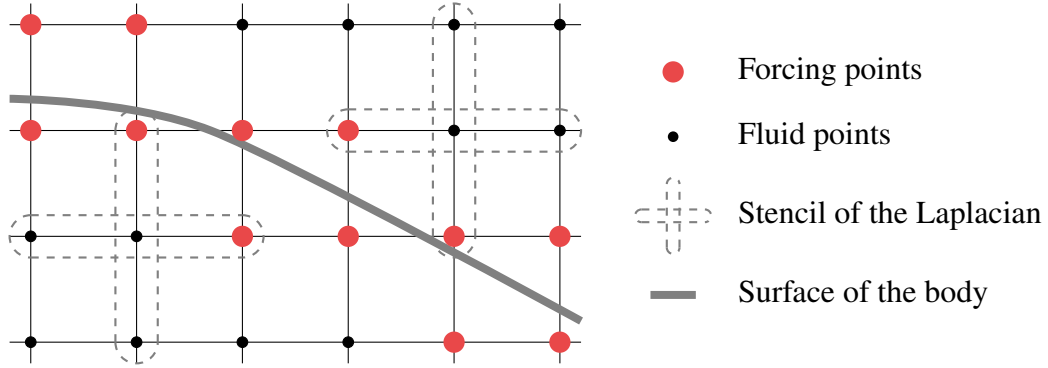
\begin{figure}[t]
    \centering
    \input{_figures/general/stencil}
    \vspace{1em}
    \caption{Stencil of the Laplacian operator near the immersed surface.
    Since on the forcing points the provisional equation is not solved, the stencil never crosses the immersed interface.}
    \label{fig:stencil}
\end{figure}
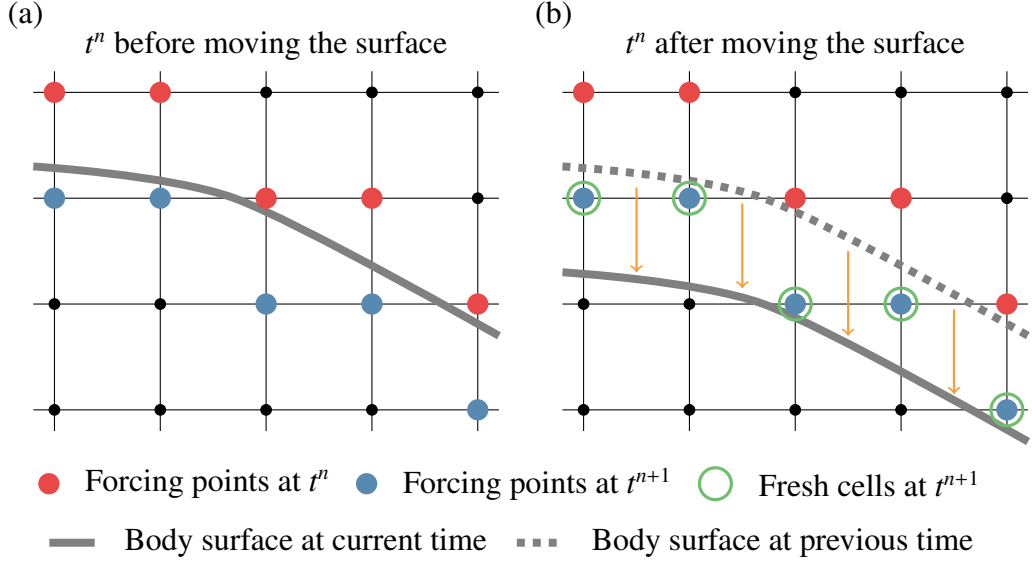
\begin{figure}[t]
    \centering
    \input{_figures/general/fresh_cell}
    \caption{Treatment of the fresh cells with the novel IBM.
    For clarity, only the forcing points on the upper side of the wall are depicted.
    Position of the moving surface at $t^n$ (a) before moving the structure and (b) after moving the structure.
    Once the structure is displaced, the time step $t^n$ is completed and the fresh cells at $t^{n+1}$ have become forcing points.}
    \label{fig:fresh_cell}
\end{figure}
Let us consider one of the three staggered grids of the velocity.
The first step of the proposed IB formulation is the identification of the forcing nodes, which is performed using a novel tagging algorithm summarized in Algorithm~\ref{alg:tagging} and illustrated in Figure~\ref{fig:tagging}a, which requires the immersed body surface to be discretized into $N_t$ triangular elements.

Once a point $\vb{x}^k$ is identified as a forcing node (hereafter indicated as $\vb{x}^f$), the IB forcing is performed.
A probe in the wall-normal direction $\vb{N}^f$ passing through $\vb{x}^f$ is casted at distance $h$ from the corresponding wall point $\vb{X}^f$, thus identifying the external point $\vb{x}^e$ (see the two-dimensional sketch in Figure~\ref{fig:tagging}b):
\begin{equation}
    \label{eq:probe}    
    \vb{x}^e = \vb{X}^f + h \vb{n}^f,
\end{equation}
where $\vb{n}^f = \vb{N}^f / \abs{\vb{N}^f}$ is the unit normal vector to the surface passing through $\vb{x}^f$.
The flow velocity component on the forcing point $u^f$ is reconstructed through linear interpolation, using the component of body velocity $U$ at $\vb{X}^f$ and the flow velocity component $u^e$ at point $\vb{x}^e$, as:
\begin{equation}
    \label{eq:new_IB}
    u^f = \qty(1-\alpha)U + \alpha u^e,
\end{equation}
being $\alpha = \abs{\vb{N}^f}/h$.
The velocity at the wall is given from the boundary condition (that may be known analytically or coming from the structural dynamics in FSI problems), while the velocity at the external point is interpolated from the Eulerian grid using the trilinear interpolation of the surrounding Eulerian nodes.
The same procedure is applied to the opposite side of the wall for both zero-thickness and finite-thickness bodies, as illustrated in Figure~\ref{fig:tagging}b.
The length of the probe $h$ must be sufficiently large to prevent point $\vb{x}^f$ from being inside the interpolation stencil of $u^e$.
Thus, in the following $h$ is chosen to be twice the diagonal length of the local Eulerian cell.
For a uniform grid, $h = 2\sqrt{3}\Delta$, being $\Delta$ the grid size.

The same procedure is repeated for the other velocity components defined on the corresponding staggered grids.
Once this IB step is completed, the provisional velocity on the forcing nodes is determined, and the provisional velocity on the fluid nodes is obtained by solving Equation~\eqref{eq:NS_prov} on all points but the forcing points.

The advantages of the proposed method, which is based on a two-sides Eulerian IBM, are multiple.
Firstly, the IB force is applied only to the forcing points, thus avoiding a smearing of the interface (as happens in Lagrangian IBMs) in favor of a sharp representation of the solid-fluid interface.
Then, the two-sided flow reconstruction allows to deal with zero-thickness bodies, requiring to impose the no-slip condition on both sides of the wet wall.
Consequently, in the provisional velocity the two flow regions are completely decoupled as the stencil of the differential operators never crosses the immersed boundary (see, for instance, the stencil of $L$ in Figure~\ref{fig:stencil}).
In addition, the flow derivatives at points adjacent to the forcing nodes are evaluated using velocity values that already incorporate the IB condition.
Furthermore, applying the IB forcing before Equation~\eqref{eq:NS_prov} allows to deal with fresh cells without any additional effort.
If the CFL number is sufficiently small, at most one layer of fresh cells can emerge between the end of the $n$-th time step and the beginning of the $(n+1)$-th due to the displacement of the immersed surface.
Hence, the fresh cells are also forcing points for which the provisional flow velocity at time $t^{n+1}$ is determined by Equation~\eqref{eq:new_IB} (see Figure~\ref{fig:fresh_cell}).
Therefore, if the velocity field at time-step $n-1$ is not required for the evaluation of the non-linear terms (as happens in a self-starting time scheme, like the RK2+CN described above since $\rho^1 = 0$), the fresh cells are automatically handled by the proposed IBM.

As a final remark, the proposed tagging and forcing procedures are inherently parallel, since each Eulerian point is treated independently.
Furthermore, as in Lagrangian IBMs, the main computational loop is performed over the surface triangles rather than on the Eulerian points.
Once a forcing node is identified, the IB step is carried out "on-line", without requiring storage of the wall-point locations.
Consequently, the computational cost is comparable to that of existing Lagrangian IBMs.

\subsection{Modified elliptic equation}\label{sec:mod_elliptic}
\begin{figure}[t]
    \centering
    \input{_figures/general/fluid_and_boundary_cells}
    \caption{Identification of fluid cells $\mathcal{C}_F$ and boundary cells $\mathcal{C}_B$.
    The truncation error $R_F$ is defined on the centers of the fluid cells, while $R_B$ on the boundary cells.}
    \label{fig:fluid_and_boundary_cells}
\end{figure}
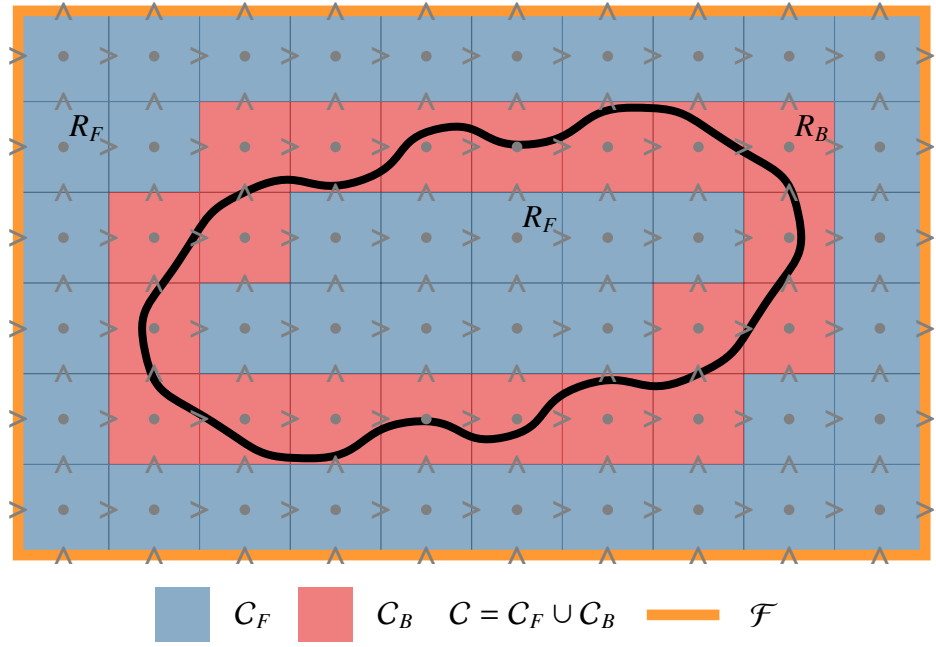
\begin{figure}[t]
    \centering
    \input{_figures/general/elliptic_cell}
    \caption{Identification of the geometrical properties of a boundary cell whose center is $\vb{x}^b$.}
    \label{fig:elliptic_cell}
\end{figure}
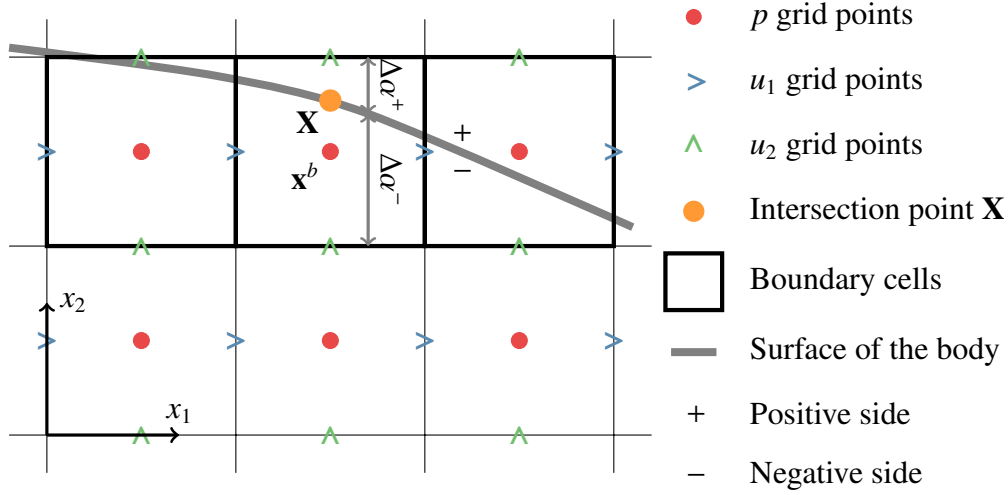
To obtain the solenoidal velocity field, the provisional velocity $\hat{\vb{u}}$ is projected on a divergence-free space using Equation~\eqref{eq:projection}.
However, since $G\varphi$ is nonzero at the forcing points, this projection alters the immersed boundary treatment, and the resulting velocity 
$\vb{u}^{s+1}$ may no longer satisfy the no-slip condition.
As discussed in the introduction, this is a well known issue arising when IBMs are coupled with the fractional step method, referred to as splitting error.

Furthermore, we note that the projection step yields another source of error.
With reference to the two-dimensional sketch of Figure~\ref{fig:fluid_and_boundary_cells}, let us consider a surface immersed in the staggered grid described in Section~\ref{sec:method}.
A cell is tagged as \emph{boundary cell} if all of its faces contain forcing nodes, and the set of boundary cells is denoted as $\mathcal{C}_B$;
The set of the remaining \emph{fluid cells} is called $\mathcal{C}_F$. Thus, $\mathcal{C} = \mathcal{C}_F \cup \mathcal{C}_B$ is the set of all the cells composing $\Omega$.
Let us consider a smooth flow $\vb{u}$ resulting from the exact solution of the incompressible Navier-Stokes equations on the box $\Omega$.
For the sake of clarity, we consider a uniform grid of constant size $\Delta$.
The discrete divergence operator $D$ applied to $\vb{u}$ is defined at the cell center $\vb{x}^k$, $k \in \mathcal{C}$ (where the pressure is located), and is given by:
\begin{equation}
    \label{eq:discrete_div}
    D\vb{u}^k =\sum_{i =1}^3 \frac{u_i\qty(\vb{x}^k + \frac{1}{2}\Delta \vb{e}_i) - u_i\qty(\vb{x}^k- \frac{1}{2}\Delta \vb{e}_i)}{\Delta},
\end{equation}
being $\vb{e}_i$ the axis unit vectors.
For a fluid cell the terms of the previous expression can be expanded in Taylor series around $\vb{x}^k$ as follows:
\begin{equation}
    \label{eq:taylor}
    u_i\qty(\vb{x}^k \pm \frac{1}{2}\Delta \vb{e}_i) = u_i\qty(\vb{x}^k) \pm \frac{1}{2}\pdv{u_i}{x_i}\bigg|_k\Delta + \frac{1}{8}\pdv[2]{u_i}{x_i}\bigg|_k\Delta^2 \pm \frac{1}{48}\pdv[3]{u_i}{x_i}\bigg|_k\Delta^3+ \order{\Delta ^4}.
\end{equation}
Inserting Equation~\eqref{eq:taylor} into the definition of $D\vb{u}$ leads to the equivalent continuous equation:
\begin{equation}
    \label{eq:res_f}
    D\vb{u}^k = \underbrace{\div{\vb{u}^k}}_{=0} + \underbrace{\frac{1}{24}\sum_{i=1}^3\pdv[3]{u_i}{x_i}\bigg|_k\Delta ^2 + \order{\Delta ^3}}_{R^k_F = \order{\Delta ^2}} \qq{on} \mathcal{C}_F,
\end{equation}
where $R_F$ is the truncation error of the discrete divergence operator on the fluid cells.
Let us now consider the boundary cell whose center is $\vb{x}^b$, highlighted in Figure~\ref{fig:elliptic_cell}.
The discrete divergence of $\vb{u}$ is again given by Equation~\eqref{eq:discrete_div}, but the Taylor series expansion in direction $x_2$ cannot be performed around the same point $\vb{x}^b$ and the $+$ and $-$ sides of the wall must be treated separately.
In particular, in direction $x_1$ and $x_3$ the expansion in Equation~\eqref{eq:taylor} can be employed, while the Taylor expansion in direction $x_2$ centered at the wall is given by:
\begin{equation}
    u_2\qty(\vb{x}^b \pm \frac{1}{2}\Delta\vb{e}_2) = u_2\qty(\vb{X} \pm \alpha^{\pm}\Delta\vb{e}_2) = u_2\qty(\vb{X}) \pm \pdv{u_2}{x_2}\bigg|_\pm\alpha^{\pm}\Delta + \order{\Delta^2},
\end{equation}
being $\pdv*{u_2}{x_2}|_{\pm}$ the derivative of $u_2$ along the $x_2$ direction evaluated on the $\pm$ sides of the wall, $\vb{X}$ the coordinate of the intersection between the solid surface and the $x_1 = x_1^b$ line, and $\alpha^{\pm}$ the fractions of cell in the $\pm$ side of the wall.
Therefore, using the continuity of $\vb{u}$ across the immersed surface, the discrete divergence can be rewritten as:
\begin{equation}
    \label{eq:truncation_int}
    D\vb{u}^b = \underbrace{\pdv{u_1}{x_1}\bigg|_b + \qty(\alpha^+\pdv{u_2}{x_2}\bigg|_+ + \alpha^-\pdv{u_2}{x_2}\bigg|_-) +\pdv{u_3}{x_3}\bigg|_b + \order{\Delta }}_{R_B^b = \order{1}}  \qq{on} \mathcal{C}_B,
\end{equation}
being $R_B$ the truncation error on the boundary cells.
The last equality shows that $D\vb{u} = 0$ is not a consistent approximation of $\div{\vb{u}} = 0$ at the boundary cells as the right-hand-side of Equation~\eqref{eq:truncation_int} (hence, the truncation error) may not vanish for $\Delta \rightarrow 0$.
In other words, the presence of an immersed interface yields a $\order{1}$ sink/source in the boundary cells when the discrete continuity operator is applied to an exact flow solution around a solid interface:
\begin{equation}
    \label{eq:cont_sinks}
    D\vb{u} =\begin{cases}
        R_F = \order{\Delta^2} &\qq{on} \mathcal{C}_F,\\
        R_B = \order{1} &\qq{on} \mathcal{C}_B.
    \end{cases}
\end{equation}
Importantly, even if the numerical scheme yields a spatial distribution of sinks/sources, it can be demonstrated that the total mass is conserved exactly at a discrete level.
Indeed, integrating $D\vb{u}$ over the whole domain and using the discrete divergence theorem (see \ref{app:discrete_divergence_theorem}), it results:
\begin{equation}
    \label{eq:constraint}
    0 = \sum_{f\in \mathcal{F}} \vb{u}^f \vdot \vb{n}^f \Delta S^f = \sum_{k\in\mathcal{C}} D \vb{u}^k \Delta V^k =  \sum_{k\in\mathcal{C}_F}R_F^k \Delta V^k + \sum_{k\in\mathcal{C}_B}R_B^k \Delta V^k,
\end{equation}
being $\Delta V^k$ the volume of the $k$-th cell, $\mathcal{F}$ the set of boundary faces of $\mathcal{C}$ (see Figure~\ref{fig:fluid_and_boundary_cells}), $\Delta S^f$ their surfaces and $\vb{n}^f$ their outer normals.
The first equality holds due to the well-posedness of the incompressible boundary value problem, with the net discrete flux through $\partial \Omega$ equal to zero.
Hence, even if the mass is not locally conserved on the boundary cells $\mathcal{C}_B$, the mass imbalance is satisfied in integral form.
It follows that, although $\order{1}$ sinks/sources are present in the boundary cells, they are balanced by the distribution of the $\order{\Delta^2}$ truncation errors on the fluid cells, and the conservation of mass over the whole domain is automatically guaranteed.
Therefore, in order to obtain a consistent approximation of the divergence of the velocity field on the boundary cells, the truncation errors $R_F$ and $R_B$ should be added to the discrete conservation of mass $D\vb{u}^{s+1}$.
However, this approach is infeasible, as by definition the truncation errors cannot be evaluated.

Therefore, Equation~\eqref{eq:cont_sinks} applied to the discrete solution $\vb{u}^{s+1}$ is rewritten as:
\begin{equation}
    \label{eq:mod_div}
    D\vb{u}^{s+1} = q =\begin{cases}
        q_F \qq{on} \mathcal{C}_F,\\
        q_B \qq{on} \mathcal{C}_B,
    \end{cases}
\end{equation}
where the sinks/sources $q_F$ and $q_B$ are some proxy of the exact $R_B$ and $R_F$ to be defined.
Thus, the modified elliptic equation (to be used in the fractional step in place of Equation~\eqref{eq:elliptic}) becomes:
\begin{equation}
    \label{eq:elliptic_mod}
    L_c\varphi = \frac{D\hat{\vb{u}}-q}{\alpha^s \Delta t^n}.
\end{equation}
$q_B$ is estimated from the provisional velocity $\hat{\vb{u}}$ evaluated at the faces of the boundary cells:
\begin{equation}
    \label{eq:q_boundary}
    q_B = D\hat{\vb{u}} \qq{on} \mathcal{C}_B.
\end{equation}
Indeed, $\hat{\vb{u}}$ incorporates the boundary condition through the IB forcing, providing a good approximation of the flow field in the vicinity of the boundary, and hence of the associated sink/source terms.
Moreover, since $R_B$ is $\order{1}$, $q_B$  converges to $R_B$ for $\Delta \rightarrow 0$ if the numerical solution converges to the exact one.
It should be noted that the form of $q_B$ sets the right-hand side of the modified elliptic Equation~\eqref{eq:elliptic_mod} to zero on the boundary cells.
As a result, $q_B$ implicitly prevents to enforce $D\vb{u}^{s+1} = 0$ on the boundary cells as usually done for IBMs, consistently with Equation~\eqref{eq:cont_sinks}.

On the other hand, the distribution of $q_F$ should follow the one of $R_F$, which is proportional to the third derivative of the exact solution of the problem, which is clearly unavailable and it cannot be estimated with sufficient accuracy given the spatially second-order accuracy of the numerical scheme.
Nevertheless, for the discrete divergence theorem (see Equation~\eqref{eq:constraint}), $q_F$ should satisfy:
\begin{equation}
    \label{eq:constraint_q}
    \sum_{k\in\mathcal{C}_F}q_F^k \Delta V^k + \sum_{k\in\mathcal{C}_B}q_B^k \Delta V^k = 0.
\end{equation}
A simple shape for $q_F$ satisfying the previous constraint is given by the following expression:
\begin{equation}
    q_F  =-\frac{1}{\abs{\mathcal{C}_F}}\sum_{k\in \mathcal{C}_B} D\hat{\vb{u}}^k \Delta V^k ,
\end{equation}
namely a uniform distribution of $q_F$ among the fluid cells, being $\abs{\mathcal{C}_F}$ the total volume of the fluid cells.
In addition, such an expression for $q_F$ corresponds to the solution of the constraint~\eqref{eq:constraint_q} having the minimal $L_2$ norm (see \ref{app:shape_q}) and it can be demonstrated that $q_F = \order{\Delta ^2}$ (see \ref{app:scaling}), which is consistent with the numerical scheme.

Importantly, since Equation~\eqref{eq:constraint_q} holds true to machine precision, the total mass of the flow is conserved during the time integration \cite{morinishiFullyConservativeHigher1998}, and, since only the right hand side of Equation~\eqref{eq:elliptic} has been altered with the novel method, the structure of the Laplacian system has not changed, allowing the use of fast FFT-based algorithm for solving Equation~\eqref{eq:elliptic_mod}.
Hence, the computational effort required to obtain the pseudo-pressure $\varphi$ is practically unaltered: the only additional step with respect to the standard case is evaluating $q$, while the cost for the direct solution of Equation~\eqref{eq:elliptic_mod} remains the same as Equation~\eqref{eq:elliptic}.

Equation~\eqref{eq:constraint_q} ensures mass conservation within the computational domain to machine precision.
Nevertheless, the distribution of $q_F$ can be shaped to also achieve kinetic energy conservation in the inviscid limit, as described in \ref{app:shape_q}.

Consistently with the modified continuity equation at the boundary cells (see Equation~\eqref{eq:mod_div}) the pressure field is not defined therein, since in those cells $\varphi$ is not obtained from the enforcement of the incompressibility constraint, but rather from a requirement on the velocity field.
Therefore, the pressure on $\mathcal{C}_B$ is reconstructed from the surrounding pressure field using the same strategy employed for the reconstruction of the velocity field for the IB forcing.
This is just a post-processing step as the pressure evaluated at the boundary cells does not enter directly in the flow solution at next steps.

\subsection{Smooth evaluation of the hydrodynamic-loads}
The evaluation of the hydrodynamic loads acting on the body is based on the Lagrangian approach detailed in \cite{vagnoliMovingleastsquaresReconstructionHydrodynamic2026}, where the MLS kernels are used to interpolate the flow quantities on the probe points.
The use of a MLS-based interpolation allows to obtain smooth hydrodynamic stresses on the surface of the body.
Here the method is briefly summarized.

The same triangulation employed for the IB tagging is used, and the triangle centers are the Lagrangian markers $\vb{X}^l$, $l\in \qty[1,N_t]$.
For each marker, a probe in the wall normal direction is casted from the marker $\vb{X}^l$ to the bulk of the fluid $\vb{x}^e$ at distance $h$:
\begin{equation}
    \vb{x}^e = \vb{X}^l + h \vb{n}^l,
\end{equation}
being $\vb{n}^l$ the normal of the $l$-th triangle and $h$ the length of the probe.
The pressure and the gradient of $\vb{u}$ are expanded in Taylor series from $\vb{x}^e$ backwards to $\vb{X}^l$, leading to the following expressions:
\begin{subequations}
    \begin{align}
        p^l &= p^e - \pdv{p}{n}\bigg|_e\,h + \frac{1}{2} \pdv[2]{p}{n}\bigg|_e\,h^2,\\
	    \grad{\vb{u}}|_l& =\grad{\vb{u}}|_e - \pdv{}{n}	\qty(\grad{\vb{u}})\bigg|_e\,h + \frac{1}{2} \pdv[2]{}{n}\qty(\grad{\vb{u}})\bigg|_e\,h^2.
        	\end{align}
\end{subequations}
The coefficients of the Taylor series on the point $\vb{x}^e$ are evaluated through a second order MLS interpolation of the flow quantities defined on the Eulerian grid.

The traction on each marker is given by:
\begin{equation}
    \vb{t}^l = -p^l\vb{n}^l + \frac{1}{Re}\qty(\grad{\vb{u}}|_l + \grad{\vb{u}}^T|_l)\vdot \vb{n}^l.
\end{equation}
Finally, the total force and torque around given by:
\begin{equation}
    \vb{F} = \sum_{l=1}^{N_t}\vb{t}^l A^l, \quad \vb{M}^p = \sum_{l=1}^{N_t}\qty(\vb{X}^l -\vb{X}^p)\cross\vb{t}^lA^l,
\end{equation}
being $A^l$ the triangle area and $\vb{X}^p$ the pole of the torque.

\section{Numerical results}\label{sec:numerical_exp}
In this section, several benchmark flows are presented to assess the capability of the method to handle zero-thickness structures, FSI problems, and flows with large pressure drop across the interface, such as pressurized membranes.
Importantly, in addition to the macroscopic behavior of the flow, we measure the error on the no-slip condition to quantify the improvement achieved by the proposed method.
\subsection{Double lid driven cavity with a vertical immersed surface}
\begin{figure}[t]
    \centering
    \begin{subfigure}[t]{\textwidth}
        \raggedright
        \input{_figures/lid_driven_cavity/streamlines}
    \end{subfigure}\vspace{-1.5em}
    \begin{subfigure}[t]{\textwidth}
    \raggedright
    \input{_figures/lid_driven_cavity/ldc_profiles}
    \centering
    \ref{leg:ldc}
    \end{subfigure}
    \caption{a) Streamlines of the double lid driven cavity flow with a vertical immersed surface placed at $x_1 = L$.
    b) Magnification of the computational grid around the immersed surface.
    In the lower panels, the velocity profiles at (c) $x_2 = L/2$ and (d)  $x_1 = L/2$ (symbols, apex $^\pm$) are compared against a body fitted solution (solid line).
    For ease of comparison, $u^+_1$ and $u^+_2$ are shifted to $0<x_1<L+\Delta/\pi$.}
    \label{fig:lid_driven_cavity}
\end{figure}
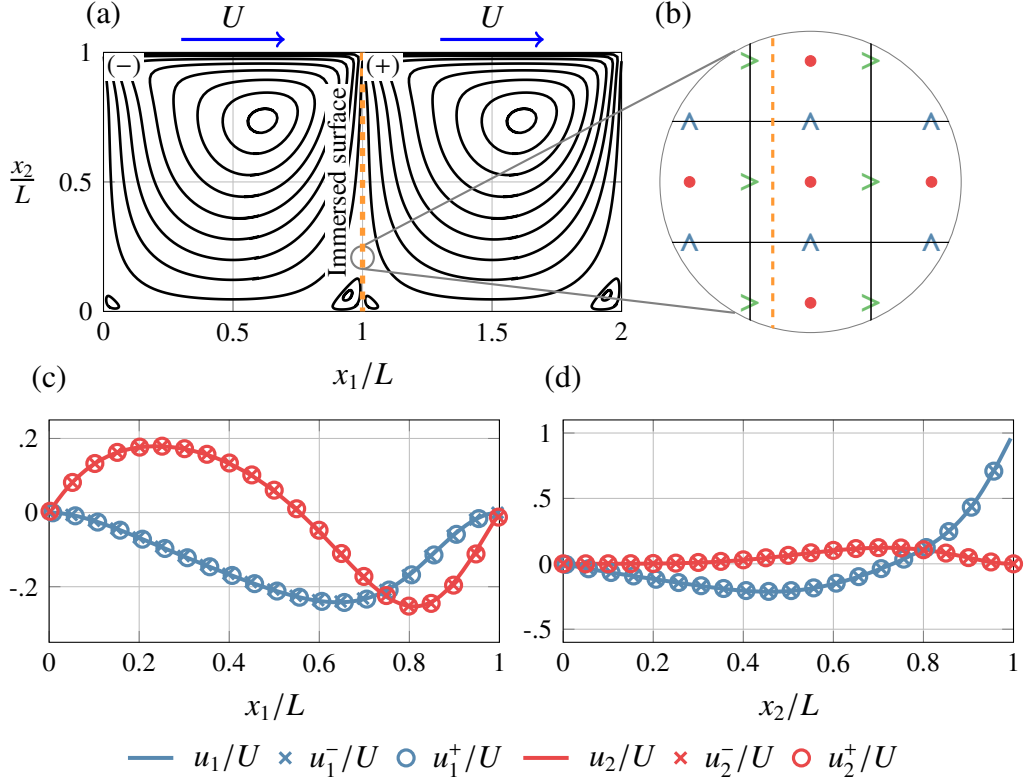
As a first test case, the flow in a two-dimensional double lid-driven cavity with a vertical immersed surface is considered, as shown in Figure~\ref{fig:lid_driven_cavity}a.
The no-slip condition on the immersed surface is enforced using the proposed IB forcing and, to avoid placing an Eulerian grid point on the immersed surface, the considered computational domain is the rectangle $\qty[0,2\qty(L+\Delta/\pi)] \times \qty[0,L]$, where $\Delta = 0.0125L$ is the uniform grid spacing.
The sketch of the Eulerian grid near the immersed surface is shown in Figure~\ref{fig:lid_driven_cavity}b.
The flow quantities are made non-dimensional using the lid velocity $U$ and the side-length $L$ of the lid, and the Reynolds number is fixed to $100$.
At the top wall, the Dirichlet boundary condition $\vb{u} =U\vb{e}_1$ is enforced, while at the other boundaries acts the homogeneous Dirichlet condition.
Thus, the same flow is expected on the left ($-$) and on the right ($+$) of the immersed surface.

The temporal integration scheme is the AB+CN, with a constant time step of $\Delta t = 0.01L/U$.
As a reference, a comparative numerical simulation made with the same temporal and spatial scheme is performed on the domain $\qty[0,L+\Delta/\pi]\times\qty[0,L]$, thus using a body-fitted discretization.

In Figures~\ref{fig:lid_driven_cavity}c and d, the velocity components on the line $x_2 = L/2$ and $x_1 = L/2$ are reported, showing a perfect agreement with the reference solution.
In particular, the $L_1$ norm of the difference between the IB solution and the body fitted solution is evaluated, yielding $\norm{u^-_1 - u_1}_1 = 0.5\times 10^{-4}U$, $\norm{u^+_1 - u_1}_1 = 1.7\times 10^{-4}U$ for the horizontal component and $\norm{u^-_2 - u_2}_1 = 3.9\times 10^{-4}U$, $\norm{u^+_2 - u_2}_1 = 1.1\times10^{-4}U$ for the vertical component.

\subsection{Poiseuille flow with a bottom immersed surface}\label{sec:pois}
\begin{figure}[t]
    \begin{subfigure}[t]{0.50\textwidth}
        \centering
        \caption{}
        \label{fig:poiseuille_flow_scheme}
        \input{_figures/poiseuille_flow/poiseuille_scheme}
    \end{subfigure}
    \hfill
    \begin{subfigure}[t]{0.40\textwidth}
        \centering
        \caption{}
        \label{fig:poiseuille_flow}
        \input{_figures/poiseuille_flow/poiseuille_flow_profiles}
    \end{subfigure}\vspace{-1em}
    \begin{subfigure}[t]{\textwidth}
        \raggedright
        \input{_figures/poiseuille_flow/poiseuille_flow_convergence}
    \end{subfigure}
    \centering
    \ref{leg:poiseuille}
   \caption{\subref{fig:poiseuille_flow_scheme}) Sketch of the Poiseuille flow with a bottom immersed surface.
   For $x_3>0$ a constant pressure gradient drives the flow, while for $x_3<0$ a forcing acts on the fluid in the wall normal direction.
   At $x_3 = 0$ an immersed surface is placed, indicated by a dashed line.
   The magnification shows the computational grid around the immersed surface on the plane $x_2 = 0$.
   \subref{fig:poiseuille_flow}) Velocity components of the Poiseuille flow along $x_3$ at $x_1 = x_2 = \delta/2$.
   c) Spatial convergence of $\norm{\qty(u_1^+ - u_e)}_1/U$ for the novel IBM.
   d) Convergence analysis of the IB error on the $(+)$ side of the immersed surface.}
   \label{fig:poiseuille}
\end{figure}
The Poiseuille flow in a three dimensional channel is investigated to address the convergence of the proposed IB method, where the lower wall is represented by an immersed surface.
The domain is the box $\qty[-\delta,\delta]\times \qty[-\delta,\delta] \times \qty[-\delta/4,2\delta]$, being $\delta$ the half-height of the channel, chosen as reference length.
The reference velocity is the maximum velocity $U$ of the plane Poiseuille flow,  and the Reynolds number based on $U$ and $\delta$ is $Re=100$.
As schematically depicted in Figure~\ref{fig:poiseuille_flow_scheme}, the domain is divided into two subregions by the immersed surface placed at $x_3 = 0$.
For $x_3>0$ ($+$ region), a constant pressure gradient along $x_1$ drives the flow, while for $x_3<0$ ($-$ region) a volumetric forcing term $F = 50\rho U^2/\delta\sin\qty[\pi\qty(x_3+\delta/4)/\qty(\delta/4)]\cos\qty(\pi x_1/\delta)$ acting in the $x_3$ direction is added to momentum equation.
On the immersed surface the no-slip condition is enforced using the IB method.

The computational grid is made of $N$ uniformly spaced points in the $x_1$ and $x_2$ directions, with uniform grid spacing $\Delta$.
Along the wall-normal direction, $x_3$, a non-uniform grid is adopted to prevent any grid node from lying on the immersed surface.
In the upper half of the channel ($x_3 > \delta$), the cells are uniformly spaced with size $\Delta$, while in the lower half the grid spacing is scaled by a factor $\alpha = 1 + \Delta/(4\delta)$ so as to avoid the placement of grid points on the immersed surface.
In the remaining part of the domain ($x_3 < -\Delta/4$), $\qty(N+1)/4 - 1$ uniformly sized cells are employed.
A close-up of the Eulerian grid in the vicinity of the immersed surface is shown in Figure~\ref{fig:poiseuille_flow_scheme}.

On the upper-half of the box the plane Poiseuille flow $u_e\qty(x_3) = -Ux_3(x_3-2\delta)/\delta^2$ is expected, while in the lower-half the forcing produces a steady structure of vortices, invariant for translation along $x_2$.

The numerical flow solution at $x_1 = \delta/2$ and $x_2 = 0$ along the $x_3$ direction is shown in Figure~\ref{fig:poiseuille_flow} for $N = 180$ (notice that $u_2=0$ everywhere).
In the lower part of the channel, both the vertical ($u_3^-$) and horizontal ($u_1^-$) components are present, but due to the IB forcing, the upper side of the channel is not influenced by the lower region, recovering the correct plane Poiseuille flow, with the only non-zero component $u_1^+$.

To assess the order of accuracy of the proposed method, the $L_1$ norm of the error $u_1^+ - u_e$, is reported in Figure~\ref{fig:poiseuille}c as a function of $N$.
The novel IB formulation exhibits clear second-order spatial accuracy.
The results are compared with those obtained using the same IB forcing but without the modification of the elliptic equation, i.e. using Equation~\eqref{eq:elliptic} instead of Equation~\eqref{eq:elliptic_mod}, which shows reduced accuracy and precision.
These results confirm that the introduction of the forcing term $q$ in Equation~\eqref{eq:elliptic_mod} is essential to recover the same level of accuracy as that attained by the underlying flow solver away from the solid boundaries.

Finally, the convergence rate associated with the enforcement of the no-slip condition is assessed.
The velocity component $u_1^+$ is interpolated using a Chebyshev polynomial of order $4$, yielding the polynomial $I^+(x_3)$ defined over the interval $0 < x_3 < 2\delta$.
The polynomial is then evaluated at the interface location $x_3 = 0$, providing the quantity $I^+(0)$, representing the Chebyshev interpolant of the velocity field at the immersed surface from the $+$ side.
In Figure~\ref{fig:poiseuille}d, the convergence of $\abs{I^+\qty(0)}$ as $N$ increases is reported: the no-slip condition enforced through the IB forcing is second-order accurate.
On the other hand, the two-sided Eulerian IB method without the modified elliptic equation exhibits a convergence rate $\order{N^{-1}}$, indicating that the accuracy of the bulk field $\norm{u_1^+ - u_e}_1$ is limited by the reduced accuracy in the enforcement of the no-slip condition on the immersed surface.

\subsection{Turbulent channel flow with a lower immersed surface}
\label{sec:turbulent_channel}
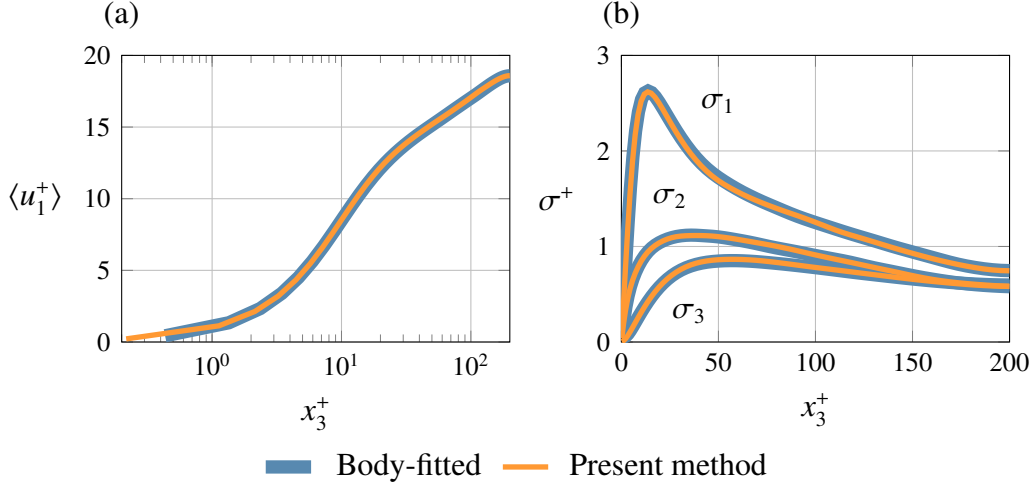
\begin{figure}[t]
    \raggedright
    \input{_figures/turbulent_channel/turbulent_channel_velocity}
    \ref{leg:turbulent_channel}
    \centering
    \caption{Turbulent channel flow with a lower immersed surface: (a) mean velocity profile and (b) turbulent intensities.}
    \label{fig:turbulent_channel}
\end{figure}
The turbulent counterpart of Section~\ref{sec:pois} is now considered.
The reference length is the channel half-height $\delta$, while the reference velocity is the friction velocity $u_\tau$.
The friction Reynolds number $Re_\tau = u_\tau\delta/\nu$ is fixed to 200 \cite{quadrioDoesChoiceForcing2016} and the domain is $\qty[-2\pi\delta,2\pi\delta]\times \qty[-\pi\delta,\pi\delta]\times \qty[-\delta/4,2\delta]$.
Similarly to the previous case, at $x_3 = 0$ an immersed rigid wall is placed, where the no-slip condition is imposed using the IB method, while for $x_3>0$, a constant pressure gradient along the $x_1$ direction drives the flow, and for $x_3<0$ the forcing $F =- 5\qty(u_3-u_{\tau})u_{\tau}/\delta$ acting in the wall-normal direction is added to Equation~\eqref{eq:NS_prov} to trigger a flow impinging on the immersed surface.

The computational grid is uniform in the $x_1$ and $x_2$ directions, both discretized with $256$ points.
As in the Poiseuille flow configuration, the wall-normal grid is constructed so as to avoid the placement of Eulerian points on the immersed surface.
Specifically, the grid on the channel side is obtained by scaling the following non-uniform grid defined on $\qty[0,2\delta]$ with $N$ points, whose $N+1$ faces are given by:
\begin{equation}
    x_3^1 = 0, \quad x_3^{N+1} = 2\delta, \quad \frac{x_3^k}{\delta} = 1 - \frac{2}{\Delta_g}\cos\qty[\frac{\pi}{2\alpha}\qty(k+\alpha-\frac{1}{2})], \quad k \in \qty[2,N],
\end{equation}
with $\alpha = 8$ and $\Delta_g = \cos\qty[\tfrac{\pi(\alpha+\nicefrac{1}{2})}{2\alpha}] - \cos\qty[\tfrac{\pi(\alpha+N-\nicefrac{1}{2})}{2\alpha}]$.
The resulting grid is then scaled by a factor $s = 1 + \Delta/(4\delta)$, where $\Delta = x_3^2 - x_3^1$, in order to ensure that no grid point lies on the immersed surface.
Finally, $\qty(N+1)/4 - 1$ uniformly spaced cells are placed in the interval $\qty[-\delta/4,-\Delta/4]$.
In the following, $N$ is fixed to $127$, for a total of 158 points in the $x_3$ direction.
The time integration is performed using  the AB+CN scheme up to convergence of the first order moments of the flow.
The simulations performed with IBM are compared against a body fitted simulation performed in the box $\qty[-2\pi\delta,2\pi\delta]\times \qty[-\pi\delta,\pi\delta]\times \qty[0,2\delta]$.

The mean velocity $\langle u_1\rangle$ for $x_3>0$ obtained with the novel IB formulation and the body-fitted simulation are shown in Figure~\ref{fig:turbulent_channel}a in wall units.
The mean velocity profile of the two simulations are superimposed.
The standard deviations $\sigma_i = \sqrt{\langle u_i - \langle u_i\rangle\rangle}$, $i=1,2,3$, of the flow for both the simulations are reported in Figure~\ref{fig:turbulent_channel}b in wall units.
The $\sigma_i$ curves obtained with the IB forcing collapse over the results of the body-fitted simulation.
Thus, the proposed method is able to recover the results of the body-fitted simulation despite the imposition of the no-slip condition on the lower wall through the IB forcing, and the impinging lower flow (in $x_3<0$) does not perturb the upper flow field.
Moreover, the addition of the source term $q$ in the conservation of mass equation does not modify the long time integration properties of the numerical scheme.

\subsection{Flow around a square plate: steady and unsteady cases}\label{sec:plate}
\begin{figure}[t]
    \centering
    \begin{subfigure}[b]{0.44\textwidth}
        \centering
        \caption{}
        \label{fig:plate_flow}
        \includegraphics[scale = 0.3]{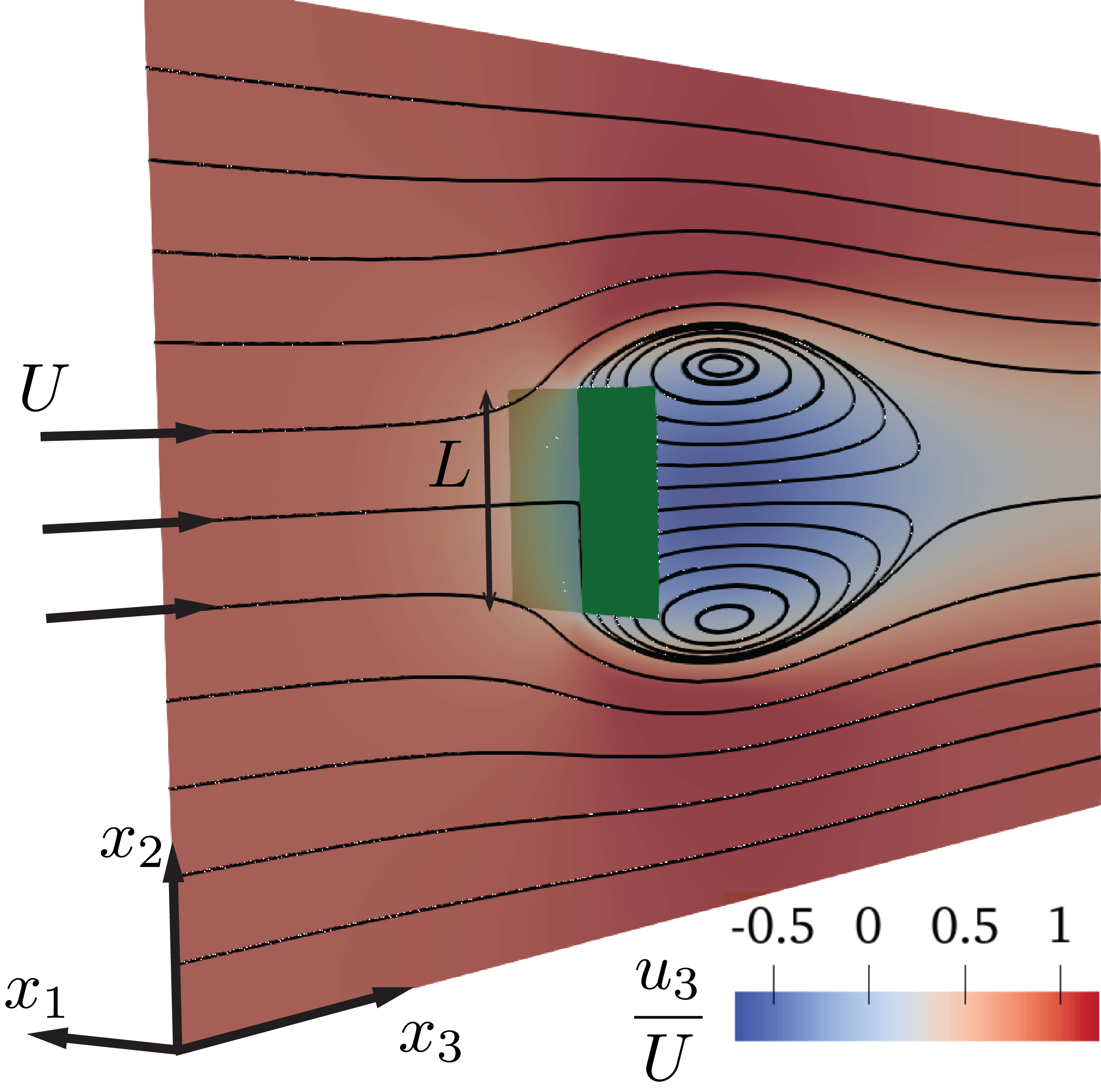}
    \end{subfigure}
    \begin{subfigure}[b]{0.55\textwidth}
        \centering
        \input{_figures/plate/plate_profiles}
    \end{subfigure}
    \begin{subfigure}[b]{\textwidth}
        \centering
        \input{_figures/plate/plate_convergence}
    \end{subfigure}
    \caption{\subref{fig:plate_flow}) Streamlines of the flow around the plate superimposed on the contour field of the $u_3$ velocity component.
    In the lateral panels, profiles of (b) $u_3$ and (c) $p$ along $x_3$ at $x_1 = x_2 = 0$.
    d) Convergence analysis of the IB error on the $(-)$ and $(+)$ sides of the plate.}
    \label{fig:plate}
\end{figure}
The next test case is the incident flow $U$ around a fixed, zero-thickness, square plate of side $L$.
Such a flow configuration is particularly challenging due to the zero thickness of the body, as impermeability and no-slip boundary conditions cannot rely on the help of the forcing in the inner points. 
The difficulty is exacerbated by the strong boundary layer separation occurring at the plate edges and the large pressure drop across the plate.
The domain is $\qty[-2L,2L]\times\qty[-2L,2L]\times\qty[-2L,4L]$, and the center of the plate is placed at the origin.
The Reynolds number is fixed to 100, based on $L$ and $U$.

The computational grid is made of $N\times N\times 3/2N$ cells, for a uniform grid size $\Delta$, and the time integration is performed using the AB+CN scheme, with a constant $\Delta t = 10^{-3}L/U$.
At the selected Reynolds number, the flow attains a steady state characterized by a recirculation region downstream of the plate, as illustrated by the streamlines on the $x_1-x_3$ plane in Figure~\ref{fig:plate}a.
The $u_3$ (streamwise) and $p$ profiles on the line $x_1 = x_2 = 0$ along $x_3$ are reported in Figure~\ref{fig:plate}b-c, respectively.
Velocity and pressure profiles show that the novel IB method is capable of capturing the fast variation of the velocity field near the plate and sustain the large pressure drop across the surface.

Then, the convergence of the IB forcing is investigated as done for the Poiseuille flow.
In particular, for the upstream ($x_3<0$) and downstream ($x_3>0$) flow regions, the normal velocity $u_3(0,0,x_3)$ is interpolated using the Chebyshev polynomial of order 64, obtaining $I^-\qty(x_3)$ and $I^+\qty(x_3)$, respectively.
The resulting polynomials are evaluated on $x_3 = 0$, and the convergence of $\abs{I^-\qty(0)}$ and $\abs{I^+\qty(0)}$ as the number of grid points is increased is depicted in Figure~\ref{fig:plate}d.
For both the sides of the plate, the imposition of the no-slip condition using the novel IBM is clearly second-order accurate.

\begin{figure}[t]
    \raggedright
    \input{_figures/heaving_plate/heaving_plate_error}
    \centering
    \ref{leg:heaving}
    \caption{Temporal evolution of $\norm{\boldsymbol{\varepsilon}}_1$ for the flow around the heaving plate.}
    \label{fig:heaving_plate}
\end{figure}
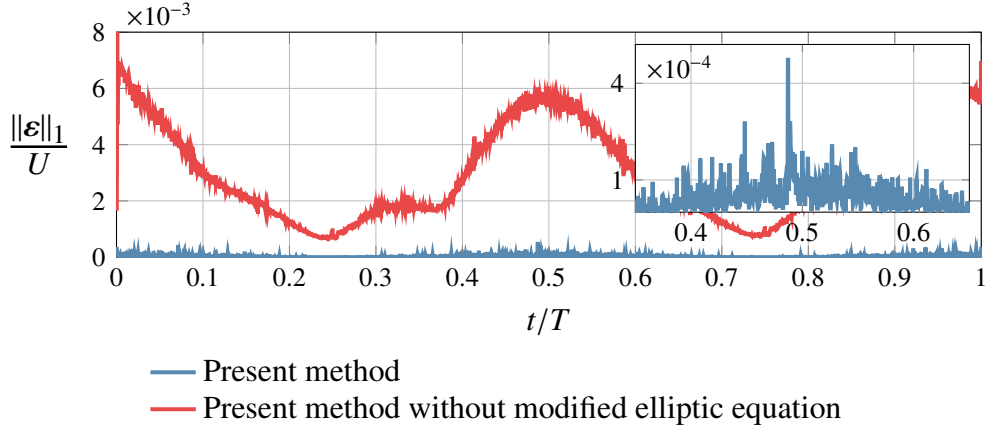
To assess the method in a flow with moving walls, for the plate is prescribed a time-dependent rigid-body motion in an otherwise quiescent fluid, superimposing a heaving motion along $\vb{e}_3$ and a pitching motion around the $x_2$ axis, yielding $\vb{X}\qty(t) = b\qty(t)\vb{e}_3 + \qty[X_1^0 \cos \theta\qty(t), X_2^0, -X_1^0\sin \theta\qty(t)]^T$, where $b\qty(t) = B\sin\qty(\omega t)$ and $\theta\qty(t) = \Theta \sin\qty(2\omega t)$.
$\vb{X}^0 = \qty[X_1^0, X_2^0, 0]^T$ is the initial configuration of the plate, $B = 6L$ and $\Theta = \pi/6$ are, respectively, the amplitude of the heaving and pitching motion of the plate, and $\omega$ is the heaving pulsation.
The reference velocity for this configuration is the maximal heaving velocity $B\omega$, and the Reynolds number $Re = \rho B\omega L/\mu$ is set to 100.
The plate is initially placed at the center of the computational domain $\qty[-L,L]\times\qty[-L,L]\times\qty[-6.5L,6.5L]$, which is discretized using $100\times100\times650$ cells, owning a constant grid size $\Delta = 0.02 L$.
At the north and south boundaries, the homogeneous Dirichlet boundary condition is enforced.
The temporal integration is performed using the RK2+CN scheme with constant time-step $\Delta t = 10^{-5}B\omega /L$.

The precision on the imposition of the no-slip condition is assessed by evaluating the mismatch $\boldsymbol{\varepsilon}$ between the velocity of the plate and the flow velocity, interpolated on the surface of the body using the same interpolator defined in Section~\ref{sec:newIB} for the IB forcing (see Equation~\eqref{eq:new_IB}).
$\norm{\boldsymbol{\varepsilon}}_1 = \sum_{i=1}^3\norm{\varepsilon_i}_1/3$ is evaluated, being $\norm{\varepsilon_i}_1$ the $L_1$ norm of $\varepsilon_i$ on the forcing points in the $i$-th direction.
The time evolution of $\norm{\boldsymbol{\varepsilon}}_1$ for one heaving period $T = 2\pi/\omega$ is reported in Figure~\ref{fig:heaving_plate}, and compared with the $\norm{\boldsymbol{\varepsilon}}_1$ obtained with the two-sided Eulerian IB introduced for the Poiseuille flow.
For this highly unsteady flow, the transpiration error of the novel IB method is of the order of $10^{-4}U$, while the two-sided Eulerian IB shows a reduction of the precision of one order of magnitude.
This result confirms the effect of the modified elliptic equation on the correct enforcing of the no-slip condition.

\subsection{Settling of buoyancy driven spheres and spheroids}
\begin{figure}[t]
    \centering 
    \begin{subfigure}[t]{0.4\textwidth}
        \centering
        \caption{}
        \input{_figures/buoyancy_driven/buoyancy_sketch}
        \label{fig:bouyancy_sketch}
    \end{subfigure}
    \hfill
    \begin{subfigure}[t]{0.59\textwidth}
        \centering
        \caption{}
        \input{_figures/buoyancy_driven/rising_sphere_vertical}
        \label{fig:rising_sphere_vertical}
    \end{subfigure}\vspace{-2em}
    \begin{subfigure}[t]{\textwidth}
        \raggedright
        \caption{}
        \input{_figures/buoyancy_driven/VP_error}
        \label{fig:VP_error}
    \end{subfigure}
    \caption{\subref{fig:bouyancy_sketch}) Sketch of the setting employed for the buoyancy-driven test cases.
    \subref{fig:rising_sphere_vertical}) Rising velocity $v_V$ of the light sphere case.
    \subref{fig:VP_error}) Comparison of the temporal evolution of $\norm{\boldsymbol{\varepsilon}}_1$ obtained with the novel IBM and a L-MLS for the VP falling regime.
    In the inset, magnification of the IBM error.
    }
\end{figure}
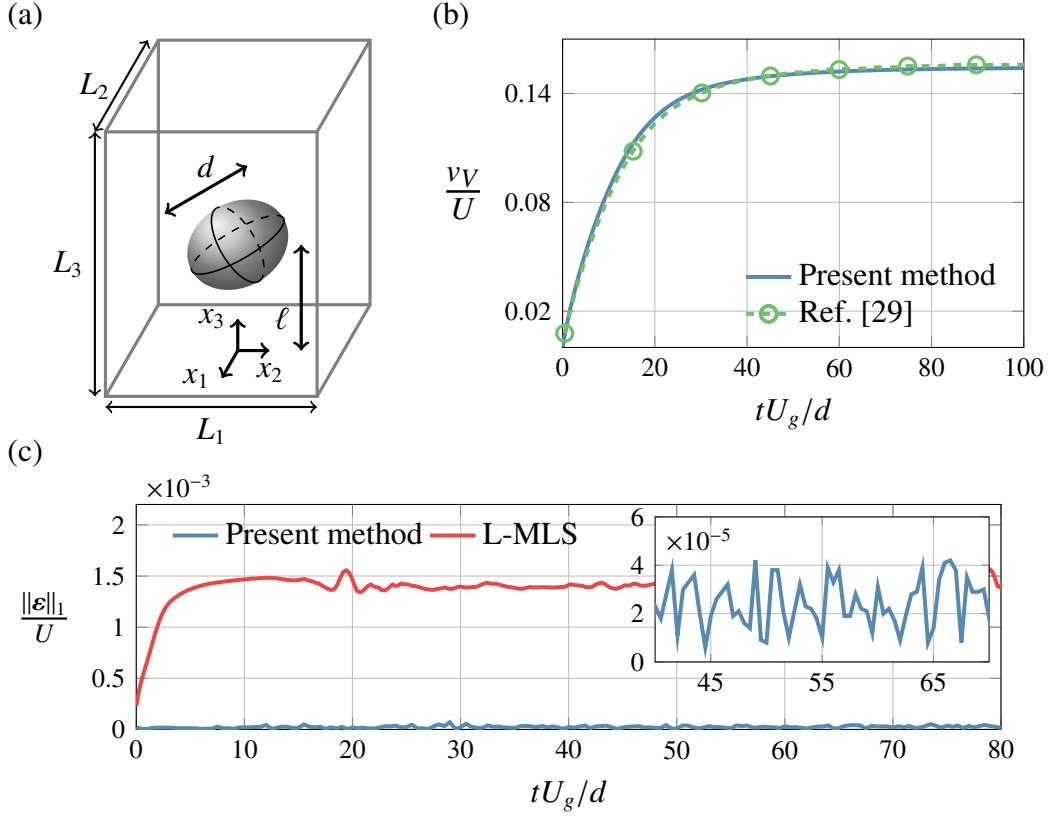

The capability of the novel IBM to handle the FSI of rigid bodies is assessed by examining the rising or falling motion of different solids of revolution under the action of gravity in a viscous fluid.
All the solids considered have a major axis of length $d$, volume $V$, and uniform density $\rho_b$ (Figure~\ref{fig:bouyancy_sketch}).
The reference velocity is the gravitational velocity $U_g = \sqrt{\abs{\rho_s-1}V^*gd}$, being $\pi_s = \rho_b/\rho$ the body-to-fluid density ratio, $V^* = V/d^3$ the non-dimensional volume and $g$ the gravitational acceleration.
The Reynolds number based on $d$ and $U_g$ is the Galileo number $Ga = \sqrt{\abs{\pi_s-1}V^*gd^3}/\nu$.

The Navier-Stokes equation for the fluid are solved in a frame of reference translating with the center of the solid, the Newton's equations for the body translation are solved in an inertial frame and the Euler equations for the body rotation are solved in the frame fixed with its principal axes, while the orientation of the solid is described using quaternions \cite{verziccoIntroductionImmersedBoundary2025}.
The gravity vector is $\vb{g} = -\operatorname{sgn}(\pi_s)g\vb{e}_3$, so that at the south boundary the unsteady Dirichlet boundary condition $\vb{u} = - \vb{v}$ is enforced, being $\vb{v}$ the velocity of the center of the body, while at the north boundary the radiative condition is imposed.
The solid is placed at the origin of the domain $\qty[-L_1/2,L_1/2]\times\qty[-L_2/2,L_2/2]\times\qty[-\ell,L_3-\ell]$, being $L_1$, $L_2$ and $L_3$ the length of the computational domain in the corresponding direction and $\ell$ the distance of the origin from the south wall.

A loose-coupling strategy is adopted to for the FSI: The fluid equations are advanced in time using the RK2+CN scheme, whereas the solid equations are integrated using the AB scheme.
In all the cases, a fixed time step of $\Delta t = 10^{-4} d / U_g$ is employed.

The first test case considers the rise of a light sphere with $Ga = 66$, $\pi_s = 0.8$, and $V^* = \pi/6$.
For this configuration, the sphere undergoes an initial acceleration and subsequently reaches a steady vertical trajectory \cite{yangNoniterativeDirectForcing2015}.
For this test case, $L_1 = L_2 = 6d$, $L_3 = 8d$ and $\ell = 3d$ and the computational domain is discretized using $300\times300\times400$ cells, for a constant grid size $\Delta = 0.02d$.
The time evolution of the rising vertical velocity $v_V$ is shown in Figure~\ref{fig:rising_sphere_vertical}.
The present IB method accurately captures the initial acceleration phase and the terminal rising velocity, with an error on the estimation of the terminal $v_V$ around 1\%.

Then, the settling of a heavy oblate spheroid is considered, for which $\pi_s = 2.14$ and $V^* = \pi/(6\chi)$, being $\chi = 1.5$ the aspect ratio of the oblate spheroid.
In the following, two values of the Galileo number are considered to obtain different falling regimes: For $Ga=110$ the spheroid settles down to a steady oblique (SO) path, while for $Ga = 150$ the solid reaches a vertical periodic (VP) motion \cite{moricheSingleOblateSpheroid2021}.
For these test cases, $L_1 = L_2 = 4d$, $L_3 = 6d$ and $\ell = 2d$ and the computational domain is discretized using $200\times200\times300$ cells, for a constant grid size $\Delta = 0.02d$.

For the SO case ($Ga = 110$), the steady state metrics of the system are reported in Table~\ref{tab:spheroid_results}.
\begin{table}[t]
    \centering
    \caption{Metrics of the SO and VP falling regimes for the oblate spheroid obtained with the novel IBM (the velocities are normalized by $U_g$).
    The reference values are extracted from \cite{moricheSingleOblateSpheroid2021}.}
    \begin{tabularx}{\textwidth}{l *{9}{>{\centering\arraybackslash}X}}
        \toprule
            & \multicolumn{3}{c}{Steady oblique} & \multicolumn{5}{c}{Vertical periodic} \\
                        &$v_H$ & $\abs{v_V}$ & $Re_V$ & $\alpha \qty(\degree)$ & $St$ & $v_H'$ & $\abs{\overline{v}_V}$ & $v_V'/U_g$ & $Re_V$\\
        \midrule
        Present method  & 0.111 & 1.645 & 181 & 3.845&0.198 & 0.202 &  1.742 & 0.0036 & 261\\
        L-MLS  & 0.117 & 1.639 & 180 & 4.097& 0.196 &0.197& 1.738 & 0.0035 &261 \\
        Ref.~\cite{moricheSingleOblateSpheroid2021}& 0.113 & 1.682 & 185 & 3.842& 0.198 & 0.194 & 1.740 & 0.0038 & 261\\
        \bottomrule
    \end{tabularx}
    \label{tab:spheroid_results}
\end{table}
The horizontal $v_H$ and vertical $v_V$ velocities of the center of the spheroid compare well with the literature results, as well as the Reynolds number based on the terminal falling velocity $Re_V = Ga\abs{v_V}$.
Also the terminal geometrical configuration of the spheroid obtained with our method agrees with the literature data, as confirmed by the angle of trajectory $\alpha = \arctan\qty(u_H/\abs{u_V})$.
The same test case is performed using the Lagrangian IBM based on the MLS kernels introduced in \cite{vanellaMovingleastsquaresReconstructionEmbeddedboundary2009}, keeping an average size of the triangles edges to $0.7 \Delta$.
Hereinafter, such Lagrangian IBM will be referred to as L-MLS.
The steady state metrics obtained with L-MLS are also reported for comparison in Table~\ref{tab:spheroid_results}.

Finally, the VP case ($Ga = 150$) is analyzed.
The metrics of the limit-cycle attained by the system evaluated using our method are reported in Table~\ref{tab:spheroid_results}.
The novel IBM correctly captures the limit cycle of the system, and all the quantities of interest are evaluated with an error below 5\% with respect to data from the literature.
For reference, the limit-cycle metrics obtained with L-MLS are reported on the same table.

For this falling regime, the transpiration errors $\boldsymbol{\varepsilon}$ of the novel IBM and L-MLS are compared.
In particular, $\boldsymbol{\varepsilon}$ is defined for the IBM as in Section~\ref{sec:plate}, while for L-MLS is defined as the mismatch between the flow velocity interpolated on the Lagrangian markers using the MLS kernels and the velocity of the Lagrangian markers: in each case the error is defined based on the corresponding IB interpolation rule, which are both second order.
In Figure~\ref{fig:VP_error}, the temporal evolution of $\norm{\boldsymbol{\varepsilon}}_1$ is depicted for the novel IBM and L-MLS: In the limit cycle, the mean value of $\norm{\boldsymbol{\varepsilon}}_1$ obtained with our method is approximately $2\times10^{-5}U$, whereas L-MLS yields a mean error of about $1.5\times10^{-3}U$.

\subsection{Flow around a flapping flag}
\begin{figure}[t]
    \centering
    \begin{subfigure}[t]{0.49\textwidth}
        \caption{}
        \input{_figures/flag/flag_sketch}
        \label{fig:flag_sketch}
    \end{subfigure}
    \hfill
    \begin{subfigure}[t]{0.40\textwidth}
        \centering
        \caption{}
        \input{_figures/flag/flag_plot}
        \label{fig:flag_plot}
    \end{subfigure}\vspace{-2em}
    \begin{subfigure}[t]{\textwidth}
        \raggedright
        \caption{}
        \input{_figures/flag/flag_error}
        \label{fig:flag_error}
    \end{subfigure}
    \caption{\subref{fig:flag_sketch}) Sketch of the flapping flag case.
    The red vertical line indicates the fixed leading edge of the flag.
    \subref{fig:flag_plot}) Initial dynamics of $x_2/L$ of the trailing edge point $Q$.
    \subref{fig:flag_error}) Comparison of the temporal evolution of $\norm{\boldsymbol{\varepsilon}}_1$ obtained with the novel IBM and L-MLS for the flapping flag for $\Delta t_1 = 10^{-4}L/U$ (solid line) and $\Delta t_2 = 10^{-5}L/U$ (dashed line).
    In the inset, magnification of the IBM error.}
\end{figure}
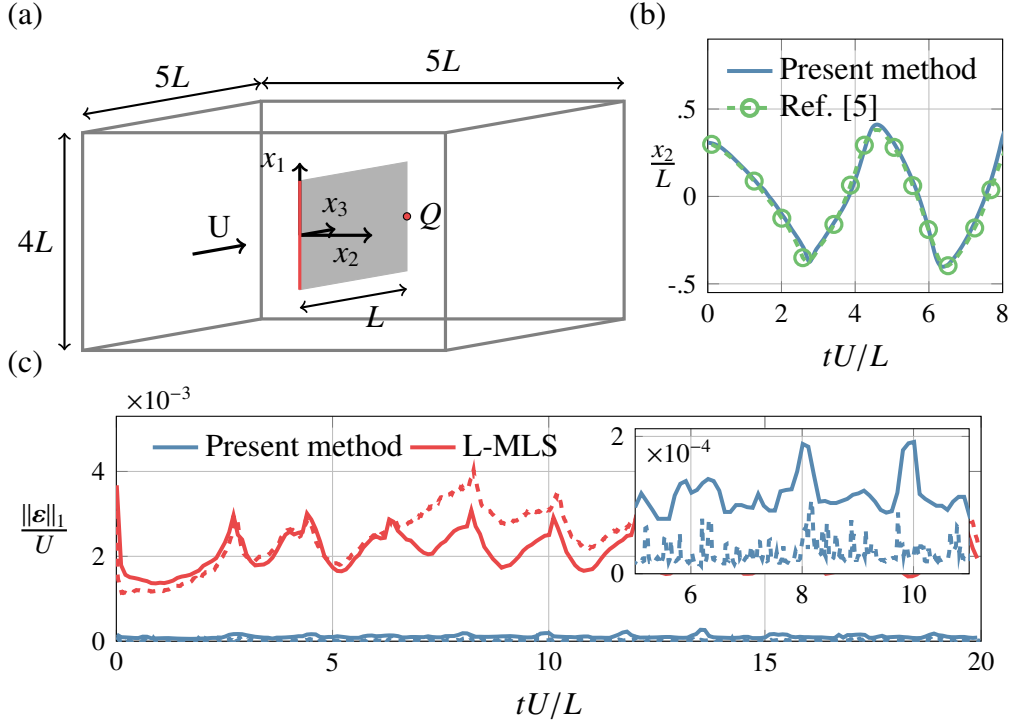
The flow around a square flapping flag hinged at the leading edge is investigated to test the novel IB method in a deformable FSI problem.
A sketch of the problem is depicted in Figure~\ref{fig:flag_sketch}, where the free-stream velocity $U$ and the length of the flag edge $L$ are the reference velocity and length, respectively, and the Reynolds number is fixed to $Re = 200$.

The flag has thickness $s = 0.01L$, and it is made of an inextensible linear-elastic uniform material of density $\rho_b = 100\rho$ with bending modulus $B = 10^{-4}\rho U^2L^3$.
The flag is treated as a two-dimensional surface within a three-dimensional flow, and the spring-mass network approach detailed in \cite{detullioMovingleastsquaresImmersedBoundary2016} is used to describe its structural dynamics, prescribing a large Young modulus to the spring-mass model to replicate the inextensibility of the flag, namely $E = 10^3\rho U^2 L/s$.
The flag is initially planar and tilted by an angle $0.1\pi$ with respect to the $x_1-x_3$ plane, and its leading edge is placed at the center of the domain $\qty[-2L,2L]\times\qty[-2.5L,2.5L]\times\qty[-2L,3L]$.
The computational domain is discretized using $200\times250\times250$ cells, for a uniform grid size $\Delta  = 0.02L$.
The fluid is advanced in time using the RK2+CN scheme, loosely coupled with the AB method for the advancement of the solid equation, with a constant time step $\Delta t = 10^{-4}L/U$.

The onset of the oscillation of the trailing edge point $Q$ obtained with the proposed IB method is depicted in Figure~\ref{fig:flag_plot} and compared against literature \cite{huangThreedimensionalSimulationFlapping2010}.
The corresponding peak-to-peak amplitude and Strouhal number of the limit-cycle method are reported in Table~\ref{tab:flag_results} for the same value of $\Delta t$.
For comparison, on the same table the metrics obtained with L-MLS are listed as well.
\begin{table}[t]
    \centering
    \caption{Normalized peak-to-peak amplitude and Strouhal number of the $x_2$ coordinate of the trailing-edge point $Q$ shown in Figure~\ref{fig:flag_plot} obtained with $\Delta t = 10^{-4}L/U$.}
    \begin{tabular}{cccc}
        \toprule
            &Present method & L-MLS & Ref.~\cite{huangThreedimensionalSimulationFlapping2010}\\
        \midrule
        Peak-to-peak amplitude & 0.784 & 0.767 &  0.780 \\
        Strouhal number        & 0.277 & 0.264 &  0.260 \\
        \bottomrule
    \end{tabular}
    \label{tab:flag_results}
\end{table}

The no-slip error $\boldsymbol{\varepsilon}$ is evaluated for $\Delta t_1 = 10^{-4}L/U$ and $\Delta t_2 = \Delta t_1/10 = 10^{-5}L/U$. 
The corresponding $\norm{\boldsymbol{\varepsilon}}_1$ values are shown in Figure~\ref{fig:flag_error}, from the initial time up to the onset of the limit cycle.
Using the novel IBM, the average $\norm{\boldsymbol{\varepsilon}}_1$ is $10^{-4}U$ for $\Delta t_1$ and $3\times 10^{-5}U$ for $\Delta t_2$, whereas L-MLS yields a mean error of $2\times 10^{-3}U$ for both time-steps.
Thus, although both methods accurately capture the macroscopic behavior of the system, such as the peak-to-peak amplitude and the oscillation frequency, the Lagrangian method yields a no-slip error that is orders of magnitude larger than the one of the novel IBM.
Moreover, for this test case, although the macroscopic flow features have reached temporal convergence, reducing the time-step size does not improve the accuracy of the L-MLS method. 
In contrast, the proposed IBM further reduces the transpiration error by one order of magnitude when the time step is decreased by a factor of ten.

\subsection{Bubble drop}
\begin{figure}[t]
    \centering
    \begin{subfigure}[t]{0.39\textwidth}
        \centering
        \caption{} 
        \input{_figures/bubble/fixed_bubble_sketch}
        \label{fig:fixed_bubble_sketch}
    \end{subfigure}
    \hfill
    \begin{subfigure}[t]{0.59\textwidth}
        \centering
        \caption{} 
        \includegraphics[scale = 0.28]{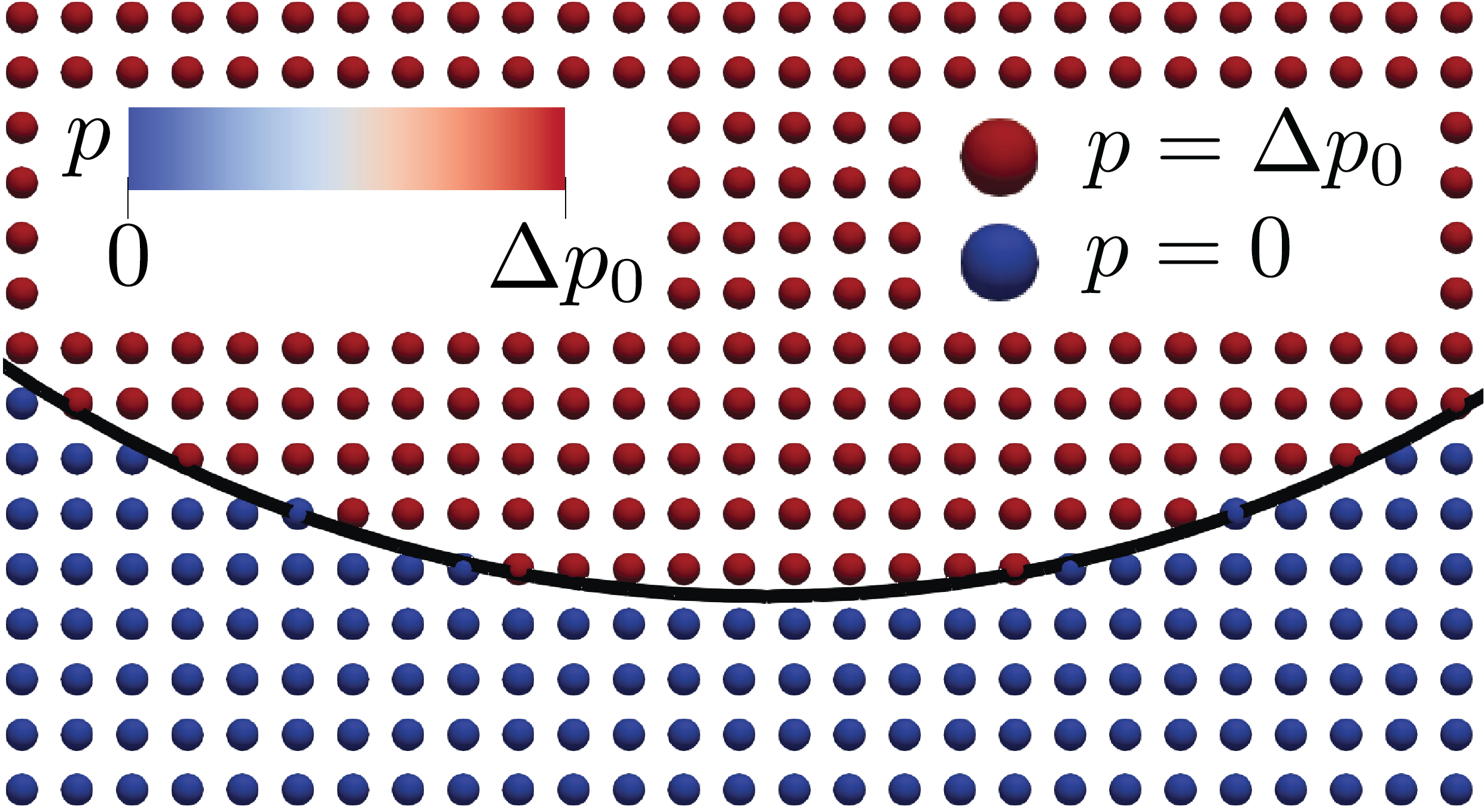}
        \label{fig:fixed_bubble_field}
    \end{subfigure}\vspace{-0.5em}
    \begin{subfigure}[b]{\textwidth}
        \raggedright
        \input{_figures/bubble/bubble_res}
        \centering
        \ref{leg:bubble_fixed_comparison}
    \end{subfigure}
    \caption{\subref{fig:fixed_bubble_sketch}) Sketch of the computational setting for the steady bubble case.
    \subref{fig:fixed_bubble_field}) Slice of the pressure field around the bubble on the $x_3=0$ plane.
    The black line denotes the trace of the bubble surface on the plane.
    The red dots denotes the cells for which $p = \Delta p_0$, while the blue dots are the cells with $p = 0$.
    On the lower panels, variation of (c) volume $V$, (d) surface area $S$ and (e) pressure drop $\Delta p$ of the fixed bubble with respect to the analytical values.
    In (f) - (h) the same quantities are reported for the oscillating bubble.}
    \label{fig:bubble_comparison}
\end{figure}
In the next test case the capability of the method to sustain large pressure drops across membranes is tested.
In particular, a bubble of diameter $d$ and surface tension $\gamma$ is considered.
The reference length is the diameter of the bubble, while the reference velocity is the capillary velocity $U_\gamma = \sqrt{\gamma/\qty(\rho d)}$.
The spherical bubble is initially placed in a quiescent fluid with uniform pressure fixed to 0.
After a transient in which the internal fluid pressurizes the bubble, the system reaches a steady regime in which the bubble keeps its spherical shape, with volume $V_0 = \pi d^3/6$, surface area $S_0=\pi d^2$ and a pressure drop across the interface given by the Young-Laplace law $\Delta p_0 = 4\gamma/d$.

The bubble is placed at the center of the fluid domain $\qty[-d,d]^3$, as depicted in Figure~\ref{fig:fixed_bubble_sketch}.
The surface of the bubble is treated as a two-dimensional surface immersed in a three-dimensional fluid, and the same structural solver of the flag case is employed.
The internal forces arising from the surface tension are introduced in the spring-mass network using a surface potential $V = \gamma S$, being $S$ the surface area of the sphere (see \ref{app:surface_tension} for a detailed derivation).
The computational domain is discretized using $120^3$ cells, for a constant grid size $\Delta  = 0.02d$.
The fluid equations are advanced in time up to $t = 15 d/U_{\gamma}$ with the RK2+CN method, loosely coupled with the AB method for the structural solver, using a constant time step $\Delta t = 10^{-4}d/U_{\gamma}$.

The steady state pressure field around the bubble obtained with the novel IBM is depicted in Figure~\ref{fig:fixed_bubble_field}.
The pressure drop across the surface is the one predicted by the Young-Laplace law, and its variation is sharp across the bubble.
The temporal evolution of the volume $V$, surface and pressure drop $\Delta p$ across the membrane obtained with our method and L-MLS are reported in Figure~\ref{fig:bubble_comparison}c-e.
As the system evolves, the proposed IB method preserves the initial bubble volume, whereas in L-MLS it decreases at a constant rate.
The same trend is observed in the evolution of the surface area, as shown in Figure~\ref{fig:bubble_comparison}d.
Finally, Figure~\ref{fig:bubble_comparison}e reports the pressure jump across the interface for both IB approaches.
After the initial transient, the proposed IBM yields the correct $\Delta p$ across the membrane, while L-MLS predicts a progressively increasing pressure drop.

Then, a non-inertial observer oscillating in the $x_3$ direction with constant amplitude $d$ and frequency $U_{\gamma}/\qty(2\pi d)$ is introduced, while keeping the bubble fixed.
Thus, in the non-inertial frame of reference, the bubble performs a sinusoidal motion with the same amplitude and frequency of the observer, but the pressure drop across its membrane is still equal to $\Delta p_0$.
Nevertheless, from the computational point of view, the bubble and the pressure field move over the grid, changing cells during the motion.

For this test case, the computational domain is modified to $\qty[-0.6d, 0.6d]\times\qty[-0.6d, 0.6d]\times \qty[-0.6d, 2.6d]$, and the bubble is initially placed in the origin of the computational domain.
The grid is made of $120\times120\times320$ cells, for a uniform grid size $\Delta = 0.01d$.
Also in this case, the same physical setting is simulated using L-MLS.
A video of the pressure field is provided in the Supplementary Material.

As in the fixed configuration, the novel IBM preserves the bubble volume and surface, whereas L-MLS exhibits a continuous volume/surface loss, see Figures~\ref{fig:bubble_comparison}f,g.
In Figure~\ref{fig:bubble_comparison}h, the pressure jump across the bubble interface is reported: the novel IBM yields the correct $\Delta p$, while L-MLS produces an increasing pressure jump due to the small curvature radius which, in turn, yields larger transpiration error.

\subsection{Over-pressurized deformable aorta}
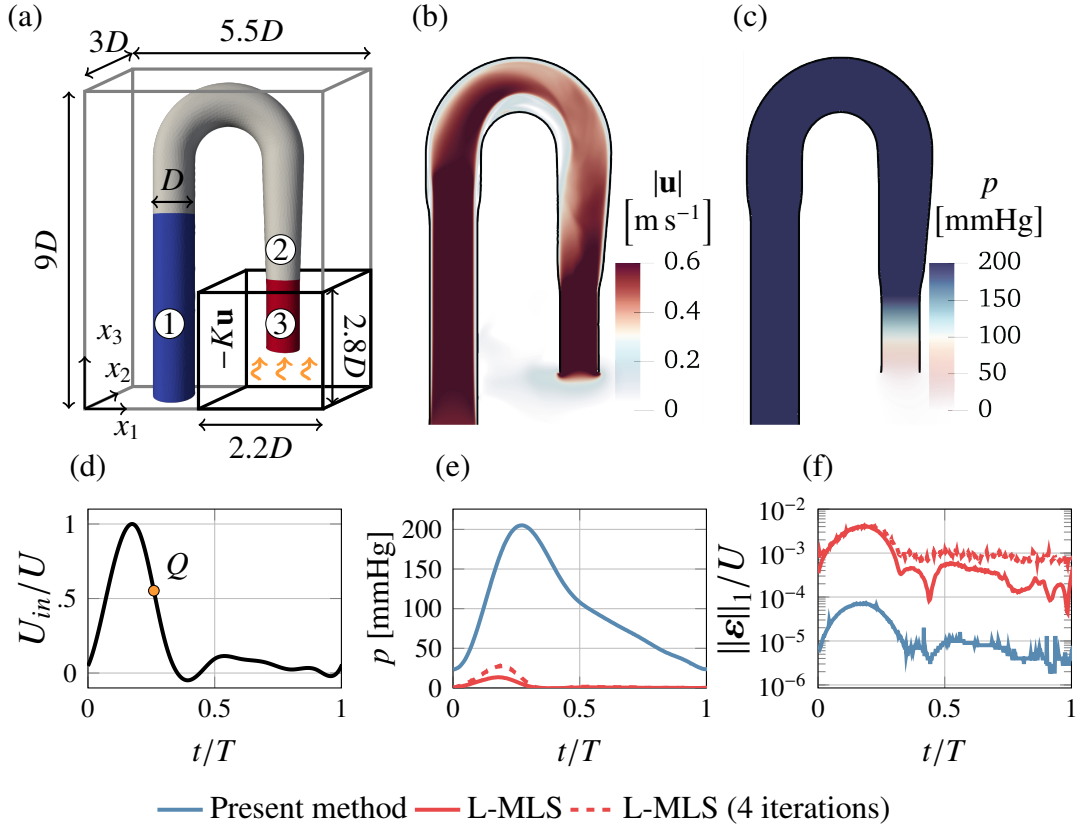
\begin{figure}[t]
    \centering
    \begin{subfigure}[t]{0.38\textwidth}
        \centering
        \caption{}
        \label{fig:aorta_sketch}
        \vspace{-1em}
        \input{_figures/aorta/aorta_sketch}
    \end{subfigure}
    \hfill
    \begin{subfigure}[t]{0.3\textwidth}
        \centering
        \caption{}
        \label{fig:aorta_velocity}
        \input{_figures/aorta/aorta_field1}
    \end{subfigure}
    \hfill
    \begin{subfigure}[t]{0.3\textwidth}
        \centering
        \caption{}
        \label{fig:aorta_pressure}
        \input{_figures/aorta/aorta_field2}
    \end{subfigure}\vspace{-1em}
    \begin{subfigure}[t]{\textwidth}
        \raggedright
        \input{_figures/aorta/aorta_plots_2}
        \centering
    \end{subfigure}\vspace{-1.0em}
    \ref{leg:aorta}
    \caption{\subref{fig:aorta_sketch}) Sketch of the computational setting for the pressurized aorta case.
    \subref{fig:aorta_velocity}) Slice of the velocity field at point $Q$.
    \subref{fig:aorta_pressure}) Slice of the pressure field at point $Q$.
    d) Inflow profile $U_{in}\qty(t)$ for one period $T$.
    Point $Q$ denotes the time instant in which panels (b) and (c) are shown.
    e) Temporal evolution of the pressure within the aorta.
    f) Transpiration error $\norm{\boldsymbol{\varepsilon}}_1$ as a function of time.}
    \label{fig:aorta}
\end{figure}
The last test case examined is inspired by the hemodynamics within the human aorta, whose simplified geometry is shown in Figure~\ref{fig:aorta_sketch}.
The aorta has an inlet diameter $D = \SI{21}{\milli \meter}$ and begins with a rigid cylindrical segment (region 1) that allows the flow to develop.
The aortic arch (region 2) is deformable and gradually tapers toward the outlet.
A final rigid cylindrical segment (region 3) guides the flow towards the outlet of the aorta.
The initial (free-stress) geometry of the aorta is provided in the Supplementary Material.

The inflow profile to the aorta is given by $u_{in}\qty(t,r) = U_{in}(t)\tanh\qty[\alpha\qty(1-2r/D)]/\tanh\qty(\alpha)$, being $r$ the radial coordinate of the inflow region, $\alpha = 20$ and $U_{in}\qty(t)$ is the $T$-periodic temporal profile depicted in Figure~\ref{fig:aorta}d, whose maximum is $U = \SI{0.5}{\meter\per\second}$ (also provided in the Supplementary Material).
The blood density is set to $\rho = \SI{1000}{\kilo\gram\per\meter^3}$, while the dynamic viscosity is $\mu = \SI{3.5}{\milli\pascal\second}$, corresponding to a Reynolds number based on $U$ and $D$ of $Re = 3000$, and the heart-rate is fixed to $\SI{60}{\bpm}$.
Region 2 is modeled as a two-dimensional, linearly elastic body with Young's modulus $E = \SI{0.6}{\mega\pascal}$, bending stiffness $B = \tfrac{E s^3}{6\sqrt{3}\qty(1-\nu^2)}$, where $\nu = 0.4$ is the Poisson ratio of the structure, and density $\rho_b = 2\rho$, thickness $s = \SI{3.2}{\milli\meter}$.

The computational domain is $\qty[-1.3D, 4.2D]\times\qty[-1.5D, 1.5D]\times\qty[0,9D]$, where the porous region $\qty[2D,4.2D]\times\qty[-1.5D, 1.5D]\times\qty[0,2.8D]$ is introduced, in which a Darcy-like term $-K \vb{u}$ is added to Equation~\eqref{eq:NS_prov} to model the outlet impedance of the aorta, with $K = 80 U/D$, that has been chosen to get an over-pressurized aorta.
The inflow profile is enforced at the south wall using the unsteady Dirichlet boundary condition, while at the northern wall the radiative boundary condition is prescribed.
The computational domain is discretized using $275\times150\times450$ cells, yielding a constant grid size $\Delta = 0.02D$.
The fluid equations are advanced using the RK2+CN scheme, loosely coupled with the same structural solver used for the flag case, which is advanced with the AB scheme.
The size of the time step is kept fixed to $\Delta t = 5\times 10^{-5}D/U$.

The velocity field at peak pressure (point $Q$ of Figure~\ref{fig:aorta}d) is presented in Figure~\ref{fig:aorta_velocity}.
The flow remains relatively uniform along the aorta, with a localized low-velocity region in the lower portion of the aortic arch.
Remarkably, no spurious transpiration velocity is observed.
The corresponding pressure field at the same instant is shown in Figure~\ref{fig:aorta_pressure}.
Within the aorta, the pressure is nearly uniform at approximately \SI{200}{\milli\meter Hg} (\SI{26600}{\pascal}), whereas in the porous region it exhibits a linear decrease toward the bulk pressure ($\approx \SI{0}{\milli \meter Hg}$), in agreement with Darcy’s law.

The time evolution of the pressure within the aorta is shown in Figure~\ref{fig:aorta}e.
For the chosen value of $K$, the peak pressure within the aorta is about $\SI{200}{\milli\meter Hg}$, which is well above the values observed even under pathological conditions ($\approx \SI{120}{\milli\meter Hg}$).
Nevertheless, the transpiration error remains below $8\times 10^{-5}U$ for all the time steps considered, as shown in Figure~\ref{fig:aorta}f.
The same test case is analyzed using both L-MLS and an iterative L-MLS approach \cite{vagnoliLocalExplicitForcing2025}, in which the IB forcing is applied four times per sub-stage to enhance the accuracy of the no-slip condition.
The Lagrangian simulations predict a peak pressure in the aorta which is below \SI{50}{\milli \meter Hg}, and the transpiration across the immersed surface is two order of magnitude larger than the one obtained with the proposed Eulerian IBM.
Interestingly, the L-MLS method with four iterations yields a slightly higher pressure within the aorta.
This increase amplifies the pressure drop across the immersed surface and, despite the use of four iterations, results in a stronger flux across the aortic surface, as shown in Figure~\ref{fig:aorta}d.

\section{Conclusions}
\label{sec:conclusion}
In the present work, a novel sharp-interface immersed boundary method is proposed.
The method exhibits the following features:
\begin{itemize}[itemsep=0pt, parsep=0pt]
    \item Sharp flow reconstruction in the vicinity of the interface, based on an Eulerian IBM;
    \item Applicability to zero-thickness bodies by enforcing the Eulerian IBM on both sides of the immersed interface;
    \item Applicability to moving or deformable bodies with a novel efficient tagging algorithm;
    \item Automatic handling of fresh cells, as the IBM forcing step is done before the predictor of the fluid nodes;
    \item Numerical consistency at boundary cells through the inclusion of source and sink terms arising from the numerical scheme;
    \item Small transpiration error with second-order accuracy, due to the proposed elliptic correction;
    \item Computational efficiency, as the Laplacian system matrix is preserved by the consistent pressure correction, enabling the use of FFT-based direct Poisson solvers without iterative multigrid methods;
    \item Lagrangian evaluation of hydrodynamic loads using MLS kernels, yielding smooth load distributions.
\end{itemize}
These properties make it well suited for the simulation of complex flows involving moving boundaries, zero-thickness structures, and FSI.
It should be noted that, the proposed method does not attempt to improve mass conservation in the boundary cells (as in \cite{kimImmersedBoundaryFiniteVolumeMethod2001, mittalVersatileSharpInterface2008}), but rather avoids enforcing it in order to get a numerically consistent approximation of the exact solution.

The proposed IBM has been extensively assessed across a series of representative flow configurations, including turbulent flows, and both rigid and deformable FSI.
The test cases further include flows characterized by a large pressure drop across immersed surfaces, as well as complex biologically inspired configurations.
Overall, the numerical experiments demonstrate second-order accuracy in the imposition of the no-slip condition on solid walls and highlight the versatility of the method in addressing diverse flow regimes.

The method can be readily integrated into existing numerical frameworks based on the fractional-step method with second-order finite differences.
Although the forcing strategy is Eulerian, the main computational loop for tagging and forcing Eulerian points is performed over the surface triangles, making its cost comparable to that of existing Lagrangian IBMs. 
Moreover, the additional cost associated with the modification of the elliptic equation is limited to the direct evaluation of a single Eulerian field, $q$, whereas the cost of solving the elliptic system remains unchanged.
Therefore, the overall computational cost of the proposed IBM is similar to that of Lagrangian IBMs; for instance, in the test cases here reported, a single time step of the L-MLS method with four iterations has the same cost of the proposed method.

%% file: _figures/general/problem_and_grid.tex
\begin{tikzpicture}
    \def\ang{30}
    \def\Lx{6.0} \def\Ly{4.0} \def\Lz{2.0}
    
    \def\dx{0.3} \def\dy{0.3} \def\dz{0.2}
    
    \coordinate (CF_G) at (\ang:-0.5*\Lz); 
    \coordinate (CB_G) at (\ang:0.5*\Lz);  
    
    \coordinate (CornerG_Back) at ($(CB_G) + (0.5*\Lx, 0.5*\Ly)$);
    \coordinate (CornerG_Front) at ($(CF_G) + (0.5*\Lx, 0.5*\Ly)$);
    \coordinate (CornerG_BottomBack) at ($(CB_G) + (0.5*\Lx, -0.5*\Ly)$);

    \draw[thick, gray, dashed] ($(CB_G)+(-0.5*\Lx,-0.5*\Ly)$) rectangle ($(CB_G)+(0.5*\Lx,0.5*\Ly)$);
    \draw[thick, gray, dashed] ($(CF_G)+(-0.5*\Lx,-0.5*\Ly)$) rectangle ($(CF_G)+(0.5*\Lx,0.5*\Ly)$);
    \draw[thick, gray, dashed] ($(CF_G)+(-0.5*\Lx,-0.5*\Ly)$) -- ($(CB_G)+(-0.5*\Lx,-0.5*\Ly)$); 
    \draw[thick, gray, dashed] ($(CF_G)+(0.5*\Lx,-0.5*\Ly)$) -- ($(CB_G)+(0.5*\Lx,-0.5*\Ly)$);  
    \draw[thick, gray, dashed] ($(CF_G)+(-0.5*\Lx,0.5*\Ly)$) -- ($(CB_G)+(-0.5*\Lx,0.5*\Ly)$);  
    \draw[thick, gray, dashed] ($(CF_G)+(0.5*\Lx,0.5*\Ly)$) -- (CornerG_Back); 

    \coordinate (ShiftSub_Back) at ($(CornerG_Back) - (0.5*\dx, 0.5*\dy) - (\ang:0.5*\dz)$);
    
    \foreach \mx/\my/\mz in {
    0.5/0.5/0.5,
    3.5/0.5/0.5, 2.5/0.5/0.5,2.5/1.5/0.5,1.5/1.5/0.5,1.5/2.5/0.5,
    0.5/1.5/0.5,0.5/2.5/0.5,0.5/3.5/0.5,0.5/0.5/1.5,0.5/0.5/2.5,0.5/0.5/3.5,0.5/1.5/1.5,0.5/2.5/1.5,0.5/1.5/2.5} {
    
        \coordinate (CurrentShift) at ($(CornerG_BottomBack) + (-\mx*\dx, \my*\dy) - (\ang:\mz*\dz)$);

        \begin{scope}[shift={(CurrentShift)}]
            \coordinate (CF_s) at (\ang:-0.5*\dz);
            \coordinate (CB_s) at (\ang:0.5*\dz);
            
            \draw[black] ($(CB_s)+(-0.5*\dx,-0.5*\dy)$) rectangle ($(CB_s)+(0.5*\dx,0.5*\dy)$);
            \draw[black, fill = white] ($(CF_s)+(-0.5*\dx,-0.5*\dy)$) rectangle ($(CF_s)+(0.5*\dx,0.5*\dy)$);
            
            \draw [black, fill = white] plot coordinates {($(CF_s)+(-0.5*\dx,0.5*\dy)$) ($(CB_s)+(-0.5*\dx,0.5*\dy)$) ($(CB_s)+(0.5*\dx,0.5*\dy)$)($(CF_s)+(0.5*\dx,0.5*\dy)$) };
            
            \draw [black, fill = white] plot coordinates {($(CF_s)+(0.5*\dx,0.5*\dy)$) ($(CB_s)+(0.5*\dx,0.5*\dy)$) ($(CB_s)+(0.5*\dx,-0.5*\dy)$)($(CF_s)+(0.5*\dx,-0.5*\dy)$) };
        \end{scope}
    }

    \coordinate (ZoomCenter) at (7.0, 0.8); 
    
    \def\zScale{9.0}

    \draw[black, dashed] ($(CornerG_BottomBack) + (\ang:-\dz) + (0,3*\dy) $) -- ($(ZoomCenter) + (\ang:-0.5*\dz*\zScale) + (-0.5*\dx*\zScale, -0.5*\dy*\zScale)$);
    \draw[black, dashed] ($(CornerG_BottomBack) + (\ang:0) + (-\dx,4*\dy) $) -- ($(ZoomCenter) + (\ang:0.5*\dz*\zScale) + (-0.5*\dx*\zScale, 0.5*\dy*\zScale)$);

    \begin{scope}[shift={(ZoomCenter)}, scale=\zScale]
        \coordinate (ZCF) at (\ang:-0.5*\dz);
        \coordinate (ZCB) at (\ang:0.5*\dz);

        \draw[thick, black] ($(ZCB)+(-0.5*\dx,-0.5*\dy)$) rectangle ($(ZCB)+(0.5*\dx,0.5*\dy)$);
        
        \foreach \i in {-0.5,0.5} \foreach \j in {-0.5,0.5}
            \draw[thick, black] ($(ZCF)+(\i*\dx,\j*\dy)$) -- ($(ZCB)+(\i*\dx,\j*\dy)$);

        \draw[thick, black]  ($(ZCF)+(-0.5*\dx,-0.5*\dy)$) rectangle ($(ZCF)+(0.5*\dx,0.5*\dy)$);

        \draw[very thick, gray, dashed] (-0.5*\dx,0) -- ( 0.5*\dx,0);
        \draw[very thick, gray, dashed] (0,-0.5*\dy) -- (0,0.5*\dy);
        \draw[very thick, gray, dashed] (ZCB) -- (ZCF);

        \node[text=mycolor1, rotate=0, font=\boldmath] at ( 0.5*\dx,0) {$>$};
        \node[text=mycolor1, rotate=0, font=\boldmath] at (-0.5*\dx,0) {$>$};
        
        \node[text=mycolor3, rotate=90, font=\boldmath] at (0, 0.5*\dy) {$>$};
        \node[text=mycolor3, rotate=90, font=\boldmath] at (0,-0.5*\dy) {$>$};
        
        \node[text=mycolor4, rotate=\ang, font=\boldmath] at (ZCF) {$>$};
        \node[text=mycolor4, rotate=\ang, font=\boldmath] at (ZCB) {$>$};
        
        \filldraw[mycolor2,scale = {1/\zScale}] (0,0) circle (3pt);
        \node[anchor = south east] at (0,0) {$p$};
        \node[anchor = south] at (-0.5*\dx,0) {$u_3$};
        \node[anchor = east] at (0,-0.5*\dy) {$u_2$};
        \node[anchor = north east] at (\ang:-0.5*\dz) {$u_1$};
        
        \draw[thick, black, <->] ($(ZCB)+(-0.5*\dx,+0.5*\dy) + (0,0.1*\dy)$) -- +(\dx, 0) node [pos=0.5, anchor = south] {$\Delta x_3^k$};
        \draw[thick, black, <->] ($(ZCF)+( 0.5*\dx,-0.5*\dy) + (0.1*\dx,0) + (\ang:\dz)$) -- +(0, \dy) node [pos=0.5, anchor = west] {$\Delta x_2$};
        \draw[thick, black, <->] ($(ZCF)+(- 0.5*\dx, 0.5*\dy) + (0,0.1*\dy)$) -- +(\ang:\dz) node [pos=0.5, anchor = south east] {$\Delta x_1$};
        
        \matrix [
            matrix of nodes,
            anchor=north,
            at = {($(ZCF) + (0.3*\dx,-0.55*\dy)$)},
            nodes={anchor=west},
    ] {
        \node[text=mycolor4, font=\boldmath] {$>$}; & $u_1$ grid &
        \node[text=mycolor3, font=\boldmath] {$>$}; & $u_2$ grid \\
        \node[text=mycolor1, font=\boldmath] {$>$}; & $u_3$ grid &
        \filldraw[mycolor2] (0.3,0) circle (3pt);   & $p$ grid  \\
    };
    \end{scope}

    \draw[thick, black, <->] ($(CB_G)+(-0.5*\Lx,+0.5*\Ly) + (0,0.1*\Ly)$) -- +(\Lx, 0) node [pos=0.5, anchor = south] {$L_3$};
    \draw[thick, black, <->] ($(CF_G)+(-0.5*\Lx,-0.5*\Ly) - (0.08*\Lx,0)$) -- +(0, \Ly) node [pos=0.5, anchor = east] {$L_2$};
    \draw[thick, black, <->] ($(CF_G)+(-0.5*\Lx, 0.5*\Ly) + (0.0,0.1*\Ly)$) -- +(\ang:\Lz) node [pos=0.5, anchor = south east] {$L_1$};
    
    \draw[thick, black, ->] ($(CF_G)+(-0.5*\Lx, -0.5*\Ly)$) -- +(1.5, 0)   node [pos=1.0, anchor = south  ] {$x_3$};
    \draw[thick, black, ->] ($(CF_G)+(-0.5*\Lx, -0.5*\Ly)$) -- +(0, 1.5)   node [pos=1.0, anchor = west   ] {$x_2$};
    \draw[thick, black, ->] ($(CF_G)+(-0.5*\Lx, -0.5*\Ly)$) -- +(\ang:1.0) node [pos=1.0, anchor = south  ] {$x_1$};

    \node[yslant=tan(\ang)] at ( 0.5*\Lx,0.0) {North};
    \node[yslant=tan(\ang)] at (-0.5*\Lx,0.0) {South};
    \node[anchor = north east] at ($(0.5*\Lx,0.5*\Ly) - (\ang:0.5*\Lz)$) {Periodic};
    \node[xslant=1/tan(\ang),yscale=sin(\ang), anchor = south east] at ($(0.5*\Lx,0.5*\Ly)-(\ang:0.5*\Lz)$) {Periodic};
    
    
    \node[] at ($(-0.7*\Lx,0.9*\Ly)$) {(a)};
    \node[] at ($( 0.8*\Lx,0.9*\Ly)$) {(b)};

    \pgfmathsetmacro\a{0.8}
    \pgfmathsetmacro\b{0.5}
    \pgfmathsetmacro\rot{30}
    \pgfmathsetmacro{\tilt}{70}
    \pgfmathsetmacro{\flatten}{cos(\tilt)}

    \coordinate (CC) at ($(0, -0.05*\Ly)$); 
    
    \begin{scope}[rotate around={\rot:(CC)}]
        \shade[ball color=gray!60] (CC) ellipse ({\a} and {\b});
        \draw[black, line width=0.6pt] (CC) ++(-\a,0) arc[start angle=180, end angle=360, x radius=\a, y radius=0.35*\b];
        \draw[black, dashed, line width=0.6pt] (CC) ++(\a,0) arc[start angle=0, end angle=180, x radius=\a, y radius=\flatten*\b];
        \draw[black, line width=0.6pt] (CC) ++(0,\b) arc[start angle=90, end angle=270, x radius=\flatten*\a, y radius=\b];
        \draw[black, dashed, line width=0.6pt] (CC) ++(0,-\b) arc[start angle=270, end angle=450, x radius=\flatten*\a, y radius=\b];
        \draw[very thick, <->] ($(CC) - (0:0.9*\a) - (0,0.8) $)--($(CC) + (0:0.9*\a) - (0,0.8)$) node [pos = 0.6, anchor = north] {$L$};
        \node[anchor = south west, inner sep = 18pt] at (CC) {$\mathcal{S}$};
    \end{scope}

    \draw[very thick, blue, ->] ($(CF_G) +(-0.2*\Lx, 0)$) -- +(0.8, 0)   node [pos=0.5, anchor = south, color = black] {$U$};

    \node[anchor = south east] at ($(0.5*\Lx,-0.5*\Ly) - (\ang:0.5*\Lz)$) {$\Omega$};
\end{tikzpicture}

%% file: _figures/general/tagging_double.tex
\begin{tikzpicture}
    \input{_figures/general/grid_and_surface_double}

    \fill [mycolor3,very thick] ( 0*\step, 1*\step) node [anchor = south east,black] {$\vb{x}^k$}     circle (4pt);
    \draw [gray    ,very thick] ( 1*\step, 1*\step) node [anchor = south west,black] {$\vb{x}^{k,n=1}$} circle (4pt);
    \draw [gray    ,very thick] ( 0*\step, 2*\step) node [anchor = south west,black] {$\vb{x}^{k,n=2}$} circle (4pt);
    \draw [gray    ,very thick] (-1*\step, 1*\step) node [anchor = south east,black] {$\vb{x}^{k,n=3}$} circle (4pt);
    \draw [gray    ,very thick] ( 0*\step, 0*\step) node [anchor = north west,black] {$\vb{x}^{k,n=4}$} circle (4pt);
    
    \fill[mycolor4] (Wt) circle (4pt) node[anchor = north east, color = black] {$\vb{X}^k$};
    \node[anchor = south east ]at (-2*\step, 3.5*\step) {(a)};

    \begin{scope}[xshift = 7cm]
        \node[anchor = south east ]at (-2*\step, 3.5*\step) {(b)};
        \input{_figures/general/grid_and_surface_double}
        \fill[mycolor2] (-\step  , \step  )  circle (4pt);
        \fill[mycolor2] (-\step  , 3*\step)  circle (4pt);
        \fill[mycolor2] (-2*\step, 3*\step)  circle (4pt);
        \fill[mycolor2] (-\step  , 2*\step)  circle (4pt);
        \fill[mycolor2] (-2*\step, 2*\step)  circle (4pt);
        \fill[mycolor2] ( 0.0    , \step  )  circle (4pt);
        \fill[mycolor2] ( \step  , 0.0    )  circle (4pt);
        \fill[mycolor2] ( 0.0    , 0.0    )  circle (4pt);
        \fill[mycolor2] ( 2*\step, 0.0    )  circle (4pt);
        \fill[mycolor2] ( 2*\step, \step  )  circle (4pt);
        \fill[mycolor2] ( \step  , \step  )  circle (4pt);
        
        \fill[mycolor4] (Wt) circle (4pt) node[anchor = north east, color = black] {$\vb{X}^f$};
        \fill[mycolor4] (Wb) circle (4pt);
        
        \node[anchor = south east, inner sep = 4pt,black] at (Fp) {$\vb{x}^f$};
        
        \coordinate (Ep) at ($(Wt)!4.0!(Fp)$);
        \draw [gray, very thick, dashed] (Wt) node[circle, fill = mycolor4, minimum size=8pt, inner sep=0pt]{} -- (Ep)
        node[circle, fill = mycolor1, minimum size=8pt, inner sep=0pt]{}
        node[anchor = south west, inner sep = 4pt, black]{$\vb{x}^e$};
        \coordinate (Em) at ($(Wb)!4.0!(Fm)$);
        \draw [gray, very thick, dashed] (Wb) node[circle, fill = mycolor4, minimum size=8pt, inner sep=0pt]{} -- (Em)
        node[circle, fill = mycolor1, minimum size=8pt, inner sep=0pt]{};
        
        \draw[very thick,black] (\step, 2*\step)  circle (6pt);
        \draw[very thick,black] (0*\step, 2*\step)  circle (6pt);
        \draw[very thick,black] (0*\step, 3*\step)  circle (6pt);
        \draw[very thick,black] (\step, 3*\step)  circle (6pt);
        
        \draw[very thick,black] (-1*\step, -2*\step)  circle (6pt);
        \draw[very thick,black] (0*\step, -2*\step)  circle (6pt);
        \draw[very thick,black] (0*\step, -1*\step)  circle (6pt);
        \draw[very thick,black] (-1*\step, -1*\step)  circle (6pt);
        
        \coordinate (Wt3) at ($(Wt) + (10:0.3*\step)$);
        \coordinate (Ep3) at ($(Ep) + (10:0.3*\step)$);
        \draw[very thick, black, <->] (Wt3) -- (Ep3) node [black, pos = 0.5, anchor = west] {$h$};
    \end{scope}

    \matrix(legA) [
        matrix of nodes,
        fill=white,
        anchor=north,
        at={(current bounding box.south)}, 
        row sep=0.2cm,
        column sep=0.2cm,
        nodes={anchor=west}
    ] {
        \fill[mycolor3,very thick] (0,0) circle (4pt); & Analyzed point & \fill[mycolor4] (0,0) circle (4pt); & Wall points  & \fill[mycolor2] (0,0) circle (4pt); & Forcing points\\ \fill[black] (0,0) circle (2pt); \draw [gray,very thick] ( 0, 0) circle (4pt); & Neighbor points  & \fill[black] (0,0) circle (2pt); \draw [very thick] ( 0, 0) circle (6pt);  & Interpolation points &\fill[mycolor1] (0,0) circle (4pt); & External points\\
    };
    \matrix [
        matrix of nodes,
        fill=white,
        anchor=north,
        at={(legA.south)},
        yshift=0.2cm,
        row sep=0.2cm,
        column sep=0.2cm,
        nodes={anchor=west}
    ] {
        \draw[gray, line width=3pt] (-0.3,0) -- (0.3,0); & Surface of the body \\
    };

\end{tikzpicture}

%% file: _figures/general/stencil.tex
\begin{tikzpicture}[scale=1, transform shape=false]
    \def\step{1.4}   
    \def\thck{0.2*\step}   

    \coordinate (OW) at (-2*\step  -0.2*\step, 1.3*\step);
    \coordinate (BW) at (-0.3*\step, 1.0*\step);
    \coordinate (CW) at ( 3*\step + 0.2*\step, -0.8*\step);
    
    \draw[step=\step, shift={({\step}, {0.0*\step})}] (-3*\step - 0.2*\step, -1.*\step - 0.2*\step) grid (2*\step + 0.2*\step, 2*\step + 0.2*\step);
    \foreach \i in {-2, -1, 0, 1, 2,3} {
        \foreach \j in {-1, 0, 1, 2} {
            \filldraw[black]  (\i*\step, \j*\step) circle (2pt);
        }
    }
    
    \draw [gray , line width = 3pt] plot [smooth] coordinates {(OW) (BW) (CW)};

    \fill[mycolor2] (-2*\step, 1*\step)  circle (4pt);
    \fill[mycolor2] (-1*\step, 1*\step)  circle (4pt);
    \fill[mycolor2] ( 0*\step, 0*\step)  circle (4pt);
    \fill[mycolor2] ( 1*\step, 0*\step)  circle (4pt);
    \fill[mycolor2] ( 2*\step,-1*\step)  circle (4pt);
    \fill[mycolor2] (-2*\step, 2*\step)  circle (4pt);
    \fill[mycolor2] (-1*\step, 2*\step)  circle (4pt);
    \fill[mycolor2] ( 0*\step, 1*\step)  circle (4pt);
    \fill[mycolor2] ( 1*\step, 1*\step)  circle (4pt);
    \fill[mycolor2] ( 2*\step, 0*\step)  circle (4pt);
    \fill[mycolor2] ( 3*\step,-1*\step)  circle (4pt);
    \fill[mycolor2] ( 3*\step, 0*\step)  circle (4pt);

    \coordinate (Pn) at (-1*\step,0*\step);
    \coordinate (Pp) at ( 2*\step,1*\step);

    \draw[gray, dashed, thick] ($(Pn) + (-\step,  \thck) $) -- +(2*\step,0);
    \draw[gray, dashed, thick] ($(Pn) + (-\step, -\thck) $) -- +(2*\step,0);
    \draw[gray, dashed, thick] ($(Pn) + ( \thck, -\step) $) -- +(0,2*\step);
    \draw[gray, dashed, thick] ($(Pn) + (-\thck, -\step) $) -- +(0,2*\step);
    
    \draw[gray, dashed, thick] ($(Pp) + (-\step,  \thck) $) -- +(2*\step,0);
    \draw[gray, dashed, thick] ($(Pp) + (-\step, -\thck) $) -- +(2*\step,0);
    \draw[gray, dashed, thick] ($(Pp) + ( \thck, -\step) $) -- +(0,2*\step);
    \draw[gray, dashed, thick] ($(Pp) + (-\thck, -\step) $) -- +(0,2*\step);

    \draw[gray, dashed, thick] ($(Pn)+(-\step,0)$) ++(90:\thck) arc[start angle=90,end angle=270,radius=\thck];
    \draw[gray, dashed, thick] ($(Pn)+(\step,0)$) ++(90:\thck) arc[start angle=90,end angle=-90,radius=\thck];
    \draw[gray, dashed, thick] ($(Pn)+(0,\step)$) ++(180:\thck) arc[start angle=180,end angle=0,radius=\thck];
    \draw[gray, dashed, thick] ($(Pn)+(0,-\step)$) ++(180:\thck) arc[start angle=180,end angle=360,radius=\thck];
    \draw[gray, dashed, thick] ($(Pp)+(-\step,0)$) ++(90:\thck) arc[start angle=90,end angle=270,radius=\thck];
    \draw[gray, dashed, thick] ($(Pp)+(\step,0)$) ++(90:\thck) arc[start angle=90,end angle=-90,radius=\thck];
    \draw[gray, dashed, thick] ($(Pp)+(0,\step)$) ++(180:\thck) arc[start angle=180,end angle=0,radius=\thck];
    \draw[gray, dashed, thick] ($(Pp)+(0,-\step)$) ++(180:\thck) arc[start angle=180,end angle=360,radius=\thck];

    \matrix [
        matrix of nodes,
        fill=white,
        anchor=west,
        at={(current bounding box.east)}, 
        xshift=0.5cm,
        row sep=0.2cm,
        column sep=0.2cm,
        nodes={anchor=west, align=left}
    ] {
        \fill[mycolor2] (0,0) circle (4pt); & Forcing points \\ 
        \fill[black   ] (0,0) circle (2pt); & Fluid points \\
        \draw[gray, dashed, thick] (-0.3*\step,  0.3*\thck) -- +(0.6*\step,0); \draw[gray, dashed, thick] (-0.3*\step, -0.3*\thck) -- +(0.6*\step,0); \draw[gray, dashed, thick] ( 0.3*\thck, -0.3*\step) -- +(0,0.6*\step); \draw[gray, dashed, thick] (-0.3*\thck, -0.3*\step) -- +(0,0.6*\step);\draw[gray, dashed, thick] (-0.3*\step,0) ++(90:0.3*\thck) arc[start angle=90,end angle=270,radius=0.3*\thck]; \draw[gray, dashed, thick] (0.3*\step,0) ++(90:0.3*\thck) arc[start angle=90,end angle=-90,radius=0.3*\thck]; \draw[gray, dashed, thick] (0,0.3*\step) ++(180:0.3*\thck) arc[start angle=180,end angle=0,radius=0.3*\thck]; \draw[gray, dashed, thick] (0,-0.3*\step) ++(180:0.3*\thck) arc[start angle=180,end angle=360,radius=0.3*\thck];& Stencil of the Laplacian\\
        \draw[gray, line width=3pt] (-0.3,0) -- (0.3,0); & Surface of the body \\
    };
\end{tikzpicture}

%% file: _figures/general/fresh_cell.tex
\begin{tikzpicture}[scale = 1, transform shape=false]
    \def\step{1.4}   
    \def\wshift{1*\step}   
    \def\xshift{7cm}   

    \coordinate (OW) at (-2*\step  -0.2*\step, 0.3*\step);
    \coordinate (BW) at (-0.3*\step, -0.0*\step);
    \coordinate (CW) at ( 2*\step + 0.2*\step, -1.3*\step);
    
    \coordinate (OWn) at ($(OW) + (0,\wshift)$);
    \coordinate (BWn) at ($(BW) + (0,\wshift)$);
    \coordinate (CWn) at ($(CW) + (0,\wshift)$);
    
    \coordinate (OWs) at ($(OW) + (\xshift,0)$);
    \coordinate (BWs) at ($(BW) + (\xshift,0)$);
    \coordinate (CWs) at ($(CW) + (\xshift,0)$);
    
    \coordinate (OWns) at ($(OWs) + (0,\wshift)$);
    \coordinate (BWns) at ($(BWs) + (0,\wshift)$);
    \coordinate (CWns) at ($(CWs) + (0,\wshift)$);

    \draw[step=\step, shift={({\step}, {0.0*\step})}] (-3*\step - 0.2*\step, -1.*\step - 0.2*\step) grid (1*\step + 0.2*\step, 2*\step + 0.2*\step);
    \foreach \i in {-2, -1, 0, 1, 2} {
        \foreach \j in {-1, 0, 1, 2} {
            \filldraw[black]  (\i*\step, \j*\step) circle (2pt);
        }
    }
    
    \draw [gray , line width = 3pt] plot [smooth] coordinates {(OWn) (BWn) (CWn)};

    \fill[mycolor1] (-2*\step, 1*\step)  circle (4pt);
    \fill[mycolor1] (-1*\step, 1*\step)  circle (4pt);
    \fill[mycolor1] ( 0*\step, 0*\step)  circle (4pt);
    \fill[mycolor1] ( 1*\step, 0*\step)  circle (4pt);
    \fill[mycolor1] ( 2*\step,-1*\step)  circle (4pt);
    
    \fill[mycolor2] (-2*\step, 2*\step)  circle (4pt);
    \fill[mycolor2] (-1*\step, 2*\step)  circle (4pt);
    \fill[mycolor2] ( 0*\step, 1*\step)  circle (4pt);
    \fill[mycolor2] ( 1*\step, 1*\step)  circle (4pt);
    \fill[mycolor2] ( 2*\step, 0*\step)  circle (4pt);

    \node[anchor = south east ]at (-2.0*\step, 2.5*\step) {(a)};
    \node[anchor = south ]at (0*\step, 2.2*\step) {$t^n$ before moving the surface};

    \begin{scope}[xshift = \xshift]
        \node[anchor = south east ]at (-2.0*\step, 2.5*\step) {(b)};
        \node[anchor = south ]at (0*\step, 2.2*\step) {$t^n$ after moving the surface};
        \draw[step=\step, shift={({\step}, {0.0*\step})}] (-3*\step - 0.2*\step, -1.*\step - 0.2*\step) grid (1*\step + 0.2*\step, 2*\step + 0.2*\step);
        \foreach \i in {-2, -1, 0, 1, 2} {
            \foreach \j in {-1, 0, 1, 2} {
                \filldraw[black]  (\i*\step, \j*\step) circle (2pt);
            }
        }
        \draw [gray,dashed, line width = 3pt] plot [smooth] coordinates {(OWns) (BWns) (CWns)};
        
        \draw [gray, line width = 3pt] plot [smooth] coordinates {(OWs) (BWs) (CWs)};
        
        \draw[thick, mycolor4, <-] (-1.5*\step,  0.3 *\step) -- +(0,0.8*\step);
        \draw[thick, mycolor4, <-] (-0.5*\step,  0.15*\step) -- +(0,0.8*\step);
        \draw[thick, mycolor4, <-] ( 0.5*\step, -0.3 *\step) -- +(0,0.8*\step);
        \draw[thick, mycolor4, <-] ( 1.5*\step, -0.85*\step) -- +(0,0.8*\step);
        
        \fill[mycolor1] (-2*\step, 1*\step)  circle (4pt);
        \fill[mycolor1] (-1*\step, 1*\step)  circle (4pt);
        \fill[mycolor1] ( 0*\step, 0*\step)  circle (4pt);
        \fill[mycolor1] ( 1*\step, 0*\step)  circle (4pt);
        \fill[mycolor1] ( 2*\step,-1*\step)  circle (4pt);
        
        \fill[mycolor2] (-2*\step, 2*\step)  circle (4pt);
        \fill[mycolor2] (-1*\step, 2*\step)  circle (4pt);
        \fill[mycolor2] ( 0*\step, 1*\step)  circle (4pt);
        \fill[mycolor2] ( 1*\step, 1*\step)  circle (4pt);
        \fill[mycolor2] ( 2*\step, 0*\step)  circle (4pt);

        \draw[mycolor3, very thick] (-2*\step, 1*\step)  circle (6pt);
        \draw[mycolor3, very thick] (-1*\step, 1*\step)  circle (6pt);
        \draw[mycolor3, very thick] ( 0*\step, 0*\step)  circle (6pt);
        \draw[mycolor3, very thick] ( 1*\step, 0*\step)  circle (6pt);
        \draw[mycolor3, very thick] ( 2*\step,-1*\step)  circle (6pt);
        
    \end{scope}

\matrix (legA) [
    matrix of nodes,
    fill=white,
    anchor=north,
    at={(current bounding box.south)},
    row sep=0.2cm,
    column sep=0.2cm,
    nodes={anchor=west}
] {
    \fill[mycolor2] (0,0) circle (4pt); & Forcing points at $t^n$ &
    \fill[mycolor1] (0,0) circle (4pt); & Forcing points at $t^{n+1}$ &
    \draw[mycolor3, very thick] (0,0) circle (6pt); & Fresh cells at $t^{n+1}$ \\
};

\matrix [
    matrix of nodes,
    fill=white,
    anchor=north,
    at={(legA.south)},
    yshift=0.2cm,
    row sep=0.2cm,
    column sep=0.2cm,
    nodes={anchor=west}
] {
    \draw[gray, line width=3pt] (-0.3,0) -- (0.3,0); &
    Body surface at current time &
    \draw[gray, dashed, line width=3pt] (-0.3,0) -- (0.3,0); &
    Body surface at previous time\\
};

\end{tikzpicture}

%% file: _figures/general/fluid_and_boundary_cells.tex
\begin{tikzpicture}[scale=1.0, transform shape=false]
    \def\step{1.2}
    \def\opacity{0.7}

    \draw[step=\step] (-5*\step,-3*\step) grid (5*\step,3*\step);

    \foreach \i in {-2, -1, 0, 1, 2} {
        \draw[fill=mycolor1,draw=none,fill opacity=\opacity] (\i*\step, 0*\step) rectangle ++(\step,\step);
    }
    \foreach \i in {-3, -2, -1, 0, 1} {
        \draw[fill=mycolor1,draw=none,fill opacity=\opacity] (\i*\step,-1*\step) rectangle ++(\step,\step);
    }
    
    \foreach \i in {-3, -2, -1, 0, 1, 2, 3} {
        \draw[fill=mycolor2,draw=none,fill opacity=\opacity] (\i*\step, 1*\step) rectangle ++(\step,\step);
    }
    \foreach \i in {-4, -3, -2, -1, 0, 1, 2} {
        \draw[fill=mycolor2,draw=none,fill opacity=\opacity] (\i*\step,-2*\step) rectangle ++(\step,\step);
    }
    
    \draw[fill=mycolor2,draw=none,fill opacity=\opacity] (-3*\step, 0*\step) rectangle ++(\step,\step);
    \draw[fill=mycolor2,draw=none,fill opacity=\opacity] ( 3*\step, 0*\step) rectangle ++(\step,\step);
    \draw[fill=mycolor2,draw=none,fill opacity=\opacity] ( 3*\step,-1*\step) rectangle ++(\step,\step);
    \draw[fill=mycolor2,draw=none,fill opacity=\opacity] ( 2*\step,-1*\step) rectangle ++(\step,\step);
    \draw[fill=mycolor2,draw=none,fill opacity=\opacity] (-4*\step,-1*\step) rectangle ++(\step,\step);
    \draw[fill=mycolor2,draw=none,fill opacity=\opacity] (-4*\step, 0*\step) rectangle ++(\step,\step);
    
    \foreach \i in {-5, -4, -3, -2, -1, 0, 1, 2, 3, 4} {
        \draw[fill=mycolor1,draw=none,fill opacity=\opacity] (\i*\step, 2*\step) rectangle ++(\step,\step);
    }
    \foreach \i in {-5, -4, -3, -2, -1, 0, 1, 2, 3, 4} {
        \draw[fill=mycolor1,draw=none,fill opacity=\opacity] (\i*\step,-3*\step) rectangle ++(\step,\step);
    }
    \foreach \j in {-2, -1, 0, 1}{
        \draw[fill=mycolor1,draw=none,fill opacity=\opacity] (-5*\step,\j*\step) rectangle ++(\step,\step);
    }
    \foreach \j in {-2, -1, 0, 1}{
        \draw[fill=mycolor1,draw=none,fill opacity=\opacity] ( 4*\step,\j*\step) rectangle ++(\step,\step);
    }
    \draw[fill=mycolor1,draw=none,fill opacity=\opacity] (-4*\step, 1*\step) rectangle ++(\step,\step);
    \draw[fill=mycolor1,draw=none,fill opacity=\opacity] ( 3*\step,-2*\step) rectangle ++(\step,\step);

    \draw[line width=4pt, mycolor4] (-5*\step,-3*\step) rectangle (5*\step,3*\step);

    \begin{scope}[rotate=12, scale=1.15*\step, yscale=1.20*\step]
      \draw[black, line width = 3pt, smooth cycle, tension=1.0]
        plot coordinates {
          (3.20,  0.00)
          (2.70,  0.55)
          (1.85,  0.92)
          (0.85,  0.82)
          (0.00,  1.05)
          (-0.85, 0.82)
          (-1.85, 0.92)
          (-2.70, 0.55)
          (-3.20, 0.00)
          (-2.70,-0.55)
          (-1.85,-0.92)
          (-0.85,-0.82)
          (0.00, -1.05)
          (0.85, -0.82)
          (1.85, -0.92)
          (2.70, -0.55)
        };
    \end{scope}

    \node[anchor=south west, inner sep=2.0pt] at ( 0.5*\step,0.5*\step) {$R_F$};
    \node[anchor=south west, inner sep=2.0pt] at (-4.5*\step,1.5*\step) {$R_F$};
    \node[anchor=south west, inner sep=2.0pt] at ( 3.5*\step,1.5*\step) {$R_B$};

    \foreach \j in {-3, ..., 2}{
        \foreach \i in {-5, ..., 5}{
            \node[text=gray, rotate=0, font=\boldmath] at (\i*\step, \j*\step + 0.5*\step) {$>$};
        }
    }
    \foreach \j in {-3, ..., 3}{
        \foreach \i in {-5, ..., 4}{
            \node[text=gray, rotate=90, font=\boldmath] at (\i*\step+ 0.5*\step, \j*\step) {$>$};
        }
    }
    \foreach \j in {-3, ..., 2}{
        \foreach \i in {-5, ..., 4}{
            \draw[fill = gray, draw = none] (\i*\step + 0.5*\step, \j*\step + 0.5*\step) circle (2pt);
        }
    }

    \matrix [
        matrix of nodes,
        fill=white,
        anchor=north,
        at={(current bounding box.south)}, 
        row sep=0.2cm,
        column sep=0.2cm,
        nodes={anchor=west}
    ] {
         \draw[fill=mycolor1,draw=none,fill opacity=\opacity] (-0.3*\step, -0.3*\step) rectangle ++(0.6*\step,0.6*\step); & $\mathcal{C}_F$ &
         \draw[fill=mycolor2,draw=none,fill opacity=\opacity] (-0.3*\step, -0.3*\step) rectangle ++(0.6*\step,0.6*\step); & $\mathcal{C}_B$ &
         $\mathcal{C} = \mathcal{C}_F \cup \mathcal{C}_B$&
         \draw[line width=4pt, mycolor4] (-0.4*\step,0) -- +(0.8*\step,0); & $\mathcal{F}$\\
    };
    
\end{tikzpicture}

%% file: _figures/general/elliptic_cell.tex
\begin{tikzpicture}[scale=1.0, transform shape=false]
    \def\step{2.5}   
    \def\wshift{-0.0*\step}   

    \coordinate (OW) at (-2.0*\step-0.2*\step, 0.75*\step+0.30*\step - \wshift);
    \coordinate (AW) at (-1.1*\step          , 0.60*\step+0.30*\step - \wshift);
    \coordinate (BW) at (-0.3*\step          , 0.40*\step+0.30*\step - \wshift);
    \coordinate (CW) at ( 1.1*\step          ,-0.20*\step+0.30*\step - \wshift);
    \coordinate (DW) at ( 2.3*\step          ,-\step     +0.30*\step - 0.2*\step);
    
    \draw[step=\step] (-2*\step - 0.2*\step,-1*\step - 0.2*\step) grid (1*\step + 0.2*\step,1*\step + 0.2*\step);
    
    \draw [gray , line width = 3pt] plot [smooth] coordinates {(OW) (AW) (BW) (CW)};
    
    \draw[black, ultra thick] (-2*\step,0) rectangle (-\step,\step);
    \draw[black, ultra thick] (-\step,0) rectangle (0,\step);
    \draw[black, ultra thick] (0,0) rectangle (\step,\step);
    
    \foreach \i in {-2, -1, 0} {
        \foreach \j in {-1, 0, 1} {
            \node[text=mycolor3, rotate=90, font=\boldmath] at (\i*\step + 0.5*\step, \j*\step) {$>$};
        }
    }
    \foreach \i in {-2, -1, 0,1} {
        \foreach \j in {-1,0} {
            \node[text=mycolor1, rotate=0, font=\boldmath] at (\i*\step, \j*\step + 0.5*\step) {$>$};
        }
    }
    \foreach \i in {-2, -1, 0} {
        \foreach \j in {-1,0} {
            \filldraw[mycolor2]  (\i*\step+ 0.5*\step, \j*\step+ 0.5*\step) circle (3pt);
        }
    }
    
    \draw[mycolor4,fill = mycolor4] (-0.5*\step,0.77*\step) circle (4pt) node[color = black, anchor = north east] {$\vb{X}$};

    \draw[gray, very thick, <->] ($(-0.5*\step,0) + (0.2*\step,0)$) -- +(0, 0.7*\step) node [anchor = south, color = black, pos = 0.5, rotate = -90] {$\Delta\alpha^-$};
    \draw[gray, very thick, <->] ($(-0.5*\step,\step) + (0.2*\step,0)$) -- +(0,-0.3*\step) node [anchor = south, color = black, pos = 0.5, rotate = -90] {$\Delta\alpha^+$};

    \node[anchor = north east] at (-0.5*\step,0.5*\step) {$\vb{x}^b$};
    \node[] at (0.2*\step,0.6*\step) {$+$};
    \node[] at (0.2*\step,0.4*\step) {$-$};

    \matrix [
        matrix of nodes,
        fill=white,
        anchor=west,
        at={(current bounding box.east)}, 
        row sep=0.2cm,
        column sep=0.2cm,
        nodes={anchor=west}
    ] {
        \fill[mycolor2] (0,0) circle (3pt); & $p$ grid points \\
        \node[text=mycolor1, rotate=0 , font=\boldmath] at (-0.25,0) {$>$}; &$u_1$ grid points \\
        \node[text=mycolor3, rotate=90, font=\boldmath] at (0,-0.25) {$>$}; &$u_2$ grid points \\
        \fill[mycolor4] (0,0) circle (4pt); & Intersection point $\vb{X}$ \\
        \draw[black, ultra thick] (-.15*\step,-.15*\step) rectangle (.15*\step,.15*\step); & Boundary cells \\
        \draw[gray, line width=3pt] (-0.3,0) -- (0.3,0); & Surface of the body \\
        \node[] at (-0.25,0.0) {$+$}; & Positive side \\
        \node[] at (-0.25,0.0) {$-$}; & Negative side \\
    };

    \draw[very thick,->] (-2*\step,-\step) -- +(0.7*\step,0) node[anchor = south] {$x_1$};
    \draw[very thick,->] (-2*\step,-\step) -- +(0,0.7*\step) node[anchor = west ] {$x_2$};
    
\end{tikzpicture}

%% file: _figures/lid_driven_cavity/streamlines.tex
\begin{tikzpicture}
    
\begin{axis}[
    name=myaxis,
    axis equal image,
    xmin = 0,xmax = 2,
    ymin = 0,ymax = 1,
    xlabel = {$x_1/L$},
    ylabel = {$\frac{x_2}{L}$},
    ytick={0 ,0.5,1},
    xmajorgrids,
    ymajorgrids,
    axis lines=box,
    ylabel style = {rotate = -90, font = \large, anchor = east},
    clip=false,
    tick label style={font=\footnotesize},
]
\addplot[black,line width=1.0pt,no markers,unbounded coords=jump,] table[x index=0, y index=1] {_data/lid_driven_cavity/cavity_streamlines.txt};
\draw[very thick, blue, ->] ( 0.3,1.05) -- +(0.4,0) node[anchor = south, pos = 0.5,color = black] {$U$};
\draw[very thick, blue, ->] ( 1.3,1.05) -- +(0.4,0) node[anchor = south, pos = 0.5,color = black] {$U$};
\node [anchor = south] at (0  ,1.05) {(a)};
\node [anchor = south] at (2.2,1.05) {(b)};
\end{axis}

\def\step{0.8};
\draw[thick,gray] ($(myaxis) - (0,1)$) circle (0.15);
\draw[thick,gray] ($(myaxis) - (0,1) + (0,0.15)$) -- ($(myaxis.east) + (2.5cm,0.0) + (120:2.0)$);
\draw[thick,gray] ($(myaxis) - (0,1) - (0,0.15)$) -- ($(myaxis.east) + (2.5cm,0.0) - (60:2.0)$);
\node[anchor = south, rotate = 90,color = black,font = \footnotesize, inner sep = 1pt, fill =white,yshift=5pt] at ($0.5*(myaxis.north) + 0.5*(myaxis.south)$){Immersed surface};
\draw[line width=2pt,dashed,mycolor4, fill = white] (myaxis.north) -- (myaxis.south);

\begin{scope}[shift={(myaxis.east)},xshift = 2.5cm, yshift = 0.0cm, scale=2.0]

    \clip (0,0) circle (1);
    \draw[thick,gray] (0,0) circle (1);
    \foreach \i in {-1,0,1} {
        \foreach \j in {-1,0,1} {
            \node[text=mycolor3, rotate=0, font=\boldmath] at (\i*\step + 0.5*\step, \j*\step) {$>$};
        }
    }
    \foreach \i in {-1,0,1} {
        \foreach \j in {-1,0,1} {
            \node[text=mycolor1, rotate=90, font=\boldmath] at (\i*\step, \j*\step + 0.5*\step) {$>$};
        }
    }
    \foreach \i in {-1,0,1} {
        \foreach \j in {-1,0,1} {
            \filldraw[mycolor2]  (\i*\step, \j*\step) circle (1pt);
            \draw (\i*\step-0.5*\step,\j*\step-0.5*\step) rectangle (\i*\step + 0.5*\step,\j*\step + 0.5*\step);
        }
    }
    \draw[very thick, dashed, mycolor4] (-0.25, -1) -- (-0.25, 1);
\end{scope}
\node [anchor = north west, fill = white, font = \footnotesize, inner sep = 1pt] at (myaxis.north west) {$(-)$};
\node [anchor = north west, fill = white, font = \footnotesize, inner sep = 1pt] at (myaxis.north) {$(+)$};

\end{tikzpicture}

%% file: _figures/lid_driven_cavity/ldc_profiles.tex
\begin{tikzpicture}
\begin{groupplot}[
    group style={
        group size=2 by 1,
        horizontal sep=2.em,
    },
    width = 0.55\textwidth,
    height = 0.6*0.55*\textwidth,
    trim axis left
]
\nextgroupplot[
            xmin= 0,
            xmax= 1,
            xlabel={$x_1/L$},
            ymin = -0.35,
            ymax = 0.25,
            ytick={-0.2,0,0.2},
            yticklabels={-.2,0,.2},
            xmajorgrids,
            ymajorgrids,
            axis lines=box,
            legend columns=6,
            legend style={fill=none, draw=none, column sep = 5pt},
            legend cell align={left},
            axis lines=box,
            mark options={solid},
            legend to name = {leg:ldc},
            title style={ at={(0,1)}},
            title = {(c)},
            tick label style={font=\footnotesize},
            ]
\addplot [line width=1.5pt,mycolor1] table [y=u,x=x]{_data/lid_driven_cavity/u1B_horizontal.txt};
\addplot [only marks,mycolor1,mark = x, mark repeat=4, mark size = 3, mark options={line width=1.2pt}] table [y=ul,x=xl]{_data/lid_driven_cavity/u1I_horizontal.txt};
\addplot [only marks,mycolor1,mark = o, mark repeat=4, mark size = 3, mark options={line width=1.2pt}] table [y=ur,x=xr]{_data/lid_driven_cavity/u1I_horizontal.txt};\label{plot:ldc_1_right}
\addplot [line width=1.5pt,,mycolor2] table [y=u,x=x]{_data/lid_driven_cavity/u2B_horizontal.txt};
\addplot [only marks,mycolor2,mark = x, mark repeat=4, mark size = 3, mark options={line width=1.2pt}] table [y=ur,x=xr]{_data/lid_driven_cavity/u2I_horizontal.txt};
\addplot [only marks,mycolor2,mark = o, mark repeat=4, mark size = 3, mark options={line width=1.2pt}] table [y=ur,x=xr]{_data/lid_driven_cavity/u2I_horizontal.txt};\label{plot:ldc_2_right}
\addlegendentry{$u_1/U$ }
\addlegendentry{$u^-_1/U$ }
\addlegendentry{$u^+_1/U$ }
\addlegendentry{$u_2/U$ }
\addlegendentry{$u^-_2/U$ }
\addlegendentry{$u^+_2/U$ }

\nextgroupplot[
            xmin= 0,
            xmax= 1,
            xlabel={$x_2/L$},
            ytick={-0.5,0,0.5,1},
            yticklabels={-.5,0,.5,1},
            ymin = -0.6,
            ymax = 1.1,
            xmajorgrids,
            ymajorgrids,
            axis lines=box,
            axis lines=box,
            mark options={solid},
            title style={ at={(0,1)}},
            title = {(d)},
            tick label style={font=\footnotesize},
            ]
\addplot [line width=1.5pt,mycolor1] table [y=u,x=x]{_data/lid_driven_cavity/u1B_vertical.txt};
\addplot [only marks,mycolor1,mark = o, mark repeat=4, mark size = 3, mark options={line width=1.2pt}] table [y=ul,x=xl]{_data/lid_driven_cavity/u1I_vertical.txt};
\addplot [only marks,mycolor1,mark = x, mark repeat=4, mark size = 3, mark options={line width=1.2pt}] table [y=ur,x=xr]{_data/lid_driven_cavity/u1I_vertical.txt};
\addplot [line width=1.5pt,mycolor2] table [y=u,x=x]{_data/lid_driven_cavity/u2B_vertical.txt};
\addplot [only marks,mycolor2,mark = o, mark repeat=4, mark size = 3, mark options={line width=1.2pt}] table [y=ul,x=xl]{_data/lid_driven_cavity/u2I_vertical.txt};
\addplot [only marks,mycolor2,mark = x, mark repeat=4, mark size = 3, mark options={line width=1.2pt}] table [y=ur,x=xr]{_data/lid_driven_cavity/u2I_vertical.txt};
\end{groupplot}
\end{tikzpicture}

%% file: _figures/poiseuille_flow/poiseuille_scheme.tex
\begin{tikzpicture}[scale=2.5, transform shape=false]

\pgfmathsetmacro{\lt}{-0.4};
\pgfmathsetmacro{\rt}{1.0};
\pgfmathsetmacro{\dt}{0.1};

\begin{scope}[shift={(-0.5,0)}]

    \draw[thick ,domain=0:1, smooth, variable=\t, mycolor1] plot ({-3*(\t-1)*\t}, {\t});
    
    \draw[thick, mycolor1] (0,0) -- (0,1);

    \draw[very thick,dashed,mycolor4] (\lt,0) -- (\rt,0);
    \draw[very thick] (\lt,1) -- (\rt,1);
    \draw[very thick] (\lt,-0.25) -- (\rt,-0.25);
    \draw[draw=none,pattern=north east lines,pattern color=black] (\lt,1    ) rectangle (\rt,1+\dt);
    \draw[draw=none,pattern=north east lines,pattern color=black] (\lt,-0.25) rectangle (\rt,-0.25-\dt);
    
    \draw[thick, <->] (0.8*\lt,0    ) -- +(0,1   ) node[anchor = east, pos = 0.5] {$2\delta$};
    \draw[thick, <->] (0.8*\lt,-0.25) -- +(0,0.25) node[anchor = east, pos = 0.5] {$\delta/4$};
    
    \draw[very thick, ->] (0,0) -- (0,0.3) node[anchor = south east] {$x_3$};
    \draw[very thick, ->] (0,0) -- (0.3,0) node[anchor = south west] {$x_1$};
    \draw[very thick, ->] (0,0) -- (45:0.3) node[anchor = south] {$x_2$};
    
    
    \draw[] (0.9*\rt, 0) circle (0.05);
    \draw[thin] ($(0.9*\rt,0) + ( 90:0.05)$) -- ($(1.5,0.4) + (175:0.5)$);
    \draw[thin] ($(0.9*\rt,0) + (-90:0.05)$) -- ($(1.5,0.4) + (255:0.5)$);

    \draw[mycolor2,very thick,->] (0.1,0.5) -- +(0.4,0.0) node [anchor = south, pos = 0.5,color = black] {$-\pdv{p}{x_1}$};
    
    \draw[mycolor2,very thick,->,decorate, decoration={snake, amplitude=1mm, segment length=3mm}] (-0.2,-0.22) -- +(0.0,0.19);
    \draw[mycolor2,very thick,->,decorate, decoration={snake, amplitude=1mm, segment length=3mm}] ( 0.2,-0.22) -- +(0.0,0.19);
    \draw[mycolor2,very thick,->,decorate, decoration={snake, amplitude=1mm, segment length=3mm}] ( 0.6,-0.22) -- +(0.0,0.19) node [anchor = east, pos = 0.5, color =black] {$F$};
    
    \node [anchor = south east] at (0.95*\rt,0) {$(+)$};
    \node [anchor = north east] at (0.95*\rt,0) {$(-)$};
    
\end{scope}
\def\step{0.8};
\begin{scope}[shift={(1.0, 0.4)}, scale=0.5]

    \draw[thick] (1.5*\step, -1.5*\step) -- (1.5*\step, -0.2*\step) node[anchor = south, pos = 0.1, rotate = 90] {$\Delta/4$};
    \draw[thick] (1.5*\step, -0.2) -- +(-0.2*\step, 0);
    \draw[thick] (1.5*\step, -0.5*\step) -- +(-0.2*\step, 0);
    \clip (0,0) circle (1);
    \draw[thick] (0,0) circle (1);
    \foreach \i in {-1,0,1} {
        \foreach \j in {-1,0,1} {
            \node[text=mycolor3, rotate=0, font=\boldmath] at (\i*\step + 0.5*\step, \j*\step) {$>$};
        }
    }
    \foreach \i in {-1,0,1} {
        \foreach \j in {-1,0,1} {
            \node[text=mycolor1, rotate=90, font=\boldmath] at (\i*\step, \j*\step + 0.5*\step) {$>$};
        }
    }
    \foreach \i in {-1,0,1} {
        \foreach \j in {-1,0,1} {
            \filldraw[mycolor2]  (\i*\step, \j*\step) circle (1pt);
            \draw (\i*\step-0.5*\step,\j*\step-0.5*\step) rectangle (\i*\step + 0.5*\step,\j*\step + 0.5*\step);
        }
    }
    \draw[very thick, dashed, mycolor4] (-1, -0.2) -- (1, -0.2);
\end{scope}

\end{tikzpicture}

%% file: _figures/poiseuille_flow/poiseuille_flow_profiles.tex
\begin{tikzpicture}[trim axis left]
\begin{axis}[%
            width  =1.0\linewidth,
            height =0.8\linewidth,
            xmin=-2.0,
            xmax= 2.0,
            xlabel={$u/U$},
            ymin = -0.250,
            ymax = 2,
            ylabel = {$\frac{x_3}{\delta}$},
            xmajorgrids,
            ymajorgrids,
            axis lines=box,
            legend columns=1,
            legend style={fill=none, draw=none},
            legend cell align={left},
            legend pos = {north west},
            ylabel style = {rotate = -90, font = \large, anchor = east},
            ]
\addplot [line width=1.5pt,mycolor1] table [y=z,x=u1]{_data/poiseuille_flow/u1.txt};
\addplot [line width=1.5pt,mycolor2] table [y=z,x=u3]{_data/poiseuille_flow/u3.txt};
\addplot [line width = 1.5pt,black,dashed,restrict expr to domain={\thisrow{z}}{0:inf}] table [y =z, x expr=-(\thisrow{z}-2.0)*\thisrow{z}]{_data/poiseuille_flow/u1.txt};
\addlegendentry{$u_1$}
\addlegendentry{$u_3$}
\addlegendentry{$u_e$}
\end{axis}
\end{tikzpicture}

%% file: _figures/poiseuille_flow/poiseuille_flow_convergence.tex
\begin{tikzpicture}
\begin{groupplot}[
    group style={
        group size=2 by 1,
        horizontal sep=3.5em,
    },
    width = 0.5\textwidth,
    height = 0.7*0.49*\textwidth,
    trim axis left
]
\nextgroupplot[
            xtick={40, 60, 80, 120, 180},
            xticklabels={40, 60, 80, 120, 180},
            xlabel={$N$},
            xmajorgrids,
            ymajorgrids,
            axis lines=box,
            xmode=log,
            ymode=log,
            ymin=0.8e-5,
            ymax=1e-2,
            legend columns=2,
            legend style={fill=none, draw=none},
            legend cell align={left},
            title style={ at={(0,1)}},
            title = {(c)},
            legend to name = {leg:poiseuille},
            ylabel = {$\norm{u_1^+-u^e}_1/U$},
            ylabel style = {yshift = -0.5em},
            tick label style={font=\footnotesize},
            ]
\addplot [line width = 1.5pt,dashed,mark=x, mark size=4, mark options={line width=1.2pt, solid}] table [y=L1,x=N]{_data/poiseuille_flow/convergence_flow.txt};\label{plot:poiseuille_conv_sharp}
\addplot [line width = 1.5pt,dashed,mark=o, mark size=4, mark options={line width=1.2pt, solid}] table [y=L1nc,x=N]{_data/poiseuille_flow/convergence_flow.txt};\label{plot:poiseuille_conv_no_corr}
\addplot [line width = 1.5pt] table [y=O1,x=N]{_data/poiseuille_flow/convergence_flow.txt};
\addplot [line width = 1.5pt] table [y=O2,x=N]{_data/poiseuille_flow/convergence_flow.txt};
\node[] at (axis cs:60,3.0e-5) {$N^{-2}$};
\node[] at (axis cs:180,1.0e-4) {$N^{-1}$};
\addlegendentry{Present method}
\addlegendentry{Present method without modified elliptic equation}

\nextgroupplot[
            xtick={40 ,60 ,80 ,120,180},
            xticklabels={40 ,60 ,80 ,120,180},
            xlabel={$N$},
            xmajorgrids,
            ymajorgrids,
            axis lines=box,
            xmode=log,
            ymode=log,
            title style={ at={(0,1)}},
            title = {(d)},
            legend columns=1,
            legend style={fill=none, draw=none},
            legend cell align={left},
            ylabel = {$\abs{I^+\qty(0)}/U$},
            ylabel style = {yshift = -0.5em},
            tick label style={font=\footnotesize},
            ]
\addplot [line width = 1.5pt,dashed,mark=x, mark size=4, mark options={line width=1.2pt, solid}] table [y=up,x=N]{_data/poiseuille_flow/convergence_IB.txt};
\addplot [line width = 1.5pt,dashed,mark=o, mark size=4, mark options={line width=1.2pt, solid}] table [y=up_no_ell,x=N]{_data/poiseuille_flow/convergence_IB.txt};
\addplot [line width = 1.5pt] table [y=O2,x=N]{_data/poiseuille_flow/convergence_IB.txt};
\addplot [line width = 1.5pt] table [y=O1,x=N]{_data/poiseuille_flow/convergence_IB.txt};
\node at (axis cs:80 ,8e-5) {$N^{-2}$};
\node at (axis cs:180,4e-4) {$N^{-1}$};
\end{groupplot}
\end{tikzpicture}

%% file: _figures/turbulent_channel/turbulent_channel_velocity.tex
\begin{tikzpicture}
\begin{groupplot}[
    group style={
        group size=2 by 1,
        horizontal sep=3.5em,
    },
    width = 0.49\textwidth,
    height = 0.8*0.49*\textwidth,
    trim axis left
]
\nextgroupplot[
            xmax=200,
            xmin=0.2,
            xlabel={$x_3^+$},
            ymin = 0,
            ymax = 20,
            ytick={0,5,10,15, 20},
            ylabel = {$\langle u_1^+\rangle$},
            xmajorgrids,
            ymajorgrids,
            axis lines=box,
            xmode=log,
            legend columns=2,
            legend to name = {leg:turbulent_channel},
            legend style={fill=none, draw=none, column sep = 5pt},
            legend cell align={left},
            axis lines=box,
            ylabel style = {rotate = -90},
            title style={ at={(0,1)}},
            title = {(a)},
            tick label style={font=\footnotesize},
            ]
\addplot [line width = 5pt,mycolor1] table [y=u,x=z]{_data/turbulent_channel/my_meanRe200.txt};
\addplot [line width = 2pt,mycolor4] table [y=um,x=z]{_data/turbulent_channel/IB_data_avg.txt};
\addlegendentry{Body-fitted}
\addlegendentry{Present method}

\nextgroupplot[
            xmin=0,
            xmax=200,
            xlabel={$x_3^+$},
            ymax = 3,
            ymin = 0.0,
            ytick={0,1,2,3},
            ylabel = {$\sigma^+$},
            xmajorgrids,
            ymajorgrids,
            axis lines=box,
            ylabel style = {rotate = -90},
            title style={ at={(0,1.0)}},
            title = {(b)},
            tick label style={font=\footnotesize},
            ]
\addplot [line width = 5pt, mycolor1] table [y=std,x=z]{_data/turbulent_channel/my_rms1Re200.txt};
\addplot [line width = 5pt, mycolor1] table [y=std,x=z]{_data/turbulent_channel/my_rms2Re200.txt};
\addplot [line width = 5pt, mycolor1] table [y=std,x=z]{_data/turbulent_channel/my_rms3Re200.txt};
\addplot [line width = 2pt,mycolor4] table [y=rmsu,x=z]{_data/turbulent_channel/IB_data_avg.txt};
\addplot [line width = 2pt,mycolor4] table [y=rmsv,x=z]{_data/turbulent_channel/IB_data_avg.txt};
\addplot [line width = 2pt,mycolor4] table [y=rmsw,x=z]{_data/turbulent_channel/IB_data_avg.txt};

\node at (axis cs:50,2.5) {$\sigma_1$};
\node at (axis cs:25,1.5) {$\sigma_2$};
\node at (axis cs:35,0.3) {$\sigma_3$};
\end{groupplot}
\end{tikzpicture}

%% file: _figures/plate/plate_profiles.tex
\begin{tikzpicture}
    \begin{groupplot}[
    group style={
        group size=1 by 2,
        vertical sep=3.5em,
    },
    width = 1.0\textwidth,
    height = 0.45*1.0*\textwidth,
    trim axis left
    ]

\nextgroupplot[
            xmin = -2,
            xmax =  4,
            xlabel = {$x_3/L$},
            ytick = {-0.5, 0, 0.5, 1},
            xmajorgrids,
            ymajorgrids,
            title style={ at={(0,1)}},
            title = {(b)},
            tick label style={font=\footnotesize},
            ]
\addplot [line width=1.5pt] table [y = u,x=z]{_data/plate/u3.txt};
\node at (axis cs:2,0.8) {$u_3/U$};
\draw[gray, thick] (axis cs:0,-1) -- (axis cs:0,1.2);
\nextgroupplot[
            xmin = -2,
            xmax =  4,
            xlabel = {$x_3/L$},
            ytick = {-0.5, 0, 0.5},
            yticklabels = {-.5, 0, .5},
            xmajorgrids,
            ymajorgrids,
            title style={ at={(0,1)}},
            title = {(c)},
            tick label style={font=\footnotesize},
            ]
\addplot [line width=1.5pt] table [y = u,x=z]{_data/plate/u4.txt};
\node at (axis cs:2,0.5) {$p/\rho U^2$};
\draw[gray, thick] (axis cs:0,-1) -- (axis cs:0,1);
\end{groupplot}
\end{tikzpicture}

%% file: _figures/plate/plate_convergence.tex
\begin{tikzpicture}
    \begin{groupplot}[
        group style={
            group size=1 by 1,
            horizontal sep=3em,
        },
        width = 0.80\textwidth,
        height = 0.38*0.80*\textwidth,
        trim axis left
    ]

\nextgroupplot[
            xtick={200,300,400,500},
            xticklabels = {$200$,$300$,$400$,$500$},
            xlabel={$N$},
            xmajorgrids,
            ymajorgrids,
            axis lines=box,
            xmode=log,
            ymode=log,
            ymin =  5e-4,
            ymax =  1.1e-2,
            legend pos=south west,
            legend columns=1,
            legend style={fill=none, draw=none},
            legend cell align={left},
            title style={ at={(0,1)}},
            title = {(d)},
            ylabel style = {rotate = -90, font = \large, anchor = east},
            ylabel = {$\frac{\abs{I\qty(0)}}{U}$},
            tick label style={font=\footnotesize},
            ]
\addplot [line width = 1.5pt,mycolor1,dashed,mark=x, mark size=4, mark options={line width=1.2pt, solid}] table [y=front,x=N]{_data/plate/plate_convergence_64.txt};
\addplot [line width = 1.5pt,mycolor2,dashed,mark=x, mark size=4, mark options={line width=1.2pt, solid}] table [y=rear,x=N]{_data/plate/plate_convergence_64.txt};
\addplot [line width = 1.5pt,black,] table [y=O2,x=N]{_data/plate/plate_convergence_64.txt};
\addlegendentry{$(-)$ side}
\addlegendentry{$(+)$ side}
\node [] at (axis cs: 480, 2.5e-3) {$N^{-2}$};

\end{groupplot}
\end{tikzpicture}

%% file: _figures/heaving_plate/heaving_plate_error.tex
\begin{tikzpicture}
    \begin{groupplot}[
        group style={
            group size=1 by 1,
            horizontal sep=4em,
        },
        width = 0.95\textwidth,
        height = 0.35*0.95*\textwidth,
        trim axis left
    ]

\nextgroupplot[
            xmin = 0.0,
            xmax = 1.0,
            ymin = 0.0,
            ymax = 8e-3,
            xlabel = {$t/T$},
            ylabel = {$\frac{\norm{\boldsymbol{\varepsilon}}_1}{U}$},
            xmajorgrids,
            ymajorgrids,
            axis lines=box,
            legend columns=1,
            legend style={fill=none, draw=none},
            legend cell align={left},
            legend pos=north west,
            legend to name = {leg:heaving},
            scaled y ticks = true,
            tick scale binop=\times,
            ylabel style={rotate = -90, font=\Large, anchor=east},
            tick label style={font=\footnotesize},
    ]
\addplot [line width=1.5pt,mycolor1] table [y = s,x=t]{_data/heaving_plate/error.txt};
\addplot [line width=1.5pt,mycolor2] table [y = s,x=t]{_data/heaving_plate/error_no_corr.txt};
\addlegendentry{Present method}
\addlegendentry{Present method without modified elliptic equation}
\coordinate (insetPosition) at (rel axis cs:0.60,0.2);
\end{groupplot}

\begin{axis}[
    at={(insetPosition)},
    footnotesize,
    width  = 6cm,
    height = 3.8cm,
    xmajorgrids,
    ymajorgrids,
    axis lines=box,
    axis background/.style={fill=white},
    xmin = 0.35,
    xmax = 0.65,
    xtick = {0.4, 0.5, 0.6},
    ymin = 0,
    ytick = {1e-4, 4e-4},
    tick scale binop=\times,
    every y tick scale label/.style={at={(axis description cs:0,1)},anchor=north west},
    tick label style={font=\footnotesize},
    ]
    \addplot [line width=1.5pt,mycolor1] table [y = s,x=t]{_data/heaving_plate/error.txt};
\end{axis}
\end{tikzpicture}

%% file: _figures/buoyancy_driven/buoyancy_sketch.tex
\begin{tikzpicture}[scale=1.4, transform shape=false]
    \pgfmathsetmacro{\ang}{60};
    \pgfmathsetmacro{\prof}{1};
    \pgfmathsetmacro{\a}{0.5}
    \pgfmathsetmacro{\b}{0.4}
    \pgfmathsetmacro\rot{30}
    \pgfmathsetmacro{\tilt}{70}
    \pgfmathsetmacro{\flatten}{cos(\tilt)}

    \coordinate (CC) at ($(0, 0) + (\ang:0.5*\prof)$); 
    \coordinate (CS) at ($(0,-1) + (\ang:0.5*\prof)$); 
    \coordinate (CN) at ($(0, 1.5) + (\ang:0.5*\prof)$); 
    \draw [very thick, gray] (-1,-1)rectangle (1,1.5);
    \draw [very thick, gray] ($(-1,-1) + (\ang:1)$)rectangle ($(1,1.5) + (\ang:\prof)$);

    \draw [very thick, gray] (-1,1.5 ) -- +(\ang:\prof);
    \draw [very thick, gray] ( 1,1.5 ) -- +(\ang:\prof);
    \draw [very thick, gray] (-1,-1) -- +(\ang:\prof);
    \draw [very thick, gray] ( 1,-1) -- +(\ang:\prof);
    
    \begin{scope}[rotate around={\rot:(CC)}]
        \shade[ball color=gray!60] (CC) ellipse ({\a} and {\b});
        \draw[black, line width=0.6pt] (CC) ++(-\a,0) arc[start angle=180, end angle=360, x radius=\a, y radius=0.35*\b];
        \draw[black, dashed, line width=0.6pt] (CC) ++(\a,0) arc[start angle=0, end angle=180, x radius=\a, y radius=\flatten*\b];
        \draw[black, line width=0.6pt] (CC) ++(0,\b) arc[start angle=90, end angle=270, x radius=\flatten*\a, y radius=\b];
        \draw[black, dashed, line width=0.6pt] (CC) ++(0,-\b) arc[start angle=270, end angle=450, x radius=\flatten*\a, y radius=\b];
        \draw[very thick, <->] ($(CC) - (0:0.9*\a) + (0,0.6) $)--($(CC) + (0:0.9*\a) + (0,0.6)$) node [pos = 0.5, anchor = south] {$d$};
    \end{scope}
    

    \draw[thick, <->] ($(-1,-1) - (0.1,0)$) -- +(0,2.5) node [anchor = east, pos  = 0.5] {$L_3$};
    \draw[thick, <->] ($(-1,1.5) - (0.1,0)$) -- +(\ang:\prof) node [anchor = east, pos  = 0.5] {$L_2$};
    \draw[thick, <->] ($(-1,-1) - (0,0.1)$) -- +(2.0,0) node [anchor = north, pos  = 0.5] {$L_1$};
    \draw[very thick, ->] (CS)--+(0,0.3)    node [anchor = east ] {$x_3$};
    \draw[very thick, ->] (CS)--+(0.3,0)    node [anchor = north] {$x_2$};
    \draw[very thick, ->] (CS)--+(\ang:-0.3) node [anchor = east ] {$x_1$};
    
    \draw[very thick, <->] ($(CC) - (0,1.0) + (0.6,0)$)--($(CC) +(0.6,0)$) node [anchor =east, pos = 0.3 ] {$\ell$};
    
\end{tikzpicture}

%% file: _figures/buoyancy_driven/rising_sphere_vertical.tex
\begin{tikzpicture}
\begin{axis}[%
            width  =0.95\linewidth,
            height =0.7\linewidth,
            xmin = 0,
            xmax = 100,
            xlabel={$tU_g/d$},
            ymin = 0,
            ymax = 0.17,
            ytick={0.02,0.08, 0.14},
            yticklabels={0.02, 0.08,0.14},
            ylabel={$\frac{v_V}{U}$},
            xmajorgrids,
            ymajorgrids,
            axis lines=box,
            legend columns=1,
            legend style={fill=none, draw=none},
            legend cell align={left},
            legend image post style={mark options={solid}},
            legend pos=south east,
            ylabel style={rotate = -90, font=\Large, anchor=east},
            tick label style={font=\footnotesize},
            ]
\addplot [line width = 1.5pt, mycolor1] table [y=u,x=t]{_data/rising_sphere/sharpIB.txt};
\addplot [line width = 1.5pt, mycolor3,dashed,mark=o, mark size=3, mark options={line width=1.2pt, solid},mark repeat = 20] table [y=u,x=t]{_data/rising_sphere/ref_Yang.txt};
\addlegendentry{Present method}
\addlegendentry{Ref.~\cite{yangNoniterativeDirectForcing2015}}
\end{axis}
\end{tikzpicture}

%% file: _figures/buoyancy_driven/VP_error.tex
\begin{tikzpicture}
    \begin{groupplot}[
        group style={
            group size=1 by 1,
            horizontal sep=4em,
        },
        width = 0.95\textwidth,
        height = 0.35*0.95*\textwidth,
        trim axis left
    ]

\nextgroupplot[
            xmin = 0,
            xmax = 80,
            xlabel={$tU_g/d$},
            ymax = 2.2e-3,
            ymin = 0,
            ylabel = {$\frac{\norm{\boldsymbol{\varepsilon}}_1}{U}$},
            xmajorgrids,
            ymajorgrids,
            axis lines=box,
            legend columns=2,
            legend style={fill=none, draw=none},
            legend pos = north west,
            title style={ at={(0,1)}},
            ylabel style={rotate=-90, font = \large, anchor = east},
            tick scale binop=\times,
            trim axis left,
            tick label style={font=\footnotesize},
    ]
\addplot [line width=1.5pt,mycolor1] table [y = s,x=t]{_data/buoyancy_driven/VP_error_sharpIB.txt};
\addplot [line width=1.5pt,mycolor2] table [y = s,x=t]{_data/buoyancy_driven/VP_error_MLS.txt};
\addlegendentry{Present method}
\addlegendentry{L-MLS}
\coordinate (insetPosition) at (rel axis cs:0.60,0.3);
\end{groupplot}

\begin{axis}[
    at={(insetPosition)},
    footnotesize,
    width  = 6cm,
    height = 3.5cm,
    xmajorgrids,
    ymajorgrids,
    axis lines=box,
    axis background/.style={fill=white},
    xmin = 40,
    xmax = 70,
    xtick = {45, 55, 65},
    ymin = 0,
    ymax = 6e-5,
    tick scale binop=\times,
    every y tick scale label/.style={at={(axis description cs:0,1)},anchor=north west},
    tick label style={font=\footnotesize},
    ]
    \addplot [line width=1.5pt,mycolor1] table [y = s,x=t]{_data/buoyancy_driven/VP_error_sharpIB.txt};
\end{axis}
\end{tikzpicture}

%% file: _figures/flag/flag_sketch.tex
\begin{tikzpicture}[scale=2.4, transform shape=false]
    \pgfmathsetmacro{\ang}{10};
    \pgfmathsetmacro{\prof}{1};
    \pgfmathsetmacro{\len}{0.6};
    
    \coordinate (CF) at (0, 0); 

    \coordinate (CC) at ($(0, 0) + (\ang:0.5*\prof)$); 
    \coordinate (CS) at ($(0,-1) + (\ang:0.5*\prof)$); 
    \draw [very thick, gray] (-1,-0.6)rectangle (1,0.6);
    \draw [very thick, gray] ($(-1,-0.6) + (\ang:1)$)rectangle ($(1,0.6) + (\ang:\prof)$);
        
    \draw [very thick, gray] (-1,0.6 ) -- +(\ang:\prof);
    \draw [very thick, gray] ( 1,0.6 ) -- +(\ang:\prof);
    \draw [very thick, gray] (-1,-0.6) -- +(\ang:\prof);
    \draw [very thick, gray] ( 1,-0.6) -- +(\ang:\prof);
    
    \draw [thick, <->] ($(-1,0.6) + (0,0.1)$) -- +(\ang:\prof) node [anchor = south, pos = 0.5] {$5L$};
    \draw [thick, <->] ($(-1,0.6) + (0,0.1) +(\ang:\prof)$) -- +(2.0,0.0) node [anchor = south, pos = 0.5] {$5L$};
    \draw [thick, <->] ($(-1,0.6) - (0.1,0.0) $) -- +(0.0,-1.2) node [anchor = east, pos = 0.5] {$4L$};

    \coordinate (F1) at ($(CF) + (\ang:0.2*\prof) +(0.0, 0.5*\len)$); 
    \coordinate (F2) at ($(F1) + (\ang:\len)$); 
    \coordinate (F3) at ($(F2) - (0,\len)$); 
    \coordinate (F4) at ($(F3) - (\ang:\len)$); 
    
    \filldraw [color=gray!60] plot coordinates { (F1) (F2) (F3) (F4) (F1)};
    \draw[fill = mycolor2] ($0.5*(F2) + 0.5*(F3)$) circle (0.02) node[black, anchor = west] {$Q$};
    \draw[very thick, ->] ($0.5*(F4) + 0.5*(F1)$)--+(0,0.4)    node [anchor = east ] {$x_1$};
    \draw[very thick, ->] ($0.5*(F4) + 0.5*(F1)$)--+(0.4,0)    node [anchor = north east] {$x_2$};
    \draw[very thick, ->] ($0.5*(F4) + 0.5*(F1)$)--+(\ang:0.2) node [anchor = south] {$x_3$};
    \draw [very thick, mycolor2] (F4) -- (F1);
    
    \draw [thick, <->] ($(F4)- (0,0.1)$) -- ($(F3)- (0,0.1)$) node [anchor = north, pos = 0.7] {$L$};
    
    \draw [very thick, ->] ($(CF)-(\ang:0.4)$)-- +(\ang:0.3)node [pos = 0.5, anchor = south] {U};
    
\end{tikzpicture}

%% file: _figures/flag/flag_plot.tex
\begin{tikzpicture}
\begin{axis}[
            width  = 1.0\linewidth,
            height = 0.9\linewidth,
            xmin = 0,
            xmax = 8,
            xtick={0,2,4,6,8},
            xlabel={$tU/L$},
            ylabel={$\frac{x_2}{L}$},
            ytick = {-0.5, 0, 0.5},
            yticklabels = {-.5, 0, .5},
            ymin = -0.55,
            ymax = 0.9,
            xmajorgrids,
            ymajorgrids,
            axis lines=box,
            legend columns=1,
            legend style={fill=none, draw=none},
            legend cell align={left},
            legend image post style={mark options={solid}},
            legend pos=north west,
            ylabel style={rotate=-90, font = \large,xshift = 1em},
            tick label style={font=\footnotesize},
            ]
\addplot [line width=1.5pt,mycolor1] table [y = y,x=t]{_data/flag/pos_sharpIB.txt};
\addplot [line width = 1.5pt, mycolor3,dashed,mark=o, mark size=3, mark options={line width=1.2pt, solid},mark repeat = 10] table [y = y,x=t]{_data/flag/ref200.txt};
\addlegendentry{Present method}
\addlegendentry{Ref.~\cite{huangThreedimensionalSimulationFlapping2010}}
\end{axis}
\end{tikzpicture}

%% file: _figures/flag/flag_error.tex
\begin{tikzpicture}
    \begin{groupplot}[
        group style={
            group size=1 by 1,
            horizontal sep=4.em,
        },
        width = 0.95\textwidth,
        height = 0.35*0.95*\textwidth,
        trim axis left
        ]
\nextgroupplot[
            xmin = 0,
            xmax = 20,
            xtick={0,5,10,15,20},
            xlabel={$tU/L$},
            ymin = 0,
            ymax = 5.3e-3,
            ylabel={$\frac{\norm{\boldsymbol{\varepsilon}}_1}{U}$},
            xmajorgrids,
            ymajorgrids,
            axis lines=box,
            legend columns=2,
            legend style={fill=none, draw=none},
            legend pos = north west,
            legend cell align={left},
            ylabel style={rotate=-90, font = \large, anchor = east},
            tick scale binop=\times,
            tick label style={font=\footnotesize},
    ]
\addplot [line width=1.5pt,mycolor1       ] table [y = s,x=t]{_data/flag/error_sharpIB.txt};
\addplot [line width=1.5pt,mycolor2       ] table [y = s,x=t]{_data/flag/error_MLS.txt};
\addplot [line width=1.5pt,mycolor1,dashed] table [y = s,x=t]{_data/flag/error_sharpIB_dt_small.txt};
\addplot [line width=1.5pt,mycolor2,dashed] table [y = s,x=t]{_data/flag/error_MLS_dt_small.txt};
\addlegendentry{Present method}
\addlegendentry{L-MLS}
\coordinate (insetPosition) at (rel axis cs:0.6,0.3);

\end{groupplot}

\begin{axis}[
    at={(insetPosition)},
    footnotesize,
    width  = 6cm,
    height = 3.5cm,
    xmajorgrids,
    ymajorgrids,
    axis lines=box,
    axis background/.style={fill=white},
    xmin = 5,
    xmax = 11,
    xtick = {6, 8, 10},
    ymin = 0,
    ytick = {0,2e-4, 4e-4},
    tick scale binop=\times,
    every y tick scale label/.style={at={(axis description cs:0,1)},anchor=north west},
    tick label style={font=\footnotesize},
    ]
    \addplot [line width=1.5pt,mycolor1       ] table [y = s,x=t]{_data/flag/error_sharpIB.txt};
    \addplot [line width=1.5pt,mycolor1,dashed] table [y = s,x=t]{_data/flag/error_sharpIB_dt_small.txt};
\end{axis}

\end{tikzpicture}

%% file: _figures/bubble/fixed_bubble_sketch.tex
\begin{tikzpicture}[scale=2.0, transform shape=false]
    \pgfmathsetmacro{\ang}{25};
    \pgfmathsetmacro{\prof}{0.5};
    \pgfmathsetmacro{\len}{0.6};
    
    \coordinate (CF) at (0, 0); 

    \coordinate (CC) at ($(-0.5,0) + (\ang:0.5*\prof)$); 
    \coordinate (CS) at ($(0,-0.5) + (\ang:0.5*\prof)$); 
    \coordinate (C) at ($(CF) + (\ang:0.5*\prof)-(0.05,0)$);

    \draw [very thick, gray] ($(-0.5,-0.5) + (\ang:\prof)$)rectangle ($(0.5,0.5) + (\ang:\prof)$);
    \draw [very thick, gray] (-0.5,-0.5)rectangle (0.5,0.5);
    \shade[ball color=gray!60] (C) circle (6pt);
    \draw[very thick, <->] ($(C)+ (-0.2,-0.3)$) -- ($(C)+ (0.2,-0.3)$) node [anchor = north, pos = 0.5] {$d$};
    
        
    \draw [very thick, gray] (-0.5,0.5 ) -- +(\ang:\prof);
    \draw [very thick, gray] ( 0.5,0.5 ) -- +(\ang:\prof);
    \draw [very thick, gray] (-0.5,-0.5) -- +(\ang:\prof);
    \draw [very thick, gray] ( 0.5,-0.5) -- +(\ang:\prof);


    \draw[thick, <->] ($(-0.5,-0.5) - (0.1,0)$) -- +(0,1.0) node [anchor = east, pos  = 0.5] {$2d$};
    \draw[thick, <->] ($(-0.5,0.5) - (0.1,0)$) -- +(\ang:\prof) node [anchor = south, pos  = 0.5] {$2d$};
    \draw[thick, <->] ($(-0.5,-0.5) - (0,0.1)$) -- +(1.0,0) node [anchor = north, pos  = 0.5] {$2d$};

    

    

\end{tikzpicture}

%% file: _figures/bubble/bubble_res.tex
\begin{tikzpicture}
\begin{groupplot}[
    group style={
        group size=3 by 2,
        horizontal sep=2.0em,
        vertical sep=2.5em,
    },
    width = 0.39\textwidth,
    height = 0.7*0.33*\textwidth,
    trim axis left
]

\nextgroupplot[
            xmin=0,
            xmax=15,
            ymax = 1.1,
            ymin = 0.7,
            ytick={0.8,1.0,1.2},
            xmajorgrids,
            ymajorgrids,
            legend columns=2,
            legend to name = {leg:bubble_fixed_comparison},
            legend style={fill=none, draw=none},
            axis lines=box,
            title style={ at={(0.15,0.85)}},
            title = {(c) $V/V_0$},
            tick label style={font=\footnotesize},
]
\addplot [very thick,mycolor1] table [y=v,x=t]{_data/fixed_bubble/sharpIB.txt};
\addplot [very thick,mycolor2] table [y=v,x=t]{_data/fixed_bubble/MLS.txt};
\addlegendentry{Present method}
\addlegendentry{L-MLS}
\draw[fill = white] (rel axis cs: 0.04,0.05) rectangle (rel axis cs: 0.35,0.55);
\shade[ball color=gray!60] ($(rel axis cs: 0.04,0.05)!0.5!(rel axis cs: 0.35,0.55)$) circle (7pt);

\nextgroupplot[
            xmin=0,
            xmax=15,
            ymax = 1.1,
            ymin = 0.7,
            ytick={0.8,1,1.2},
            xmajorgrids,
            ymajorgrids,
            axis lines=box,
            title style={ at={(0.15,0.85)}},
            title = {(d) $S/S_0$},
            tick label style={font=\footnotesize},
            ]
\addplot [very thick,mycolor1] table [y=s,x=t]{_data/fixed_bubble/sharpIB.txt};
\addplot [very thick,mycolor2] table [y=s,x=t]{_data/fixed_bubble/MLS.txt};
\draw[fill = white] (rel axis cs: 0.04,0.05) rectangle (rel axis cs: 0.35,0.55);
\shade[ball color=gray!60] ($(rel axis cs: 0.04,0.05)!0.5!(rel axis cs: 0.35,0.55)$) circle (7pt);

\nextgroupplot[
            xmin=0,
            xmax=15,
            ymax = 1.3,
            ymin = 0.9,
            ytick={0.8,1.0,1.2},
            xmajorgrids,
            ymajorgrids,
            axis lines=box,
            title style={ at={(0.20,0.85)}},
            title = {(e) $\Delta p/\Delta p_0$},
            tick label style={font=\footnotesize},
            ]
\addplot [very thick,mycolor1] table [y=dp,x=t]{_data/fixed_bubble/sharpIB.txt};
\addplot [very thick,mycolor2] table [y=dp,x=t]{_data/fixed_bubble/MLS.txt};
\draw[fill = white] (rel axis cs: 0.04,0.05) rectangle (rel axis cs: 0.35,0.55);
\shade[ball color=gray!60] ($(rel axis cs: 0.04,0.05)!0.5!(rel axis cs: 0.35,0.55)$) circle (7pt);

\nextgroupplot[
            xmin=0,
            xmax=15,
            xlabel={$t U_{\gamma}/d$},
            ymax = 1.1,
            ymin = 0.5,
            ytick={0.6,0.8,1.0,1.2},
            xmajorgrids,
            ymajorgrids,
            legend columns=2,
            legend style={fill=none, draw=none},
            legend cell align={left},
            axis lines=box,
            title style={ at={(0,0.85)}},
            title = {(f)},
            tick label style={font=\footnotesize},
            xlabel style={yshift=0.2cm},
            ]
\addplot [very thick,mycolor1] table [y=v,x=t]{_data/relative_bubble/sharpIB.txt};
\addplot [very thick,mycolor2] table [y=v,x=t]{_data/relative_bubble/MLS.txt};
\def\xc{0.195}
\def\dx{0.12}
\draw[fill = white] (rel axis cs: 0.04,0.05) rectangle (rel axis cs: 0.35,0.55);
\shade[ball color=gray!60] ($(rel axis cs: 0.04,0.05)!0.5!(rel axis cs: 0.35,0.55)$) circle (7pt);
\draw[<->, thick] (rel axis cs: {\xc-\dx},0.3) -- (rel axis cs: {\xc+\dx},0.3);

\nextgroupplot[
            xmin=0,
            xmax=15,
            xlabel={$t U_{\gamma}/d$},
            ymax = 1.1,
            ymin = 0.7,
            ytick={0.8,1.0,1.2},
            xmajorgrids,
            ymajorgrids,
            axis lines=box,
            title style={ at={(0,0.85)}},
            title = {(g)},
            tick label style={font=\footnotesize},
            xlabel style={yshift=0.2cm},
            ]
\addplot [very thick,mycolor1] table [y=s,x=t]{_data/relative_bubble/sharpIB.txt};
\addplot [very thick,mycolor2] table [y=s,x=t]{_data/relative_bubble/MLS.txt};
\def\xc{0.195}
\def\dx{0.12}
\draw[fill = white] (rel axis cs: 0.04,0.05) rectangle (rel axis cs: 0.35,0.55);
\shade[ball color=gray!60] ($(rel axis cs: 0.04,0.05)!0.5!(rel axis cs: 0.35,0.55)$) circle (7pt);
\draw[<->, thick] (rel axis cs: {\xc-\dx},0.3) -- (rel axis cs: {\xc+\dx},0.3);

\nextgroupplot[
            xmin=0,
            xmax=15,
            xlabel={$t U_{\gamma}/d$},
            ymax = 1.3,
            ymin = 0.9,
            ytick={1.0,1.2,1.4},
            xmajorgrids,
            ymajorgrids,
            axis lines=box,
            title style={ at={(0,0.85)}},
            title = {(h)},
            xlabel style={yshift=0.2cm},
            tick label style={font=\footnotesize},
            ]
\addplot [very thick,mycolor1] table [y=dp,x=t]{_data/relative_bubble/sharpIB.txt};
\addplot [very thick,mycolor2] table [y=dp,x=t]{_data/relative_bubble/MLS.txt};
\def\xc{0.195}
\def\dx{0.12}
\draw[fill = white] (rel axis cs: 0.04,0.05) rectangle (rel axis cs: 0.35,0.55);
\shade[ball color=gray!60] ($(rel axis cs: 0.04,0.05)!0.5!(rel axis cs: 0.35,0.55)$) circle (7pt);
\draw[<->, thick] (rel axis cs: {\xc-\dx},0.3) -- (rel axis cs: {\xc+\dx},0.3);

\end{groupplot}
\end{tikzpicture}

%% file: _figures/aorta/aorta_sketch.tex
\begin{tikzpicture}[scale=0.7, transform shape=false]
    \pgfmathsetmacro{\ang}{25};
    \pgfmathsetmacro{\prof}{1.0};
    \pgfmathsetmacro{\boxx}{4.5};
    \pgfmathsetmacro{\boxy}{6.0};
    \pgfmathsetmacro{\xpor}{2.0};
    \pgfmathsetmacro{\xin}{-0.15};
    \pgfmathsetmacro{\yin}{-3.0};
    \pgfmathsetmacro{\porx}{\boxx - \xpor +\xin};
    \pgfmathsetmacro{\pory}{2.2};

    \coordinate (box_front_b) at ($(\xin,\yin) - (\ang:0.5*\prof)$); 
    \coordinate (box_front_t) at ($(box_front_b) + (\boxx,\boxy)$); 
    \coordinate (box_back_b) at ($(box_front_b) + (\ang:\prof)$); 
    \coordinate (box_back_t) at ($(box_back_b) + (\boxx,\boxy)$); 
    \coordinate (por_front_b) at ($(\xpor,\yin) - (\ang:0.5*\prof)$); 
    \coordinate (por_front_t) at ($(por_front_b) + (\porx,\pory)$); 
    \coordinate (por_back_b) at ($(por_front_b) + (\ang:\prof)$); 
    \coordinate (por_back_t) at ($(por_back_b) + (\porx,\pory)$); 

    \draw [very thick, gray] (box_back_b) rectangle (box_back_t);
    \draw [very thick, black] (por_back_b) rectangle (por_back_t);
    \draw [very thick, black] ($(por_back_b) + (0,\pory)$) -- ($(por_front_b) + (0,\pory)$);
    \node [anchor = west] (img) at (0,0) {\includegraphics[scale = 0.105,angle = 90]{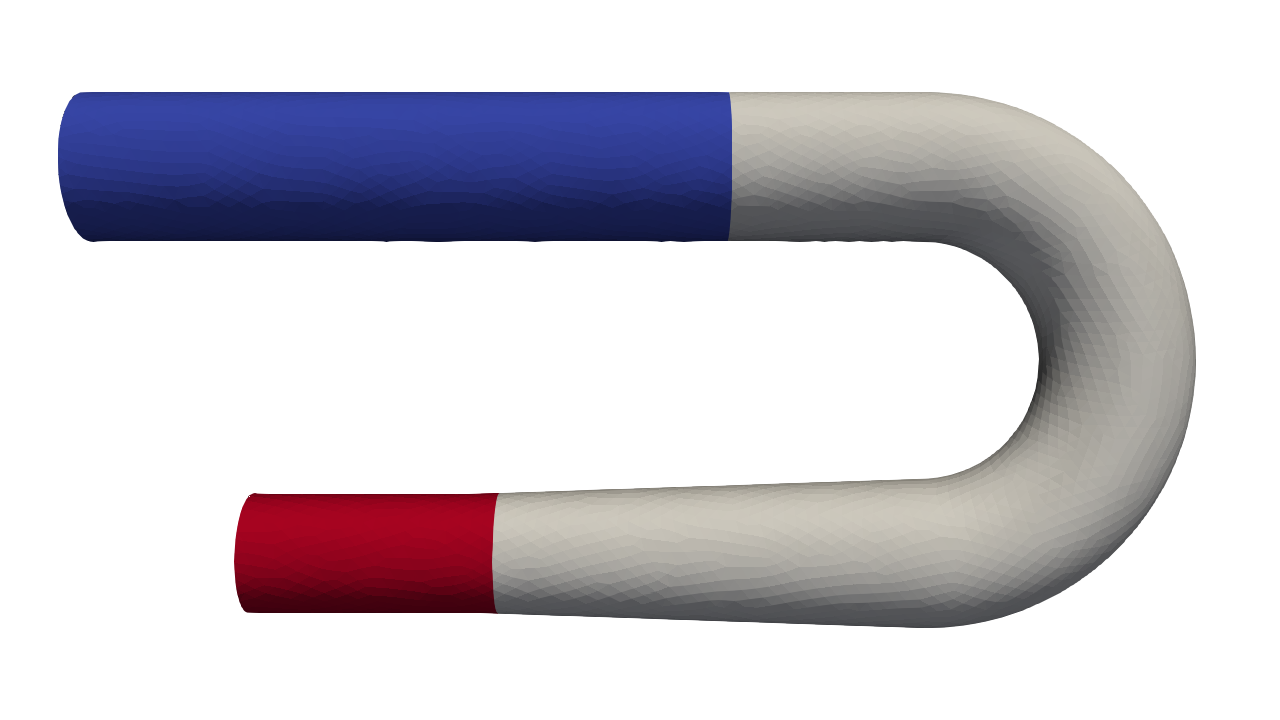}};
    \draw [very thick, gray] (box_front_b)rectangle (box_front_t);
    \draw [very thick, gray] (box_back_b) -- (box_front_b);
    \draw [very thick, gray] (box_back_t) -- (box_front_t);
    \draw [very thick, gray] ($(box_back_b) + (0,\boxy)$) -- ($(box_front_b) + (0,\boxy)$);
    \draw [very thick, gray] ($(box_back_b) + (\boxx,0)$) -- ($(box_front_b) + (\boxx,0)$);
    \draw [very thick, black] (por_front_b)rectangle (por_front_t);
    \draw [very thick, black] (por_back_b) -- (por_front_b);
    \draw [very thick, black] (por_back_t) -- (por_front_t);
    \draw [very thick, black] ($(por_back_b) + (\porx,0)$) -- ($(por_front_b) + (\porx,0)$);

   \draw[thick, black, <->]  ($(box_back_t) + (0,0.3*\prof)$) -- +(-\boxx,0) node [anchor = south, pos = 0.5] {$5.5D$};
   \draw[thick, black, <->]  ($(box_back_t) + (0,0.3*\prof) - (\boxx,0)$) -- +(\ang:-\prof) node [anchor = south, pos = 0.5] {$3D$};
   \draw[thick, black, <->]  ($(box_front_t) - (0.3*\prof,0) - (\boxx,0)$) -- +(0,-\boxy) node [anchor = south, pos = 0.5, rotate = 90] {$9D$};

   \draw[thick, black, <->]  ($(\xpor,\yin) - (\ang:0.5*\prof) - (0,0.3*\prof)$) -- +(\porx,0) node [anchor = north, pos = 0.5] {$2.2D$};
   \draw[thick, black, <->]  ($(\xpor,\yin) + (\porx, 0) - (\ang:0.3*\prof)$) -- +(0,\pory) node [anchor = south, pos = 0.5, rotate = -90] {$2.8D$};

   %
   \draw[thick, black, <->] (0.65,0.7) -- +(0.8,0) node [anchor = south, pos = 0.5] {$D$};
   \draw[fill = white] (1. ,-1.6) node {1} circle (8pt);
   \draw[fill = white] (3.1,-0.2) node {2} circle (8pt);
   \draw[fill = white] (3.1,-1.6) node {3} circle (8pt);

   \draw[mycolor4,very thick,->,decorate, decoration={snake, amplitude=1mm, segment length=3mm}] ($(por_front_b) +(\ang:0.5*\prof) + (0.3*\porx,0.15*\pory)$) -- +(0.0,0.2*\pory);
   \draw[mycolor4,very thick,->,decorate, decoration={snake, amplitude=1mm, segment length=3mm}] ($(por_front_b) +(\ang:0.5*\prof) + (0.5*\porx,0.15*\pory)$) -- +(0.0,0.2*\pory);
   \draw[mycolor4,very thick,->,decorate, decoration={snake, amplitude=1mm, segment length=3mm}] ($(por_front_b) +(\ang:0.5*\prof) + (0.7*\porx,0.15*\pory)$) -- +(0.0,0.2*\pory);
    \node[color = black, rotate = 90] at($(box_front_b) + (\boxx,0) + (\ang:0.5*\prof) + (-\porx,0.5*\pory)$) {$-K\vb{u}$};

   \draw[thick,black, ->] (box_front_b) -- +(0,1*\prof)    node[anchor = south west ] {$x_3$};
   \draw[thick,black, ->] (box_front_b) -- +(0.8*\prof,0)    node[anchor = north] {$x_1$};
   \draw[thick,black, ->] (box_front_b) -- +(\ang:0.7*\prof) node[anchor = south] {$x_2$};
\end{tikzpicture}

%% file: _figures/aorta/aorta_field1.tex
\begin{tikzpicture}[scale=1.0, transform shape=false]
    \node at (0,0) {\includegraphics[width = 0.9\linewidth]{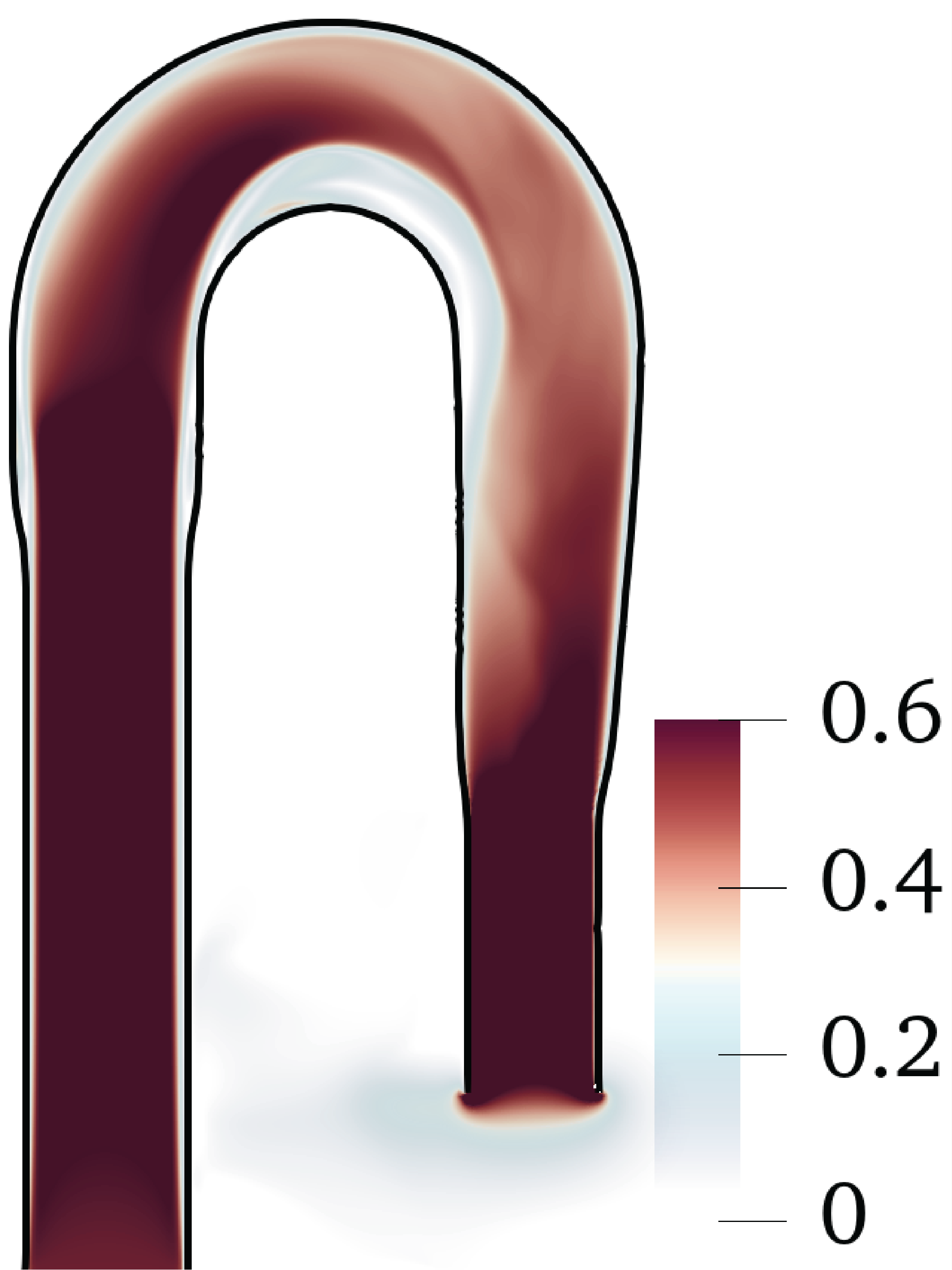}};
    \node[align=center] at (1.4,0.4) {$\abs{\vb{u}}$ \\ $\qty[\SI{}{\meter\per\second}]$};
\end{tikzpicture}

%% file: _figures/aorta/aorta_field2.tex
\begin{tikzpicture}[scale=1.0, transform shape=false]
    \node at (0,0) {\includegraphics[width = 0.9\linewidth]{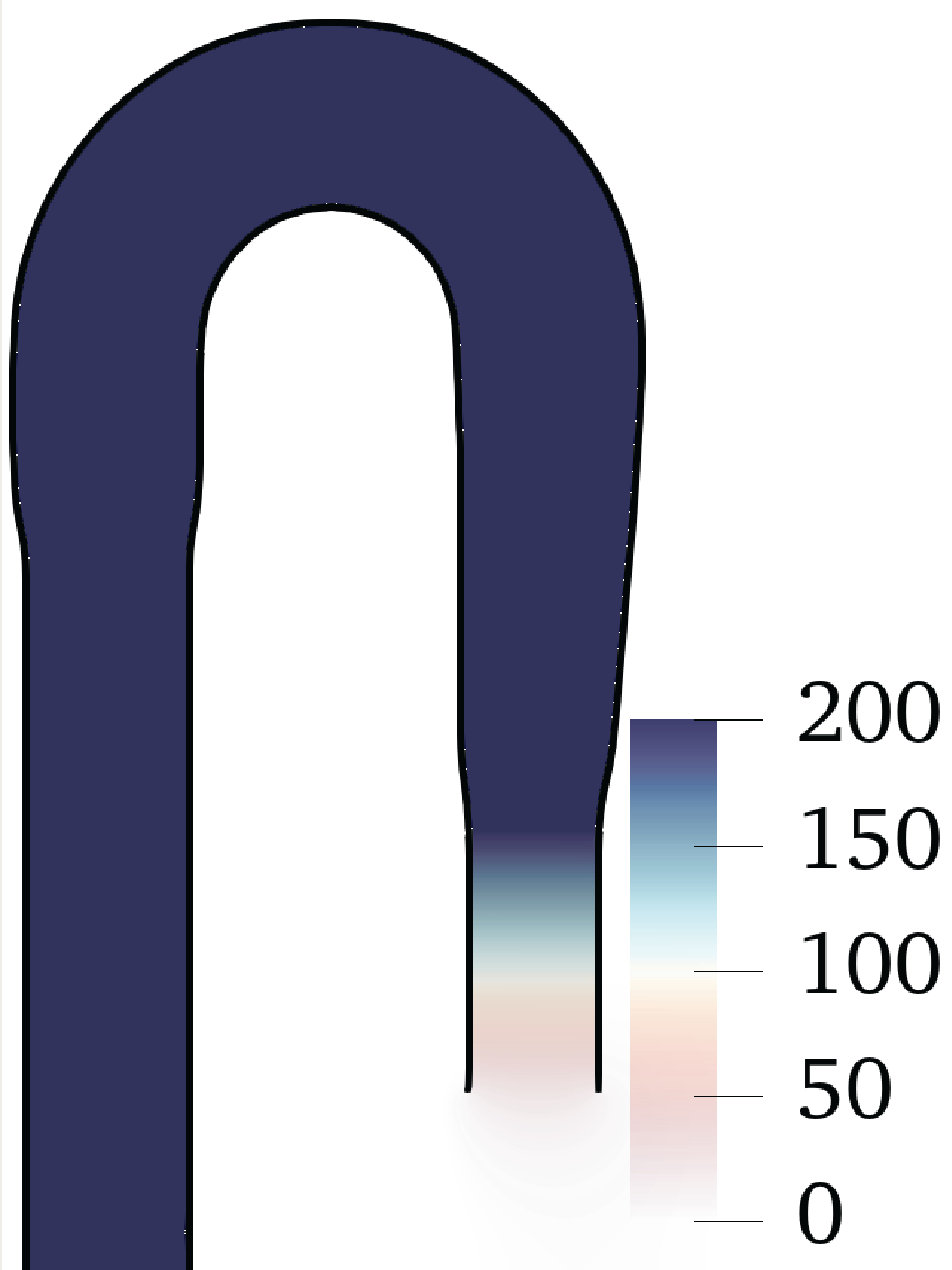}};
    \node [align = center] at (1.4,0.4) {$p$\\ $\qty[\SI{}{\milli \meter Hg}]$};
\end{tikzpicture}

%% file: _figures/aorta/aorta_plots_2.tex
\begin{tikzpicture}
\begin{groupplot}[
    group style={
        group size=3 by 1,
        horizontal sep=3.5em,
    },
    width = 0.36\textwidth,
    height = 0.8*0.36*\textwidth,
    trim axis left
]
\nextgroupplot[
                xmin = 0,
                xmax = 1,
                xtick={0,0.5,1},
                xlabel={$t/T$},
                ymin = -0.1,
                ymax = 1.1,
                xmajorgrids,
                ymajorgrids,
                axis lines=box,
                ylabel = {$U_{in}/U$},
                ytick={0,0.5,1},
                yticklabels={0,.5,1},
                ylabel style={rotate=0, font = \large, yshift = -0.2cm},
                title style={ at={(0,1)}},
                title = {(d)},
                tick label style={font=\footnotesize},
                ]            
\addplot [line width=1.5pt] table [y = u,x=t]{_data/aorta/inflow_data.txt};
\draw[fill = mycolor4]  (axis cs:0.26, 0.5525) circle(2pt) node[anchor = south west, black] {$Q$};

\nextgroupplot[
            xmin = 0,
            xmax = 1,
            xtick={0,0.5,1},
            xlabel={$t/T$},
            ymin = 0,
            ylabel = {$p$ [\SI{}{\milli\meter Hg]}},
            ytick = {0, 50, 100, 150, 200},
            xmajorgrids,
            ymajorgrids,
            axis lines=box,
            ylabel style={ yshift = -0.2cm},
            title style={ at={(0,1)}},
            title = {(e)},
            legend columns=3,
            legend to name = {leg:aorta},
            legend style={fill=none, draw=none},
            legend cell align={left},
            tick label style={font=\footnotesize},
            ]
\addplot [line width = 1.5pt,mycolor1] table [y=pr,x=t]{_data/aorta/pressure_sharpIB.txt};
\addplot [line width = 1.5pt,mycolor2] table [y=pr,x=t]{_data/aorta/pressure_mls.txt};
\addplot [line width = 1.5pt,mycolor2,dashed] table [y=pr,x=t]{_data/aorta/pressure_mls4.txt};
\addlegendentry{Present method}
\addlegendentry{L-MLS}
\addlegendentry{L-MLS (4 iterations)}

\nextgroupplot[
            xmin=0,
            xmax=1,
            xtick={0,0.5,1},
            xlabel={$t/T$},
            ymax = 1e-2,
            ytick={1e-6, 1e-5, 1e-4, 1e-3, 1e-2},
            ylabel = {$\norm{\boldsymbol{\varepsilon}}_1/U$},
            xmajorgrids,
            ymajorgrids,
            axis lines=box,
            ylabel style={rotate=0, font = \large, yshift = -0.2cm},
            title style={ at={(0,1)}},
            title = {(f)},
            tick scale binop=\times,
            ymode=log,
            tick label style={font=\footnotesize},
            ]
\addplot [line width = 1.5pt,mycolor1] table [y=s,x=t]{_data/aorta/error_sharpIB.txt};
\addplot [line width = 1.5pt,mycolor2] table [y=s,x=t]{_data/aorta/error_mls.txt};
\addplot [line width = 1.5pt,mycolor2,dashed] table [y=s,x=t]{_data/aorta/error_mls4.txt};

\end{groupplot}
\end{tikzpicture}

%% file: appendix.tex
\section{The discrete divergence theorem}
\label{app:discrete_divergence_theorem}
Let us consider a smooth vector field $\vb{u} = \qty[u_1,u_2,u_3]^T$ within the domain $\Omega$ introduced in Section \ref{sec:method}.
In this setting, the divergence theorem holds:
\begin{equation}
    \int_{\Omega}\div{\vb{u}}\dd{V} = \int_{\partial \Omega} \vb{u}\vdot\vb{n}\dd{S}.
\end{equation}
Introducing the staggered grid, and denoting with $\vb{c} = \qty(k_1,k_2,k_3)$ the Cartesian index of the cells, $k_i\in \qty[1, N_i]$, the integrals can be approximated using the midpoint rule, while the divergence operator is approximated by $D$, leading to:
\begin{equation}
    \begin{split}
        \int_{\Omega}\div{\vb{u}}\dd{V} &= \sum_{\vb{c}\in\mathcal{C}} D\vb{u}^{\vb{c}}\Delta V^{\vb{c}} +\order{\Delta_M^2},\\
         \int_{\partial \Omega} \vb{u}\vdot\vb{n}\dd{S} &= \sum_{i=1}^3\sum_{\vb{f}\in \mathcal{F}_i} u_i^{\vb{f}} n_i^{\vb{f}} \Delta S^{\vb{f}} + \order{\Delta_M^2},
    \end{split}
\end{equation}
being $\Delta_M$ the maximum grid size, $\Delta V^{\vb{c}}$ the volume of the cell, $\mathcal{F}_i$ the set of boundary faces of $\Omega$ with normal in the $i$-th direction, $n_i^{\vb{f}}$ their outer normal $i$-th component and $\Delta S^{\vb{f}}$ their surfaces.
In this setting, the \emph{discrete divergence theorem} holds true:
\begin{equation}
    \sum_{\vb{c}\in\mathcal{C}} D\vb{u}^{\vb{c}}\Delta V^{\vb{c}} = \sum_{i=1}^3\sum_{\vb{f}\in \mathcal{F}_i} u_i^{\vb{f}} n_i^{\vb{f}} \Delta S^{\vb{f}}.
\end{equation}
Importantly, the equality holds for any grid size and for any vector field up to machine precision when evaluated numerically.
Indeed, the discrete divergence on the cell $\vb{c}$ can be written as:
\begin{equation}
   D\vb{u}^{\vb{c}} = \sum_{i=1}^3\frac{u_i^{\vb{c}+\frac{1}{2}\vb{e}_i} - u_i^{\vb{c}-\frac{1}{2}\vb{e}_i}}{\Delta x_i^{\vb{c}}} = \frac{1}{\Delta V^{\vb{c}}}\sum_{i=1}^3 u_i^{\vb{c}+\frac{1}{2}\vb{e}_i}\Delta S_i^{\vb{c}+\frac{1}{2}\vb{e}_i} - u_i^{\vb{c}-\frac{1}{2}\vb{e}_i}\Delta S_i^{\vb{c}-\frac{1}{2}\vb{e}_i},
\end{equation}
being $\Delta S_i^{\vb{c}\pm\frac{1}{2}\vb{e}_i}$ the surface of the cell-face in the $i$-direction on the $\pm$ side.
Thus, integrating over the whole domain using the midpoint rule leads to:
\begin{equation}
    \begin{split}
        \sum_{\vb{c}\in\mathcal{C}} D\vb{u}^{\vb{c}}\Delta V^{\vb{c}} &=  \sum_{\vb{c}\in\mathcal{C}} \sum_{i=1}^3 u_i^{\vb{c}+\frac{1}{2}\vb{e}_i}\Delta S_i^{\vb{c}+\frac{1}{2}\vb{e}_i} - u_i^{\vb{c}-\frac{1}{2}\vb{e}_i}\Delta S_i^{\vb{c}-\frac{1}{2}\vb{e}_i} =\\
        &= \sum_{i=1}^3\sum_{\vb{c}\in\mathcal{C}} u_i^{\vb{c}+\frac{1}{2}\vb{e}_i}\Delta S_i^{\vb{c}+\frac{1}{2}\vb{e}_i} - u_i^{\vb{c}-\frac{1}{2}\vb{e}_i}\Delta S_i^{\vb{c}-\frac{1}{2}\vb{e}_i}= \sum_{i=1}^3 \sum_{\vb{f}\in \mathcal{F}_i} u_i^{\vb{f}}\Delta S_i^{\vb{f}}n_i^{\vb{f}}.
    \end{split}
\end{equation}
The last equality holds true since the sum is telescopic and only the terms on the boundary of $\Omega$ are not canceled.

\section{Evaluation of $q_F$}
\label{app:shape_q}
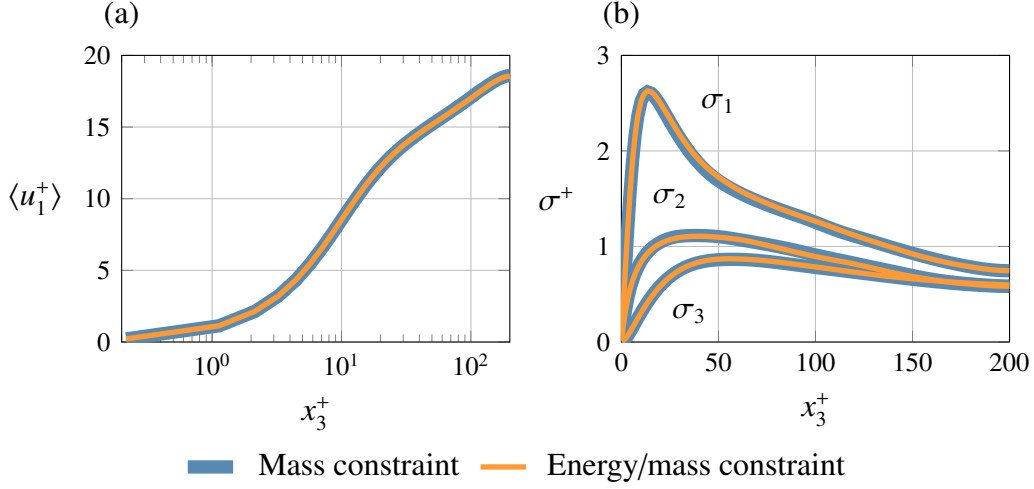
\begin{figure}[t]
    \raggedright
    \input{_figures/turbulent_channel/turbulent_channel_energy}
    \ref{leg:turbulent_channel_energy}
    \centering
    \caption{Comparison of the turbulent channel flow with a lower immersed surface with the mass and mass/energy constraints: (a) mean velocity profile and (b) turbulent intensities.}
    \label{fig:turbulent_channel_energy}
\end{figure}
As discussed in Section~\ref{sec:mod_elliptic}, the shape of $q_F$ is chosen to satisfy the constraint \eqref{eq:constraint_q} and having minimum $L_2$ norm over $\mathcal{C}_F$.
Accordingly, the Lagrangian of the problem is:
\begin{equation}
    \mathcal{L}\qty(q^k,\lambda_1) = \frac{1}{2}\sum_{k\in \mathcal{C}_F} \qty(q^k)^2\Delta V^k - \lambda_1 \qty[\sum_{k\in \mathcal{C}_F} q^k \Delta V^k  + \sum_{k\in \mathcal{C}_B} D\vb{u}^k \Delta V^k],
\end{equation}
where $\lambda_1$ is the Lagrange multiplier associated to the mass constraint and $q_F$ is determined by setting $\pdv*{\mathcal{L}}{q^k} = 0$ and $\pdv*{\mathcal{L}}{\lambda_1} = 0$, leading to:
\begin{equation}
    \label{eq:energy_shape}
    q_F  =-\frac{1}{\abs{\mathcal{C}_F}}\sum_{k\in \mathcal{C}_B} D\hat{\vb{u}}^k \Delta V^k.
\end{equation}
This expression for $q_F$ assures the conservation of mass throughout the simulation within the whole domain as in standard fractional step.

Alternatively, another constraint can be added to the previous Lagrangian to also enforce the conservation of kinetic energy in the inviscid limit within the whole domain.
The total kinetic energy equation in the inviscid limit reads:
\begin{equation}
    \dv{E}{t} = - \int_{\Omega} \vb{u}\vdot\qty(\div{\vb{u}\vb{u}})\dd{V} -  \int_{\Omega} \vb{u}\vdot\grad{p}\dd{V},
\end{equation}
being $e= \vb{u}\vdot\vb{u}/2$ the local kinetic energy and $E = \int_{\Omega} e\dd{V}$ the total kinetic energy.
Using integration by parts, the previous equation can be rewritten as follows:
\begin{equation}
    \begin{split}
        \dv{E}{t} &= - \int_{\Omega} e\div{\vb{u}}\dd{V} - \int_{\Omega} \div{\qty(e\vb{u})}\dd{V} -  \int_{\Omega} \div{\qty(p\vb{u})}\dd{V} + \int_{\Omega} p\div{\vb{u}}\dd{V} =\\
        &=\int_{\Omega} \qty(p-e)\div{\vb{u}}\dd{V} -  \int_{\Omega} \div{\qty[\qty(e+p)\vb{u}]}\dd{V}.
    \end{split}
\end{equation}
The last term is conservative and acts as a redistribution of kinetic energy.
On the other hand, if $\div{\vb{u}} = q \neq 0$, the first integral does not vanish, and kinetic energy is globally conserved only if the following constraint holds:
\begin{equation}
    \int_{\Omega} \qty(p-e)q\dd{V} = 0 \quad \forall (p-e),
\end{equation}
which in discrete form reads:
\begin{equation}
    \sum_{k\in \mathcal{C}_F} \qty(p^k-e^k)q_F^k\Delta V^k + \sum_{k\in \mathcal{C}_B} \qty(p^k-e^k)D\vb{u}^k\Delta V^k = 0.
\end{equation}
Thus, the following augmented Lagrangian is considered:
\begin{equation}
    \begin{split}
        \mathcal{L}\qty(q^k,\lambda_1,\lambda_2) &= \frac{1}{2}\sum_{k\in \mathcal{C}_F} \qty(q^k)^2\Delta V^k- \lambda_1 \qty[\sum_{k\in \mathcal{C}_F} q^k \Delta V^k  + \sum_{k\in \mathcal{C}_B} D\vb{u}^k \Delta V^k] -\\
        &- \lambda_2\qty[\sum_{k\in \mathcal{C}_F}\qty(p^k-e^k)q^k\Delta V^k + \sum_{k\in \mathcal{C}_B} \qty(p^k-e^k)D\vb{u}^k\Delta V^k],
    \end{split}
\end{equation}
being $\lambda_2$ the Lagrange multiplier associated to the energy constraint.
Setting $\pdv*{\mathcal{L}}{q^k} = 0$, $\pdv*{\mathcal{L}}{\lambda_1} = 0$ and $\pdv*{\mathcal{L}}{\lambda_2} = 0$ leads to:
\begin{equation}
    \label{eq:unif_shape}
    q_F^k  =\frac{-A S_2 + B S_1}{C} + \frac{A S_1 - B S_0}{C}\qty(p^k - e^k),
\end{equation}
being $S_i = \sum_{k\in \mathcal{C}_F} \qty(p^k - e^k)^i \Delta V^k$, $A = \sum_{k\in \mathcal{C}_B} D\vb{u}^k \Delta V^k$, $B = \sum_{k\in \mathcal{C}_B} \qty(p^k-e^k)D\vb{u}^k\Delta V^k$ and $C = S_0 S_2 - S_1^2$.

To assess the effect of the shape of $q_F$, the turbulent channel flow with a lower immersed surface described in Section~\ref{sec:turbulent_channel} is simulated using the mass-conserving (Equation~\eqref{eq:unif_shape}) and the  mass/energy-conserving (Equation~\eqref{eq:energy_shape}) shapes of $q_F$.
The resulting mean velocity profile and turbulent intensities are shown in Figure~\ref{fig:turbulent_channel_energy}.

\section{Scaling of truncation errors}\label{app:scaling}
The modified elliptic Equation~\eqref{eq:elliptic_mod} incorporates a distribution of sinks/sources on the fluid cells, which vanish as the grid is refined.
Indeed, with reference to the setting of Section~\ref{sec:newIB_meth}, the truncation error on the regular cells can be casted as follows:
\begin{equation}
    R_F^k = \frac{\Delta^2}{24}\sum_{i=1}^3\pdv[3]{u_i}{x_i}\bigg|_k + \order{\Delta^3} = \frac{\Delta^2}{24}\div{\qty(\mathcal{D}_2\vb{u})}^k + \order{\Delta^3} = \frac{\Delta^2}{24}D{\qty(\mathcal{D}_2\vb{u})^k} + \order{\Delta^3},
\end{equation}
being $\mathcal{D}_2\vb{u} = \qty[\pdv*[2]{u_1}{x_1},\pdv*[2]{u_2}{x_2},\pdv*[2]{u_3}{x_3}]^T$, and the last equality holding true since $R_F$ is defined on the regular cells.
Thus, using the discrete divergence theorem, the discrete integral of $R_F$ over the regular cells is given by:
\begin{equation}
    \sum_{k\in\mathcal{C}_F} R_F^k \Delta V^k = \frac{\Delta^2}{24} \sum_{f\in \mathcal{F}_F} \mathcal{D}_2\vb{u}^f\vdot \vb{n}^f \Delta S^f + \order{\Delta^3},
\end{equation}
being $\mathcal{F}_F$ the boundary faces of $\mathcal{C}_F$.
The term $\mathcal{D}_2\vb{u}\vdot \vb{n}$ is linked to the hydrodynamic stress acting on $\partial \mathcal{C}_F$, hence, at convergence, it is fixed by the specific flow under investigation, leading to $\mathcal{D}_2\vb{u}\vdot \vb{n} = \order{1}$.
Therefore, the mean value of $R_F$ is given by:
\begin{equation}
    \overline{R}_F = \frac{1}{\abs{\mathcal{C}_F}}\sum_{k\in\mathcal{C}_F} R_F^k \Delta V^k =\Delta ^2\underbrace{\frac{\overline{\mathcal{D}_2\vb{u}\vdot \vb{n}}\,\abs{\partial\mathcal{C}_F}}{24\abs{\mathcal{C}_F}}}_{\order{1}} = \order{\Delta^2},
\end{equation}
being $\abs{\mathcal{C}_F}$ the volume of the fluid cells and $\abs{\partial\mathcal{C}_F}$ the total area of the boundary faces of the fluid cells.
Thus, from Equation~\eqref{eq:constraint}, the following estimation for the mean value of $R_B$ holds:
\begin{equation}
    \overline{R}_B = \frac{1}{\abs{\mathcal{C}_B}}\sum_{k\in\mathcal{C}_B} R_B^k \Delta V^k = \order{\Delta}.
\end{equation}
The same order of convergence is inherited by $q_B$ and $q_F$.
Indeed, assuming that on the boundary cells $D\hat{\vb{u}}$ is a good approximation of $D\vb{u}$, then $\overline{q}_B$ is expected to scale as $\overline{R}_B$, thus, in force of Equation~\eqref{eq:constraint_q} follows that:
\begin{equation}
    q_F = \overline{q}_F = \order{\Delta ^2},
\end{equation}
showing that $q_F \rightarrow 0$ as $\Delta \rightarrow 0$ with the same order of $R_F$.

\section{Spring network formulation of surface tension} \label{app:surface_tension}
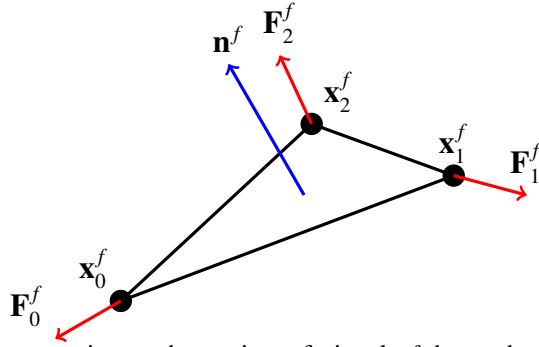
\begin{figure}
    \centering
    \input{_figures/general/surface_tension_scheme}
    \caption{Forces acting on the vertices of triangle $f$ due to the surface tension $\gamma$.}
    \label{fig:surface_tension_scheme}
\end{figure}
The potential energy associated to a body of surface area $S$ and surface tension $\gamma$ is given by:
\begin{equation}
    V_\gamma = \gamma S.
\end{equation}
Given a triangular spring network made of $N_v$ vertices, $N_f$ faces and $N_e$ edges, the potential becomes:
\begin{equation}
    V_\gamma = \gamma\sum_{f=1}^{N_f} S^f,
\end{equation}
being $S^f$ the surface area of each triangle, given by $\abs{\qty(\vb{x}^f_1-\vb{x}^f_0)\cross\qty(\vb{x}^f_2-\vb{x}^f_0)}/2$, where $\vb{x}_i^f$, $i=0,1,2$ are the coordinates of the vertices of face $f$ (see Figure~\ref{fig:surface_tension_scheme}).
The force acting on the node $i$ of triangle $f$ due to the surface tension is given by the variation of the potential $\gamma S^f$ with respect to the variation of the position of its vertices:
\begin{equation}
    \vb{F}_i^f = -\gamma\fdv{S^f}{\vb{x}^f_i}.
\end{equation}
The first variation of $S^f$ can be expressed as $\var{S^f} = \vb{n}^f\vdot \var{\vb{N}^f}/2$, being $\vb{N}^f = \qty(\vb{x}^f_1-\vb{x}^f_0)\cross\qty(\vb{x}^f_2-\vb{x}^f_0)$ and $\vb{n}^f = \vb{N}^f/\abs{\vb{N}^f}$.
From the definition of $\vb{N}^f$ follows that:
\begin{equation}
    \var{\vb{N}^f} = \qty( \var{\vb{x}^f_1}-\var{\vb{x}^f_0})\cross \qty(\vb{x}^f_2-\vb{x}^f_0) + \qty(\vb{x}^f_1-\vb{x}^f_0)\cross \qty(\var{\vb{x}^f_2}-\var{\vb{x}^f_0}) .
\end{equation}
Using the triple product properties is possible to obtain the following expression for the variations of $S^f$:
\begin{equation}
    \fdv{S^f}{\vb{x}^f_0} = \frac{1}{2} \qty(\vb{x}_1^f - \vb{x}_2^f) \cross \vb{n}^f, \quad
    \fdv{S^f}{\vb{x}^f_1} = \frac{1}{2} \qty(\vb{x}_2^f - \vb{x}_0^f) \cross \vb{n}^f,\quad
    \fdv{S^f}{\vb{x}^f_2} = \frac{1}{2} \qty(\vb{x}_0^f - \vb{x}_1^f) \cross \vb{n}^f.
\end{equation}
Thus, the force acting on a vertex is given by the sum of the contributes that each face provides to such node.

%% file: _figures/turbulent_channel/turbulent_channel_energy.tex
\begin{tikzpicture}
\begin{groupplot}[
    group style={
        group size=2 by 1,
        horizontal sep=3.5em,
    },
    width = 0.49\textwidth,
    height = 0.8*0.49*\textwidth,
    trim axis left
]
\nextgroupplot[
            xmax=200,
            xmin=0.2,
            xlabel={$x_3^+$},
            ymin = 0,
            ymax = 20,
            ytick={0,5,10,15, 20},
            ylabel = {$\langle u_1^+\rangle$},
            xmajorgrids,
            ymajorgrids,
            axis lines=box,
            xmode=log,
            legend columns=2,
            legend to name = {leg:turbulent_channel_energy},
            legend style={fill=none, draw=none, column sep = 5pt},
            legend cell align={left},
            axis lines=box,
            ylabel style = {rotate = -90},
            title style={ at={(0,1)}},
            title = {(a)},
            tick label style={font=\footnotesize},
            ]
\addplot [line width = 5pt,mycolor1] table [y=um,x=z]{_data/turbulent_channel/IB_data_avg.txt};
\addplot [line width = 2pt,mycolor4] table [y=um,x=z]{_data/turbulent_channel/IB_data_avg_energy.txt};
\addlegendentry{Mass constraint}
\addlegendentry{Energy/mass constraint}

\nextgroupplot[
            xmin=0,
            xmax=200,
            xlabel={$x_3^+$},
            ymax = 3,
            ymin = 0.0,
            ytick={0,1,2,3},
            ylabel = {$\sigma^+$},
            xmajorgrids,
            ymajorgrids,
            axis lines=box,
            ylabel style = {rotate = -90},
            title style={ at={(0,1.0)}},
            title = {(b)},
            tick label style={font=\footnotesize},
            ]
\addplot [line width = 5pt, mycolor1] table [y=rmsu,x=z]{_data/turbulent_channel/IB_data_avg.txt};
\addplot [line width = 5pt, mycolor1] table [y=rmsv,x=z]{_data/turbulent_channel/IB_data_avg.txt};
\addplot [line width = 5pt, mycolor1] table [y=rmsw,x=z]{_data/turbulent_channel/IB_data_avg.txt};
\addplot [line width = 2pt,mycolor4] table [y=rmsu,x=z]{_data/turbulent_channel/IB_data_avg_energy.txt};
\addplot [line width = 2pt,mycolor4] table [y=rmsv,x=z]{_data/turbulent_channel/IB_data_avg_energy.txt};
\addplot [line width = 2pt,mycolor4] table [y=rmsw,x=z]{_data/turbulent_channel/IB_data_avg_energy.txt};

\node at (axis cs:50,2.5) {$\sigma_1$};
\node at (axis cs:25,1.5) {$\sigma_2$};
\node at (axis cs:35,0.3) {$\sigma_3$};
\end{groupplot}
\end{tikzpicture}

%% file: _figures/general/surface_tension_scheme.tex
\begin{tikzpicture}[scale=2.0, transform shape=false]
    \pgfmathsetmacro{\angv}{160};
    \pgfmathsetmacro{\angn}{120};
    \pgfmathsetmacro{\ang}{30};
    
    \coordinate (X1) at (0, 0);
    \coordinate (X2) at ($(X1) + (\angv:1)$);
    \coordinate (XM) at ($0.5*(X1) + 0.5*(X2)$);
    \coordinate (X0) at ($(XM) - (\ang:2)$);
    \coordinate (C)  at (1.2,0.2);
    
    \draw[very thick] (X0) node[anchor = south east] {$\vb{x}_0^f$} -- (X1) node[anchor = south] {$\vb{x}^f_1$} -- (X2) node[ anchor = south west] {$\vb{x}_2^f$} -- (X0);
    
    \filldraw (X0) circle (2pt);
    \filldraw (X1) circle (2pt);
    \filldraw (X2) circle (2pt);
    
    \draw[very thick, blue, ->] ($(XM) - (\ang:0.6)$) -- +(\angn:1) node[color = black, anchor = south] {$\vb{n}^f$};
    
    \draw[very thick, red, ->] (X0) -- +(210:0.5)    node[color = black, anchor = south east] {$\vb{F}^f_0$};
    \draw[very thick, red, ->] (X1) -- +(-15:0.5)    node[color = black, anchor = south] {$\vb{F}^f_1$};
    \draw[very thick, red, ->] (X2) -- +(115 :0.5) node[color = black, anchor = south] {$\vb{F}^f_2$};
    
\end{tikzpicture}